%% file: Article_J0901_main.tex
\pdfoutput=1
\documentclass[twocolumn]{aastex63} 
\input{Input_usepackage}

\input{Input_newcommand}

\usepackage{graphicx}
\graphicspath{{Input_figures/}}

\accepted{\today}
\submitjournal{ApJ}
\shorttitle{J0901 CO and H$\alpha$ Kinematics and Ring}
\shortauthors{D. Liu et al.}

\begin{document}

\title{A $\sim$600~pc view of the strongly-lensed, massive main sequence galaxy J0901: a baryon-dominated, thick turbulent rotating disk with a clumpy cold gas ring at $z = 2.259$}

\input{Input_authors}

\begin{abstract}
We present a high-resolution kinematic study of the massive main-sequence star-forming galaxy (SFG) SDSS~J090122.37+181432.3 (J0901) at $z=2.259$, using $\sim 0.36''$ ALMA CO(3--2) and $\sim 0.1 \text{--} 0.5''$ SINFONI/VLT H$\alpha$ observations. 
J0901 is a rare, strongly-lensed but otherwise normal massive ($\log (\Mstar/\Msun) \sim 11$) main sequence SFG, offering a unique opportunity to study a typical massive SFG under the microscope of lensing. 
Through forward dynamical modeling incorporating lensing deflection, we fit the CO and \Halpha{} kinematics in the image plane out to about one disk effective radius ($\Reff \sim 4$~kpc) at a $\sim 600$~pc delensed physical resolution along the kinematic major axis. 
Our results show high intrinsic dispersions of the cold molecular and warm ionized gas ($\sigma_{0,\,\mathrm{mol.}} \sim 40$~km/s and $\sigma_{0,\,\mathrm{ion.}} \sim 66$~km/s) that remain constant out to $\Reff$; a moderately low dark matter fraction ($\fDM \sim 0.3$--0.4) within $\Reff$; and a centrally-peaked Toomre $Q$-parameter --- agreeing well with the previously established $\sig0$ vs. $z$, $\fDM$ vs. $\Sigbar$, and $Q$'s radial trends using large-sample non-lensed main sequence SFGs. 
Our data further reveal a high stellar mass concentration within $\sim 1$--2~kpc with little molecular gas, and a clumpy molecular gas ring-like structure at $R \sim 2$--4~kpc, in line with the inside-out quenching scenario. 
Our further analysis indicates that J0901 had assembled half of its stellar mass only $\sim 400$~Myrs before its observed cosmic time, and cold gas ring and dense central stellar component are consistent with signposts of a recent wet compaction event
of a highly turbulent disk found in recent simulations. 
\end{abstract}


\section{Introduction}

In galaxy formation and evolution theories, massive star-forming galaxies (SFGs) form gas-rich, turbulent disks at high redshift via cold gas stream accretion from the circumgalactic medium (e.g., \citealt{Dekel2006,Dekel2009b,Dekel2009a}). 
Cold streams penetrate through the hot dark matter halo and transport cold gas and angular momentum inwards, feeding the growth of disks, bulges and giant clumps (e.g., \citealt{Bournaud2007,Ceverino2010,Danovich2015}). 
At $z\sim2$--3, the disk intrinsic dispersion ($\sig0$) is anticipated to higher than at lower-$z$ with stronger cold streams and increased gas fraction and disk instability (e.g., \citealt{Krumholz2010c,Krumholz2016a,Krumholz2018b}). 
Observationally, the evolution of disks' $\sig0$ and instability (i.e. the Toomre $Q$-parameter; \citealt{Toomre1964}) in main sequence SFGs at $z\sim1$--3 has been mostly studied with kpc-scale kinematics of ionized gas tracers (e.g., \citealt{ForsterSchreiber2006,Genzel2006,Genzel2008,Genzel2011,Kassin2012,Wisnioski2015,Simons2017,Johnson2018,Girard2018,Ubler2019,Girard2021}). 
There are still very limited studies that have both high spatial resolution cold and ionized gas kinematics in high-$z$ massive SFGs (see compilations in \citealt{Ubler2019} and \citealt{Girard2021}), and almost none can probe down to sub-kpc scale in both gas phases.

It is critical to probe the high-$z$ massive SFG disks at a kpc (about the Toomre scale at $z\sim2$--3; see e.g. \citealt{Escala2008,Genzel2008,Genzel2011}) or even better resolution, to investigate the disk instability, star formation, feedback and quenching physics. 
Massive SFGs show the strongest signatures for cold stream accretion, mass assembly and feedback, and are prime targets for investigating internal physics of galaxy formation.

Near-IR Integral Field Unit (IFU) spectroscopic and (sub-)mm interferometric imaging (e.g., with the Atacama Large Millimeter/Submillimeter Array, ALMA, and the Northern Extended Millimeter Array, NOEMA) are important techniques to spatially-resolve the ionized and cold gas kinematics of high-$z$ SFG disks (see review by \citealt{ForsterSchreiber2020}). 
Near-IR IFU observations usually reach an angular resolution of $\sim 0.5'' \text{--} 0.7''$ ($\sim 4$--6~kpc at $z \sim 2$) under natural seeing, and $\sim 0.1'' \text{--} 0.2''$ ($\sim 1$--2~kpc at $z \sim 2$) with the assistance of Adaptive Optics (AO). 
To date, $\sim 200$ high-redshift SFGs have been observed with AO-assisted IFU spectrographs, spanning $z \sim 0.8 \text{--} 3.7$ and $\log (\Mstar/\Msun) \sim 9.5 \text{--} 11.5$ (\citealt{ForsterSchreiber2018,ForsterSchreiber2020}; and references therein). 
There are $\sim 80$ strongly-lensed SFGs among the AO samples (e.g., \citealt{Jones2010,Livermore2015,Leethochawalit2016,Sharma2018,Hirtenstein2019}).
However, strong-lensing samples tend to be intrinsically lower-mass systems ($\log (\Mstar/\Msun) \sim 8.0 \text{--} 10.5$) whose number density is orders of magnitude higher than the most massive SFGs (see stellar mass functions, e.g., \citealt{Davidzon2017}). 
Strongly-lensed SFGs that can represent massive galaxies are still very rare.

Meanwhile, increasing numbers of ALMA and NOEMA data sets now probe the cold gas kinematics in high-$z$ galaxies, 
but very few of them were obtained at resolutions of $\lesssim 0.5 \text{--} 0.6''$ and with deep integrations for massive main-sequence SFGs at $z \sim 1 \text{--} 3$ (see \citealt{Genzel2013,Ubler2018,HerreraCamus2019} for examples with also resolved ionized gas kinematics).

In this work, we present new $\sim 4$-hour on-source integration, high-resolution ($\sim 0.36''$) ALMA CO(3--2) observations of a rare, massive ($\log [\Mstar/\Msun] \sim 11$), strongly-lensed, main-sequence SFG SDSS~J090122.37+181432.3 (hereafter J0901) at $z = 2.259$ (\citealt{Diehl2009,Hainline2009}). 
Together with the AO-assisted SINFONI/VLT observations previously published by \citet{Davies2020a}, 
we study the rotation curves and velocity dispersions of both cold molecular and warm ionized gas in J0901, at $\sim 600$~pc delensed resolution, via direct image-plane kinematic fitting, and examine the Toomre stability and gas properties across the galaxy.

This paper is organized as follows. 
Target and observation properties are presented in Sect.~\ref{sec: data}. 
Methods to obtain the delensed stellar and molecular mass maps, dynamical modeling and image-plane kinematic fitting are given in Sect.~\ref{sec: analysis}. 
Main scientific results and discussions are in Sect.~\ref{sec: results}, including rotation curves, gas velocity dispersions, dark matter fractions, Toomre $Q$ distribution, and a cold gas ring in J0901. 
Finally, we conclude in Sect.~\ref{sec: summary}. 
In addition, 
Table~\ref{Table1} summarizes the key results of J0901. 
A gallery of all our data products is shown in Appendix~\ref{appx: all products}. 
More details about the astrometry correction, lens modeling, delensing method, line map extraction can be found in Appendix~\ref{appx: analysis}.

We adopt a flat $\Lambda$CDM cosmology with $H_0 = 70 \ \mathrm{km \, s^{-1} \, Mpc^{-1}}$, $\Omega_{\mathrm{M}} = 0.3$ and $\Omega_{\Lambda} = 0.7$, and a \citet{Chabrier2003} initial mass function (IMF).

\vspace{2ex}

\section{Target and Data}
\label{sec: data}

J0901 was identified in the Sloan Digital Sky Survey (SDSS) data by \cite{Diehl2009}. 
It is lensed by a foreground galaxy cluster at $z=0.346$ into three main parts: a highly-distorted, partially-lensed northeast (NE) arc, a less-distorted but completely-lensed southeast (SE) arc, and the least-distorted west (W) image (\citealt{Fadely2010,Tagore2014,Sharon2019,Davies2020a}). 

J0901's intrinsic total stellar mass and SFR are $\log (M_{\star} / \Msun) \sim 11.2$ and $\mathrm{SFR} \sim 200 \, \Msyr$ (\citealt{Davies2020a}), placing it on the star-forming main sequence. 
An active galactic nucleus (AGN) has been identified at the galaxy center by the high \NIIHA{} and \OIIIHB{} line ratios and the \Nfive{} line detection from rest-frame UV and optical spectroscopic observations (\citealt{Hainline2009,Diehl2009}), with an AGN-driven outflow rate of $\sim 25 \pm 8 \, \Msyr$ (\citealt{Davies2020a}). 

The cold molecular gas in J0901 has been studied by \citet{Saintonge2013} using Very Large Array (VLA) CO(1--0) and IRAM Plateau de Bure Interferometer (PdBI) $\sim3.5''$ CO(3--2) data, and \citet{Rhoads2014} using \textit{Herschel} HIFI spectrometer for the global \Ctwo{} emission, as well as \citet{Sharon2019} with $\sim 1.33'' \times 0.98''$ PdBI CO(3--2) data. 

Here we present new ALMA CO(3--2) data (PI: D. Lutz; project code: 2016.1.00406.S) observed at a $3 \times$ higher angular resolution than the previous CO(3--2) observation (Sect.~\ref{subsec: ALMA}). 
We also adopt the SINFONI \Halpha{}+\Ntwo{} AO and non-AO data from \citet{Davies2020a} for our ionized gas kinematic study (Sect.~\ref{subsec: SINFONI}). 
In addition, we use archival \textit{HST} images for SED fitting and lens modeling, and an $1.4'' \times 0.9''$ ALMA 1mm observation (PI: C. Sharon; project code: 2013.1.00952.S) for visual comparison of CO and dust (Appendix~\ref{appx: all products}). 

\begin{figure*}[htb]
\centering%
\includegraphics[width=\textwidth]{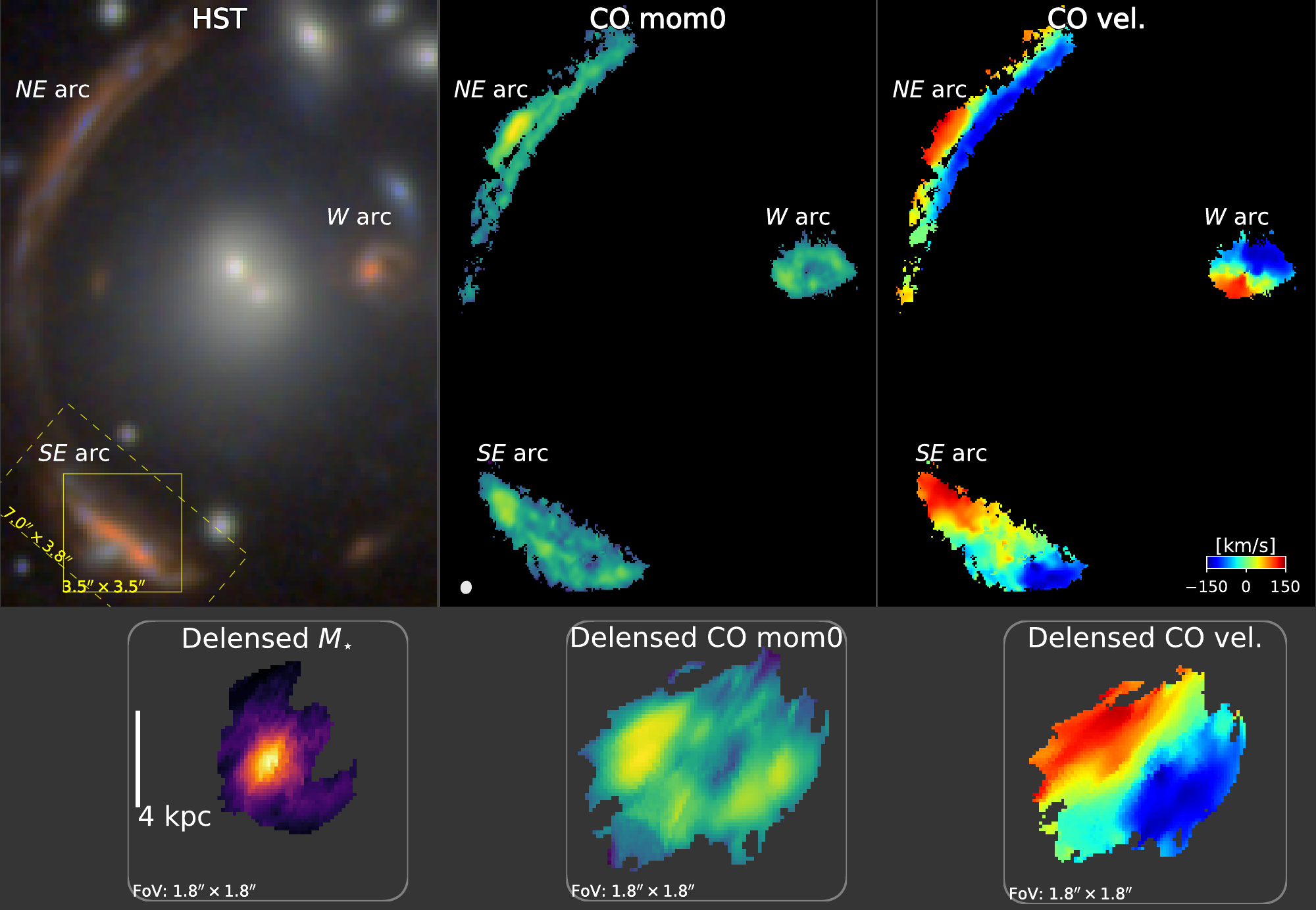}
\vspace{1ex}
\caption{%
\textit{Upper panels:} J0901 \textit{HST} RGB image (\textit{left}; F814W/F110W/F160W), ALMA CO(3--2) line integrated intensity (\textit{middle}), and CO(3--2) velocity map (\textit{right}), with the same field of view of $13 \times 18''$. 
\textit{Lower panels:} Delensed stellar mass (\textit{left}; derived from SED fitting, see Sect.~\ref{subsec: mass distribution}), CO(3--2) line integrated intensity (\textit{middle}) and velocity map (\textit{right}), with the same field of view of $1.8 \times 1.8''$. 
The delensed image is constructed with the SE arc which is the most magnified and completely lensed image of J0901. 
The solid and dashed yellow boxes at the bottom left of the upper left panel indicate the valid areas of our SINFONI/VLT AO and non-AO data, respectively. 
The small ellipse at the lower left corner of the upper middle panel shows the angular resolution of the CO data.
All panels have north to the up and east to the left. 
\label{fig: J0901 full area}
}\vspace{2ex}
\end{figure*}

In Fig.~\ref{fig: J0901 full area}, we show the \textit{HST} false-color image of the foreground lenses and J0901, and ALMA CO(3--2) line intensity and velocity maps (only from J0901) in the upper panels, together with the delensed stellar mass, CO intensity and velocity maps in the bottom panels.

\vspace{2ex}
\input{Input_table_1}
\vspace{1ex}

\subsection{ALMA CO(3--2) observations and data reduction}
\label{subsec: ALMA}

Our observing program 2016.1.00406.S was executed on Nov. 20, 2016 and during Aug. 02--17, 2017, with two array configurations corresponding to an angular resolution of $\sim 0.98''$ and $0.22''$ (baseline ranges 15--704~m and 21--3300~m), respectively. 
The raw data are reduced with the standard observatory calibration pipeline using the Common Astronomy Software Applications (CASA) software package (version 4.7.2). The calibrated visibilities are then continuum subtracted and binned to a channel width of $\sim 22$~km/s.

The imaging and primary beam correction of the visibilities were done within CASA version 5.5.0-149 using the \textsc{tclean} task. 
We produced a Briggs-weighting cube with a robust parameter of 0.5 to balance the angular resolution and sensitivity, and a natural-weighting cube to maximize the S/N but with a degraded resolution. 
We cleaned down to twice the RMS noise iteratively measured in a previously cleaned residual cube. The achieved Briggs-weighting synthesized beam is $0.40'' \times 0.336''$ at a position angle of $-11^{\circ}$, and natural-weighting synthesized beam $0.58 \times 0.51''$, with the latter having $\sim$20\% lower noise. 
We focus on the higher-resolution Briggs-weighting data in this work.

We create line integrated intensity, line center velocity and line width (dispersion) maps via pixel-by-pixel Markov Chain Monte Carlo (MCMC)-based 1D-Gaussian line profile fitting (see Appendix~\ref{appx: line fitting}).

\subsection{VLT SINFONI $K$-band IFU data}
\label{subsec: SINFONI}

The effective footprints of the SINFONI/VLT AO and non-AO data from \cite{Davies2020a} are shown as the yellow boxes in Fig.~\ref{fig: J0901 full area}. They cover the SE arc with a PSF FWHM of $\sim 0.2''$ and $\sim 0.5''$, and on-source integration time of $\sim 10$ and $\sim 9$~hours, respectively. 
The $K$-band grating was used to cover the \Halpha{} and \Ntwo{} doublet lines, which has line spread function (LSF) FWHM of $\sim 85\,\kms$. 
We refer the reader to \citet{Davies2020a} and \citet{ForsterSchreiber2009,ForsterSchreiber2014,ForsterSchreiber2018} for more details of the observation and data reduction. 

We combine the AO and non-AO data into one data cube so that our kinematic fitting can use both the sharper AO data for the inner rapidly rising rotation curve, and the wider non-AO data for the outer part. 
We tested various combination methods and found that they do not obviously affect our kinematic analysis (Appendix~\ref{appx: kinematic fitting for all}).

The line integrated intensity, velocity and velocity dispersion maps are created in a similar approach as for the CO data, 
but with broad-line outflow components subtracted (Appendix~\ref{appx: line fitting}), given the strong, marginally resolved AGN-driven outflows as characterized in \citet{Genzel2014a} and \citet{Davies2020a}. 
In the remainder of this work, we use only the narrow line component, i.e., outflow-subtracted \Halpha{}, for further analysis.
The LSF broadening is also corrected by subtracting the Gaussian $\sigma$ of the LSF, $36.1\,\kms$, in quadratic from the measured velocity dispersion along each line of sight.

\vspace{2ex}

\section{Delensed Data and Dynamical Modeling}
\label{sec: analysis}

\begin{figure*}[htb]
\centering%
\includegraphics[width=0.95\textwidth]{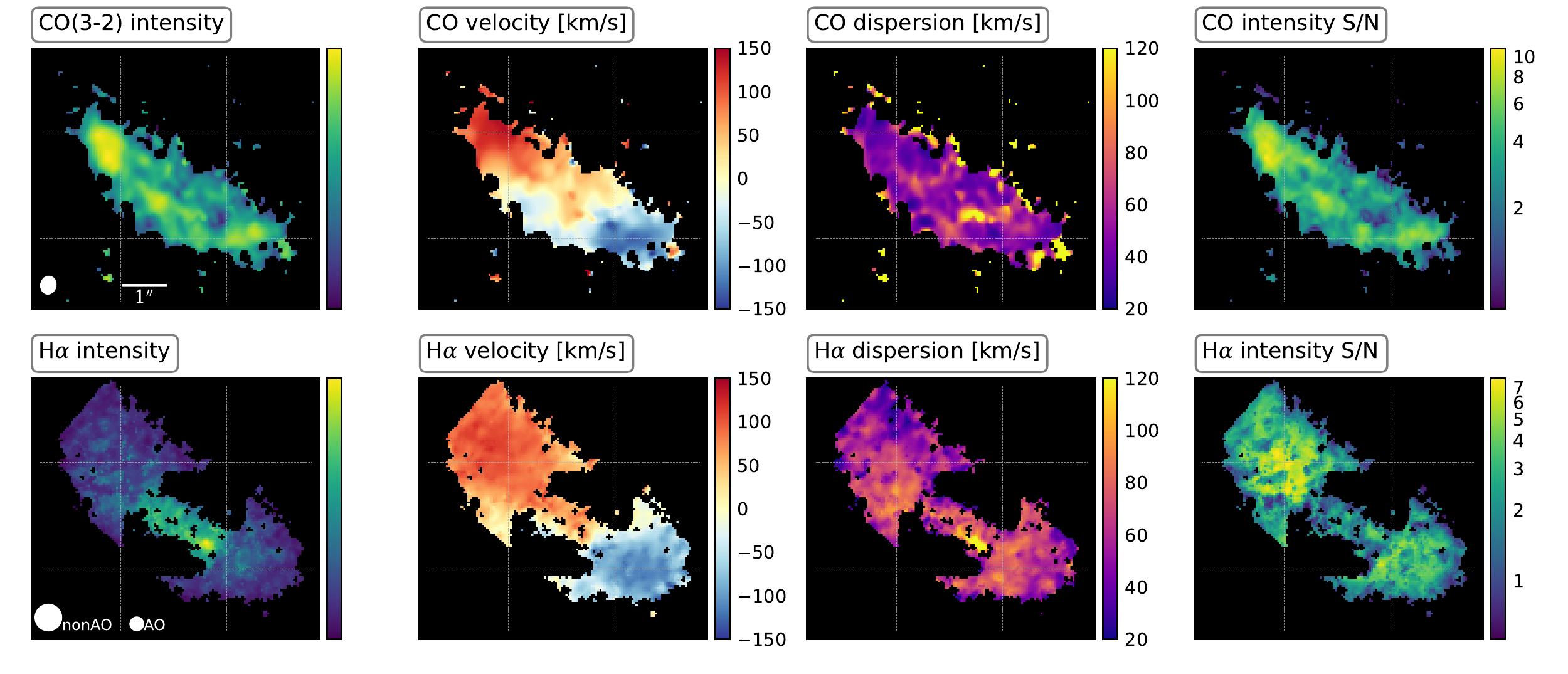}
\vspace{0.2ex}
\caption{%
Image-plane CO (\textit{upper}) and \Halpha{} (\textit{lower}) line intensity, velocity, velocity dispersion (corrected for LSF; Sect.~\ref{subsec: SINFONI}), and line intensity S/N maps of the J0901 lensed SE arc, from left to right, respectively. 
The PSF is shown at the lower left corner of each line intensity panel (for \Halpha{}, both non-AO and AO PSFs are indicated). 
The \Halpha{} maps are outflow broad-line removed via pixel-by-pixel multi-component spectral line fitting (see Appendix~\ref{appx: line fitting}). 
All panels have the same field of view of $6.775'' \times 6.135''$ and north is up. 
\label{fig: image plane images}
}
\vspace*{\floatsep}
\centering%
\includegraphics[width=0.95\textwidth]{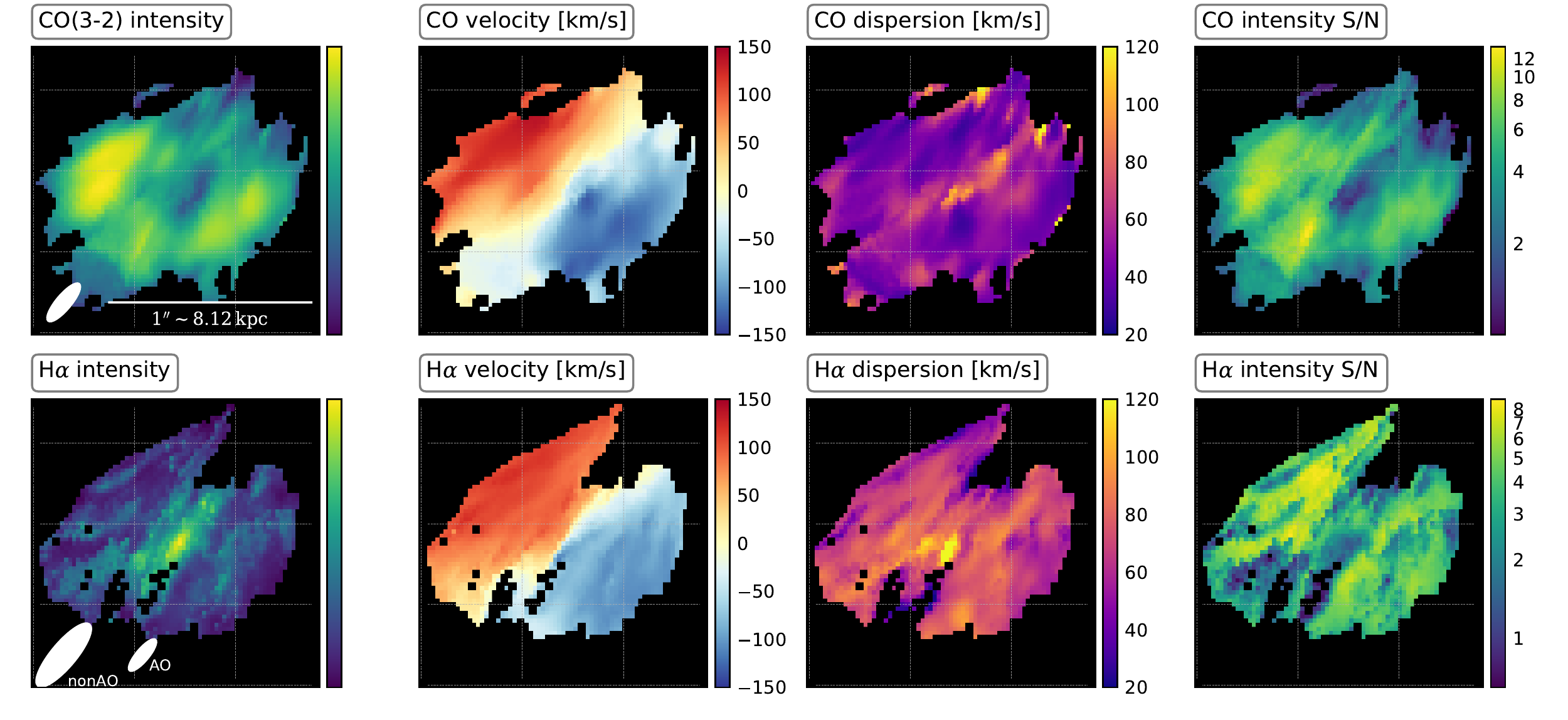}
\vspace{0.2ex}
\caption{%
Source-plane maps corresponding to the panels in Fig.~\ref{fig: image plane images}. 
All panels have the same field of view 
and north is up. 
\label{fig: source plane images}
}\vspace{2ex}
\end{figure*}

We performed detailed astrometry correction, lens modeling, and pixel-by-pixel spectral line fitting and SED fitting to obtain both image-plane and source-plane data cubes and maps of CO, \Halpha{} and stellar mass (Appendix~\ref{appx: analysis}). 
We took advantage of the highly complementary spatial distributions of the $\sim 0.08''$ \textit{HST} F814W image and the $\sim 0.36''$ ALMA CO channel maps for our new lens modeling, and did various delensing/relensing quality checks to optimize our lens model (Appendix~\ref{appx: lensing}; introducing at most 20\% uncertainty to the intrinsic source sizes/shapes).

We present the image- and source-plane maps of the SE arc, which is the most magnified and completely lensed image of J0901, in Sects.~\ref{subsec: maps} and \ref{subsec: mass distribution}. 
Our kinematic fitting then directly uses the image-plane data and our best-fit lens model in Sect.~\ref{subsec: kinematic fitting}.

\subsection{Image- and source-plane CO and \Halpha\ maps}
\label{subsec: maps}

We show the CO and \Halpha{} line intensity, velocity, velocity dispersion and intensity S/N maps of J0901's SE arc in Figs.~\ref{fig: image plane images} and \ref{fig: source plane images}, in the image- and source-plane, respectively. 

The CO and \Halpha{} emission exhibit very different spatial distributions in both image and source planes. The brightest spot in the \Halpha{} intensity map corresponds to the galaxy center and the AGN, whereas the CO emission is distributed in the disk out to a galactocentric radius of about 4~kpc and exhibits an asymmetric, ring-like structure. 

The line velocity maps of CO and \Halpha{} agree well at large scales, exhibiting a systematic disk rotation pattern in the source plane. 
At small scales, the velocity maps are affected by the different spatial resolution (see the elongated PSFs in the source plane in Fig.~\ref{fig: source plane images}), complexity of lensing caused by a nearby lens galaxy (the southern perturber, see Appendix~\ref{appx: lensing}), and possibly different higher-order kinematics of the cold and ionized gas. 

The CO and \Halpha{} velocity dispersion maps consistently peak around the galaxy center. 
However, a global difference in their velocity dispersions can be seen over the whole galactic disk, which we discuss further in Sect.~\ref{subsec: sig0}.

\begin{figure*}[htb]
\centering%
\includegraphics[width=0.95\textwidth]{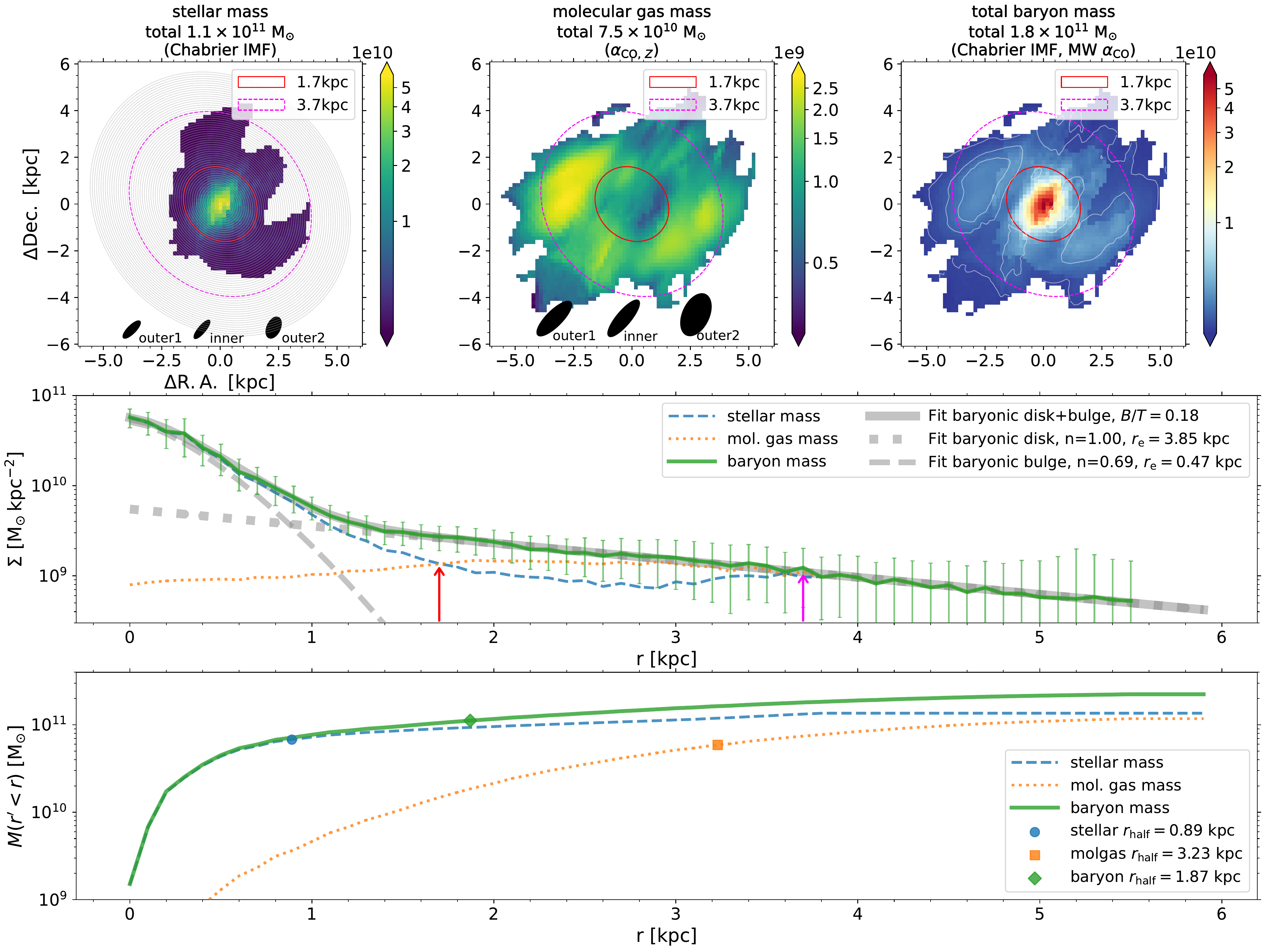}
\caption{%
\textit{Top panels:} Source-plane distributions of the stellar and molecular gas masses, and their sum as the baryon mass. The stellar mass map is derived from SED fitting to the PSF-matched, delensed \textit{HST} images as described in Appendix~\ref{appx: SED fitting}. The molecular gas mass map is converted from the source-plane CO(3--2) line intensity map as described in Sect.~\ref{subsec: mass distribution}.
The baryon mass map is taken as their sum.
Color bars indicate the per-pixel mass (in $\Msun \, \mathrm{kpc}^{-2}$), where the pixel size is $\sim 160$~pc on a side, equivalent to $0.02''$ if the source was observed unlensed. 
The varying PSF shapes across the kinematic major axis of J0901 are shown as three ellipses at the bottom of the first two panels, corresponding to ($\Delta \mathrm{RA}$, $\Delta \mathrm{Dec}$) of about ($-0.3$, $+0.2$),  ($-0.1$, $+0.0$) and ($+0.3$, $-0.4$), respectively (see also Fig.~\ref{fig: source plane psf}). 
\textit{Middle panel:} Mass density radial profiles measured from the top panel maps. The blue, orange and green lines are the radially measured stellar, molecular gas and total baryon mass surface density corresponding to the top panels. The green error bars are the coadded photometric uncertainty in \textit{HST} five-band and CO intensity maps. The gray lines represent the two-component \Sersic{} fitting, with fitted parameters listed in the legend. 
The red and magenta arrows mark the radii where the molecular gas surface density exceeds that of the stars (1.7--3.7~kpc). 
\textit{Bottom panel:} Enclosed mass as a function of radius for the three mass maps, where the half-mass radii are marked by the solid symbols and shown in the legend. 
\label{fig: mass distribution}
}\vspace{3ex}
\end{figure*}

\subsection{Stellar, cold gas and baryonic mass distributions}
\label{subsec: mass distribution}

We show the delensed stellar, cold gas and baryonic mass distributions of J0901 in Fig.~\ref{fig: mass distribution}. 
The stellar mass map is derived from \textsc{FAST} (\citealt{Kriek2009}) SED fitting (Appendix~\ref{appx: SED fitting}). 
The cold molecular gas mass map is inferred from the CO(3--2) line intensity by adopting a metallicity-dependent CO-to-H$_2$ conversion factor ($\alphaCO = 3.8 \, \mathrm{M_{\odot} \, {(K \, km \, s^{-1} \, pc^{-2})^{-1} } }$) and a global CO excitation $R_{31} \equiv I_{\mathrm{CO(3\textnormal{--}2)}} / I_{\mathrm{CO(1\textnormal{--}0)}} = 0.79$, following \citet{Sharon2019}. 
An inclination of $30^{\circ}$ and position angle of $-138^{\circ}$ are inferred from the projected axial ratio and major axis in the source plane as well as our kinematic fitting below. 

The total stellar mass in the delensed map is $1.1 \times 10^{11} \,\Msun$, agreeing with previous studies with unresolved SED fitting and independent lens modeling ($\Mstar \sim 9.5 \times 10^{10} \text{--} 3.0 \times 10^{11}\,\Msun$; \citealt{Saintonge2013,Sharon2019,Davies2020a}). 

The total intrinsic molecular gas mass in Fig.~\ref{fig: mass distribution} is $M_{\mathrm{mol\,gas}} \sim 7.5 \times 10^{10} \, \Msun$, with a corresponding intrinsic $L^{\prime}_{\mathrm{CO(3\text{--}2)}} = 1.56 \times 10^{10} \, \mathrm{K \, km \, s^{-1} \, pc^{2}}$ (or lensed $L^{\prime}_{\mathrm{CO(3\text{--}2)}} = 1.45 \times 10^{11} \, \mathrm{K \, km \, s^{-1} \, pc^{2}}$ in the SE arc). 
The uncertainty in the measured total line luminosity is very small (a few percent) given the general $\mathrm{S/N} \gtrsim 10$ in the map (Fig.~\ref{fig: source plane images}). 
Our CO(3--2) luminosity agrees well with the delensed CO(3--2) luminosity $L^{\prime}_{\mathrm{CO(3\text{--}2}} = 1.99^{+0.32}_{-0.29} \times 10^{10} \, \mathrm{K \, km \, s^{-1} \, pc^{2}}$ from \citet{Sharon2019}. 
They also reported intrinsic molecular gas mass of $M_{\mathrm{mol\,gas}} \sim 7.0 \times 10^{10} \text{--} 1.53 \times 10^{11} \, \mathrm{M_{\odot}}$, from their $\sim 1.3''$ matched-resolution and $\sim 0.7''$ native-resolution CO(1--0) data, respectively. 
They obtain a global magnification factor of $\sim 7.4 \text{--} 15.1$ for the SE arc from their two aforementioned data sets (see their Table~5). 
In comparison, we obtain a consistent magnification factor of $\sim 9.3$ for the SE arc in our $\sim 0.36''$ CO(3--2) data, 
and $\sim 8.5 \text{--} 11.8$ in our \Halpha{} data with different combination methods (Appendix~\ref{appx: combo}). 

We sum the stellar and cold molecular gas mass maps to obtain the baryonic mass map as shown in the top-right of Fig.~\ref{fig: mass distribution}. 
The atomic gas is neglected because the cold gas on the $\sim 1\,\Reff$ scale with a high gas surface density (e.g., $\sim 10^{8} \text{--} 10^{9} \, \mathrm{M_{\odot}\,kpc^{-2}}$ in our case) is likely dominated by the molecular gas for massive $z\sim2$ SFGs (e.g., \citealt{Tacconi2020}). 
The baryonic mass map shows a significant stellar mass concentration within a galactocentric radius of $\sim 1.7$~kpc. Further outside, the molecular gas starts to dominate the baryonic component and exhibits a ring-like feature at radii of $\sim $1.7--3.7~kpc. 

We show the radial profiles of the average surface density and enclosed mass in the middle and bottom panels of Fig.~\ref{fig: mass distribution}, respectively. 
We perform a two-component \Sersic{} least-$\chi^2$ fitting to the radial profile of the total baryon mass, obtaining best-fits for the following free parameters: 
$n_{\mathrm{disk}} = 1.00$, 
$n_{\mathrm{bulge}} = 0.69$, 
$r_{\mathrm{e,\,disk}} = 3.85$~kpc, 
and $r_{\mathrm{e,\,bulge}} = 0.47$~kpc. 
The innermost region or bulge component is dominated by stars whereas the outer region or disk component consists of comparable amount of stars and cold gas. 
The bulge-to-total mass ratio is $\mathrm{B/T} \sim 0.18$. 
Although the bulge is marginally resolved by the data and affected by inhomogeneous resolution, this constraint provides a sufficiently robust prior for the kinematic modeling. 

To evaluate how much the lens modeling can affect our radial profile analysis, we repeated the same analysis with various lens models, either within the 2-sigma uncertainty of the best lens model from our MCMC fitting (Sect.~\ref{appx: lensing}), or by manual inspection. We found only minor variations for the derived morphological parameters: about 10\% in $r_{\mathrm{e,\,disk}}$ and $n_{\mathrm{disk}}$, and about 20\% in $r_{\mathrm{e,\,bulge}}$ and $n_{\mathrm{bulge}}$. These variations should not affect our kinematic analysis because we allow certain variation in these parameters. 

In the bottom panel of Fig.~\ref{fig: mass distribution}, a curve-of-growth analysis of the various baryonic components gives a half-mass radius of 
$\sim 0.89$, 3.23 and 1.87~kpc 
for the stellar, cold gas and total baryon masses, respectively. As expected, this yields a total baryon half-mass radius in between the decomposed bulge and disk effective radii.
In the remainder of this paper, we take $r_{\mathrm{e,\,disk}}$ as the effective radius $\Reff$ of J0901.

\subsection{Dynamical modeling}
\label{subsec: kinematic fitting}

\subsubsection{DysmalPy+Lensing for direct image-plane fitting}

We perform forward dynamical modeling and MCMC-based kinematic fitting to each of our CO and \Halpha{} data sets using the {\sc Dysmal}/{\sc DysmalPy} software. 
{\sc Dysmal} has been used in a series of earlier studies: \citet{Genzel2006,Genzel2011,Genzel2014a,Genzel2017}, \citet{Cresci2009}, \citet{Davies2011}, \citet{Wuyts2016}, \citet{Burkert2016}, \citet{Lang2017} and \citet{Ubler2017}.
The {\sc Python} version, {\sc DysmalPy}, is recently updated by \citet{Price2021}
and used by \citet{Ubler2018,Ubler2019,Ubler2021}, \citet{Genzel2020} and \citet{Nestor2022} for various 
highest-resolution kinematic data sets
as well as simulated galaxies. 

In brief, {\sc Dysmal}/{\sc DysmalPy} is a physically-motivated, multi-component, 3D galaxy dynamical forward-modeling tool. 
It generates an intrinsic 3D+dynamics hyper model cube including baryonic and dark matter mass distributions, and computes the resulting light from baryons and kinematics (line-of-sight velocity and velocity dispersion) in the observed 3D space, fully accounting for projection, spatial and spectral resolution, and sampling effects. The fitting is performed in the observed space (``data space''), which can be either 3D (fitting data cube), or 2D (fitting velocity and velocity dispersion maps), or 1D (fitting 1D profiles extracted in a pseudo slit), all applying the identical extraction procedure as done for the data.

Currently, there are very few 3D forward-modeling kinematic fitting tool that can fit strongly-lensed galaxy kinematics (see, e.g., \citealt{Rizzo2020,Rizzo2021}; \citealt{Tokuoka2022}). 
Because of the lensing geometry, the PSF in the image plane corresponds to different shapes in the source plane, depending on the location. To properly fit the kinematics, it is needed to implement either a per-pixel-based PSF in the source plane or a lensing deflection when projecting the intrinsic model cube to the observed data space.  Without these techniques, kinematic fitting would lead to largely incorrect results (see our tests in Appendix~\ref{appx: kinematic fitting for all}). 

For this work, we developed a new lensing transformation module in {\tt C++} that can be plugged into {\sc DysmalPy}, hereafter {\sc DysmalPy+Lensing}. It enables direct image-plane kinematic fitting by implementing a computationally efficient lensing transformation when propagating the 3D model cube into the data space before convolving with the PSF and LSF. 
Here we use only the best-fit lens model's mesh grid to do the deflection (Appendix~\ref{appx: lensing}). 
Unlike galaxy-galaxy lensing which has much fewer free parameters, it is extremely time intensive when simultaneously performing the J0901's cluster lens modeling and the kinematic fitting in MCMC. 
In Appendix~\ref{appx: lensing}, we performed independent MCMC fitting to the lens modeling and found very tight posterior probability distribution functions (PDFs) for the lens parameters. This means that even when combining the lens modeling and kinematic fitting into a joint MCMC fitting, the kinematic parameters' PDFs will not be significantly broadened. 
We tested various lens models within the 2-sigma MCMC uncertainties of our best-fits or fitted by hand as mentioned in Sect.~\ref{subsec: mass distribution} and Appendix~\ref{appx: lensing}, finding that the kinematic fitting with different testing lens models led to variations within the errors of MCMC kinematic fitting. 
Therefore, we do not combine the lens modeling and kinematic fitting into one joint MCMC fitting.

\subsubsection{Model components}
\label{subsec: model components}

\textsc{DysmalPy} builds up a galaxy using several physically-motivated components, e.g., a bulge, a disk and a dark matter halo. The bulge and disk components are usually set as \Sersic{} profiles, and the dark matter halo as a Navarro-Frenk-White (NFW; \citealt{NFW}) profile. 
The key parameters for the bulge+disk components are $r_{\mathrm{e,\,disk}}$, $r_{\mathrm{e,\,bulge}}$, $n_{\mathrm{disk}}$, $n_{\mathrm{bulge}}$, $\mathrm{B/T}$, and $\sig0$ (see \citealt{Price2021} for more details). 
All except $\sig0$ are previously measured in Sect.~\ref{subsec: mass distribution} and shown in Fig.~\ref{fig: mass distribution}. For $r_{\mathrm{e,\,disk}}$, 
we adopt a Gaussian prior PDF in our MCMC sampling centered at the best-fits with a 0.2~dex sigma representing the uncertainty in lensing and photometry. For $r_{\mathrm{e,\,bulge}}$, $n_{\mathrm{disk}}$, $n_{\mathrm{bulge}}$ and $\mathrm{B/T}$, we fix them to the best-fit values. We have tested that changing the fixed parameters by 10--20\% does not obviously affect our results. 
For $\sig0$, we adopt a flat prior PDF. 
We note that its posterior distribution is tightly constrained regardless of the prior PDF shape or range. We also adopt a constant $\sig0$ profile across the galactic disk as indicated by our data (see Fig.~\ref{fig: obs vs model vel}; see also \citealt{Ubler2019}). 

The NFW profile is characterized by a virial mass ($\Mvir$) and a halo concentration. 
We adopt a Gaussian prior PDF for the $\Mvir$ centered at $\log (\Mvir^{\mathrm{Moster}} / \Msun) \sim 12.3$ with a sigma of 0.7~dex. 
This $\Mvir^{\mathrm{Moster}}$ is the average halo mass for a $\log (\Mstar / \Msun) \sim 11.0$ SFG based on the $\Mstar\,\text{--}\,\Mvir$ relation (\citealt{Moster2018,Moster2020}). 
We note that using a flat prior PDF with a wide range $\log (\Mvir / \Msun) = 10.0$--$14.0$ will not significantly change our derived $\Mvir$ by more than 0.2~dex, and the results are within the MCMC fitting derived uncertainty.

The halo concentration is fixed to 4.0, appropriate for the $z \sim 2$ of J0901 (e.g., \citealt{Bullock2001,Dutton2014,Ludlow2014,Moster2020}). 
We do not have enough constraints to explore the possibility of other halo models because the derived uncertainty in $\Mvir$ is already about 0.5~dex (Table~\ref{Table1}).

\input{Input_fig_5}

We allow the inclination ($\sin(i)$ in the modeling) and position angle (PA) to vary following Gaussian prior PDFs with sigma of $\sim$~0.2 and $\sim$30~deg, respectively. 
The spatial and velocity kinematic center coordinates are also allowed to vary within small ranges under flat prior PDFs considering the uncertainty brought about by lensing.
Following previous work, we only fit the velocity and dispersion profiles not the flux distribution. This is because the \Halpha{} and CO emission individually do not trace the overall mass distribution. 

The asymmetric drift (pressure support) is corrected as
$v_{\mathrm{rot}}^{2}(r) = v_{\mathrm{circ}}^{2}(r) - 2 \, \sigma_{0}^{2} \left({r}/{r_{d}}\right)$
for a turbulent disk with isotropic and radially constant velocity dispersion (\citealt{Burkert2010,Burkert2016,Genzel2020,Price2021}), where $r$ is the galactocentric radius, and $r_{d}$ is the disk scale length ($\Reff = 1.68 \, r_{d}$ for an exponential profile).
Also following previous work, we neglect the effects of adiabatic contraction of the dark matter halo (see also discussion in \citealt{Burkert2010}). 

Our key best-fit parameters are given in Table~\ref{Table1}, and a direct comparison of the image-plane data and best-fit model convolved with PSF and LSF is given in Appendix~\ref{appx: image-plane kinematic fitting plots}. Below we focus on the scientific results which are discussed in the source-plane.

\input{Input_fig_6}

\vspace{3ex}

\section{Results and Discussion}
\label{sec: results}

\subsection{Rotation curve and velocity dispersion profiles}
\label{subsec: rotation curve}

In Fig.~\ref{fig: obs vs model vel}, we present the source-plane 2D velocity map and 1D rotation curve of J0901, extracted consistently from the data and our best-fit kinematic model, respectively, with 1D rotation curve extracted in a pseudo slit along the kinematic major axis. 
Error bars of the data points are the uncertainties from our pixel-by-pixel MCMC line fitting, which are larger in \Halpha{} than in CO partially because of the outflow removal. 
At large radii, the CO rotation curve appears to have a larger $\vrot$ than \Halpha{}. This is because the intrinsic disk dispersion is higher for \Halpha{}, leading to a stronger asymmetric drift bending down the curve.

In Fig.~\ref{fig: obs vs model disp}, we show the CO and \Halpha{}'s velocity dispersion in 2D and 1D. 
Both tracers have a dispersion peaking consistently at the galaxy center because of the rapidly rising inner rotation curve smeared by the PSF. 
The dispersion peak is somewhat still seen in the residual maps, but the error bars are also large. The large uncertainties near the center come from the outflow removal for \Halpha{} and the rather low S/N ($\sim 1$--2) for CO. 
Up to a galactocentric radius of about 4~kpc or $\sim 1 \, \Reff$, the CO and \Halpha{} dispersions do not show an obvious decrease from the inner to the outer disk. 
This supports our assumption of a constant disk dispersion.

\input{Input_fig_7}

\vspace{1ex}

\subsection{Different velocity dispersions of the cold and ionized gas in J0901}
\label{subsec: sig0}

From our kinematic modeling, the best-fit intrinsic disk dispersion for the CO and \Halpha{} traced molecular and ionized gas are:
$\sigma_{0,\,\mathrm{mol.}} = 37.3_{-2.8}^{+2.5} \,\kms$, 
and 
$\sigma_{0,\,\mathrm{ion.}} = 64.3_{-5.1}^{+5.0} \,\kms$, 
respectively (corrected for the LSF). 
Their difference is
$\sim 27.0 \pm 3.0 \,\kms$. 
We compare these to other massive $z\sim 1$--3 SFGs in Fig.~\ref{fig: disk dispersion}.

There are still very few massive SFGs at $z \gtrsim 2$ that have both ionized and cold gas dispersion measurements. 
J0901, interestingly, follows the empirical evolution trends derived by \citet{Ubler2019}. 
In the upper panel of Fig.~\ref{fig: disk dispersion}, J0901's $\sigma_{0,\,\mathrm{ion.}}$ is slightly above the \citet{Ubler2019} trend but is consistent with other massive SFGs.
There is a large scatter in the ionized gas disk dispersion at all redshifts, but the mean trend is increasing with redshift.

J0901's $\sigma_{0,\,\mathrm{mol.}}$ is about 25~km/s higher than that of $z \sim 0.1$ DYNAMO galaxies (\citealt{Girard2021}), and is about 10--20~km/s higher than most of the $z \sim 0.6$--1.5 main-sequence galaxies from the PHIBSS survey (\citealt{Tacconi2013,Tacconi2018}), with CO dispersion measurements compiled by \citet{Girard2021}.

\citet{Girard2021} also included lower-mass, strongly-lensed galaxies at $z \sim 1.0$ (from \citealt{Patricio2018} and \citealt{Girard2019}), which show cold gas dispersions as low as 11--20~km/s and have much smaller gas disk sizes. 
There is also an extreme emission-line selected galaxy in their compilation at $z \sim 1.5$, originally from \citet{Molina2019} and having a very high $\fgas \sim 0.8$ and $\sigma_{0,\,\mathrm{mol.}} \sim 91$~km/s. 
It is likely that such an outlier has entered a starbursting phase and thus deviates from the mean trend. 
J0901, as a representative of the massive main sequence SFG, robustly confirms the disk dispersion trends of other main sequence SFGs.

\input{Input_fig_8}

\input{Input_fig_9}

\subsection{Dark matter fraction within the disk}
\label{subsec: fdm}

We obtain a dark matter fraction $\fDM$ of
$0.44_{-0.16}^{+0.15}$ 
from the CO, and 
$0.36_{-0.14}^{+0.18}$ 
from the \Halpha{} kinematics, within $\Reff$ in J0901. 
In Fig.~\ref{fig: intrinsic rotation curve}, we show the intrinsic circular velocity curve of our best-fit model. 
The relative contributions from baryon and dark matter are shown as the stacked blue and yellow areas, respectively. 
The dark matter starts to dominate over the baryons only at about $> 1 \, \Reff$. 
The CO and \Halpha{} curves show overall very good consistency.

In Fig.~\ref{fig: fDM Sigbar}, we compare J0901's $\fDM$ and kinematically-determined baryon surface density $\Sigbar$ within $1 \Reff$ to that of a hundred massive $z\sim1$--3 SFGs from \citet{Nestor2022}. 
The empirical trend derived by \citet{Wuyts2016} using the KMOS$^\mathrm{3D}$ seeing-limited survey data is overlaid, which covers most of these SFGs within $\sim$0.2~dex.

With better constraints from both CO and \Halpha{} kinematics at twice or higher physical resolution, the derived $\fDM(<\Reff)$ of J0901 is in excellent agreement with results from other (unlensed) massive galaxy samples at similar redshift. 
The baryon dominance in the inner regions requires a flatter inner dark matter halo profile than the assumed cuspy NFW one in order to remain consistent with the global $\Mstar\,\text{--}\,\Mvir$ relation. 
Mechanisms for such a coring process could be AGN/star-formation feedback, dynamical friction by the dark matter on merging satellites, and giant baryonic clumps (e.g., \citealt{ElZant2001,Dekel2003b,Martizzi2012,Peirani2017,Dekel2021,Ogiya2022}; see also discussions in \citealt{Genzel2020} and \citealt{Dekel2021}).
These mechanisms, especially the energetic AGN feedback in the hot dark matter halo, may have well happened in J0901 over its last billion years.

In addition, there is a marginal discrepancy between the kinematically and photometrically-derived total baryon masses in J0901, at $\sim 0.5$~dex. 
This is likely due to well-known uncertainties in the photometric mass estimation, i.e., SED fitting with only five-band \textit{HST} data up to the $H$-band, 
Initial Mass Function (IMF), 
star formation history, 
and dust attenuation, etc. 
These uncertainties add up to also about 0.5~dex, if considering the delensing and IMF variations (e.g., \citealt{Cappellari2012,Zhang2018,Hopkins2018Review}).

\input{Input_fig_10}

\vspace{1ex}

\subsection{Disk instability and inside-out quenching}
\label{subsec: quenching}

We compute the Toomre $Q$ (\citealt{Toomre1964}) map of the molecular gas in Fig.~\ref{fig: Toomre Q} following the Eq.~2 of \citet{Genzel2014a}: $Q_{\mathrm{gas}} = \frac{\kappa (r) \sigma_{0}}{\pi G \Sigma_{\mathrm{gas}}(r)}$, where $\kappa$ is the epicyclic frequency depending on the rotation curve, $\sigma_{0}$ is the intrinsic velocity dispersion, $G$ is the gravitational constant, and $\Sigma_{\mathrm{gas}}(r)$ is the molecular gas surface density at radius $r$. 
A threshold $Q$-value, $Q_{\mathrm{crit}} = 0.67$, describes a single gas phase, thick disk (\citealt{Goldreich1965a}). 
This $Q_{\mathrm{crit}}$ is slightly larger for a mixture of gas and stars. 
When $Q$ is below the $Q_{\mathrm{crit}}$, gas becomes gravitationally unstable and is subject to collapse and/or fragments and forms stars locally. 
J0901's molecular gas shows a central peak of high $Q$-value within about 1~kpc, then a significant instability with $Q < 0.5 \, Q_{\mathrm{crit}}$ at larger radii. 

\citet{Genzel2014a}, using \Halpha{} kinematics and cold gas surface densities estimated from \Halpha-based SFRs, reported a central peak of $Q$ in the majority of massive SFGs among their sample. 
They pointed out that the rapidly rising inner rotation curve (the gradient of epicyclic frequency $\kappa$) contributes more to the $Q$-gradient than the gas surface density distribution. 
Their finding, along with our direct cold gas based result, are consistent with an inside-out quenching scenario in which the global gravitational instability is suppressed from the inside out during the secular evolution of SFGs.

The $Q$-parameter can be alternatively expressed as a function of the gas fraction $\fgas$ when substituting $\kappa$ with a combination of the enclosed total mass and radius (see Eq.~3 of \citealt{Genzel2014a}), leading to $Q = a \cdot \fgas^{-1} \cdot (\vrot / \sig0)^{-1}$, where $a = \sqrt{2}$ for a flat rotation curve (see also \citealt{Genel2008,Law2009,Genzel2011,Wisnioski2015,Turner2017}). 
J0901's baryonic $\fgas$ is $\sim 0.42$ 
as indicated from our mass maps (Fig.~\ref{fig: mass distribution}), agreeing well with those derived from the empirical scaling relations (e.g., the \citealt{Tacconi2018} scaling relation gives $\fgas = 0.51$ and the \citealt{Liudz2019b} scaling relation predicts $\fgas = 0.48$). 
With the cold gas $\vrot / \sig0 = 6.5 \pm 1.0$ at $\Reff$ in J0901
(see Table~\ref{Table1}), the formula predicts a $Q$-value of about 0.5, confirming the generally unstable $Q$-values in Fig.~\ref{fig: Toomre Q}.

The cold gas depletion time $\tau_{\mathrm{depl.}} \equiv \Mgas / \SFR$ is about $420 \, \mathrm{Myr}$, 
agreeing with the general trend at $z \sim 2$ for a massive ($\log \Mstar/\Msun = 11$), main sequence SFG (e.g., \citealt{Tacconi2013,Tacconi2018,Genzel2015,Liudz2019b}). 
The amount of cold gas and depletion timescale, if assuming a constant SFR in the rest of the time and ignoring the gas accretion from halo, indicate that J0901 will run out of its cold gas fuel in about 4.5 orbital periods.

Assuming that J0901 stays on the star-forming main sequence (i.e., following the evolution of \citealt{Speagle2014}) since its formation time $t_{\mathrm{form}}$ with an initial stellar mass $M_{\star,\,\mathrm{init}}$, then we can compute its mass assembly history till reaching its current  $\Mstar$ and SFR at $z=2.259$ ($t_{\mathrm{cosmic\,age}}=2.85$~Gyr). 
We find that a $t_{\mathrm{form}} \sim 1.45$~Gyr ($z_{\mathrm{form}} \sim 4.1$) and $\log M_{\star,\,\mathrm{init}} / \Msun \sim 9.0$ are needed to match its current properties. 
This assembly history also means that J0901 had half of its current $\Mstar$ at $z \sim 2.61$, only about 400~Myrs before. 
At that time, its molecular gas within the inner $\sim 1.7$~kpc could be about $5 \times 10^{10} \, \Msun$ with a mean gas surface density at least twice the current peak value (Fig.~\ref{fig: mass distribution}). 
If further considering a mass loading factor of $\sim 0.5$ due to galactic outflows, then either a massive radial transport of the cold gas inward or an even larger $\Sigma_{\mathrm{mol.\,gas}} \gtrsim 10^{10} \, \mathrm{M_{\odot}\,kpc^{-2}}$ within the inner $\sim 1.7$~kpc radius is required. The latter would indicate a $\Sigma_{\mathrm{mol.\,gas}}$ much higher than in the most starbursty, merger-driven ultra-luminous infrared galaxies (e.g., \citealt{Ward2003,Privon2017}). 
Therefore, it is very likely that a significant radial transport has played a role in J0901's assembly history, with ensuing star formation and possibly outflows then exhausting its inner gas reservoir, 
then the central bulge is starved by disk gas having too high 
angular momentum to keep feeding central star formation (\citealt{PengYJ2020}).

The AGN-driven ionized gas outflow at a rate of
$25 \pm 8 \ \Msyr$ (\citealt{Davies2020a}) is a probable reason for the shutdown of further gas inflow into the central kpc. 
This outflow is still much lower than the current SFR, but could have been stronger in the past. Even maintaining its current rate, it can still cancel out the remaining gas inward transport which is likely becoming weaker and weaker with cosmic time.

\vspace{1ex}

\subsection{Cold gas ring formation and longevity}

The ring-like cold gas structure revealed in this study raises interesting questions, e.g., how common is a cold gas ring found in massive SFGs at high-$z$, and what is its origin. 
The first question is hard to answer statistically because the number of high-resolution cold gas kinematic measurements in main-sequence SFGs is still very small. 
Possibly due to either resolution or sensitivity, none of the previous studies with 
CO mapping has revealed a cold gas ring (e.g., \citealt{Tacconi2006,Tacconi2008,Bolatto2015,Barro2017,CalistroRivera2018,HerreraCamus2019,Rybak2019,Kaasinen2020}). 
In comparison, with systematic deep surveys with AO, \Halpha{} rings have been commonly found in massive SFGs at $z \sim 1$--3 at 1--4~kpc scales (\citealt{Genzel2008,Genzel2011,Genzel2014a,ForsterSchreiber2018}; e.g., $\gtrsim 50\%$ in the \citealt{Genzel2014a} sample of 19 SFGs). 

In theory, CO and \Halpha{} rings both can indicate that there is less star formation in the inner than the outer disk, and they are different from the stellar mass rings found in some studies which may originate from minor mergers (e.g., \citealt{Elmegreen2006rings,Elagali2018,Yuan2020}). 
Whether CO and \Halpha{} rings are two phases in the same evolutionary path or are two distinct populations is still an open question. 

On the second question, 
the formation of cold gas rings in the secular evolution of massive SFGs has been seen in high-resolution numerical simulations (\citealt{Danovich2015,Dekel2020b}).
In these simulations, most massive SFGs have experienced a wet compaction event event when their masses reach certain threshold ($M_{\mathrm{DM}} \sim 10^{11.5} \, \Msun$ or $\Mstar \sim 10^{9.5} \, \Msun$; \citealt{Dekel2014,Dekel2020b}). 
\citet{Dekel2020b} showed that such a wet compaction includes following phases: 
\textit{i)} a highly turbulent rotating disc develops from cold gas streams; 
\textit{ii)} a central blob of high gas density builds up; 
\textit{iii)} central gas depletes via star formation and outflows; and 
\textit{iv)} an extended, clumpy cold gas ring forms, which is continuously fed by incoming cold streams and will live for several billion years without an inward migration. 
Taking these simulated galaxies as an example, phases \textit{i--iii} happened during $z=3.3$--2.7, with giant clumps and inter-clump gas exist and migrate inwards. Then, phase transition \textit{iii--iv} happens rapidly from $z=2.7$--2.4 and forms the post-compaction ring, roughly matching the redshift of J0901. Such a cold gas ring in these simulations appears together with the central mass concentration and lasts for about two billion years from $z=2.4$--1.2.

J0901's CO ring is consistent with the physical scenario demonstrated in the above simulations. 
Our ring radius ($\sim 4$~kpc) is also consistent with some of the simulated galaxies, e.g., ``V20'' in \citet{Dekel2020b}, but not their ``V07'' galaxy. The variation in the ring radius of their simulated galaxies is $\sim 4$--10~kpc. 
Comparing to J0901's properties, the conditions of the simulated galaxy ``V07'' match J0901 well, with $\Mstar = 10^{10.5\text{--}10.8} \, \Msun$ and $M_{\mathrm{DM}} = 10^{11.8\text{--}12.1} \, \Msun$ during it's phases \textit{iii--iv} at $z=2.4$--1.2. 
The conditions for ``V20'' are not mentioned in their paper, but its total virial mass is only slightly (0.15~dex) smaller than that of ``V07''. 
It could be gas accretion history or other stochastic properties that caused the different final ring sizes.

\vspace{2ex}

\section{Summary}
\label{sec: summary}

In this work, we present an analysis of the currently highest-resolution CO and \Halpha{} data sets in the strongly-lensed, representative massive main sequence SFG J0901 at $z=2.259$, achieving a delensed physical resolution of $\sim 600$~pc for our major-axis kinematic study. 
We derived a new lens model utilizing the highly complementary \textit{HST} and CO data, and examined the uncertainty of lensing via MCMC fitting (Appendix~\ref{appx: lensing}). 
We developed a {\tt C++} {\sc Lensing} module for the 3D forward-modeling kinematic fitting software {\sc DysmalPy} (\citealt{Price2021}), to enable direct image-plane kinematic fitting for our data sets. 

The resulting CO and \Halpha{} kinematics show that J0901 is a baryon-dominated rotating disk within its $\sim 1 \, \Reff$ ($\sim 4$~kpc). 
Its dark matter fraction inside $\sim 1 \, \Reff$ is fully consistent with the general trend established for $z \sim 1$--3 massive main sequence SFGs by large-sample, non-lensed kinematic studies (e.g., \citealt{Wuyts2016,Genzel2020,Nestor2022} and simulations (\citealt{Moster2018,Moster2020}). 

We find that J0901's intrinsic velocity dispersion ($\sigma_{0}$) is roughly constant from the inner to the outer part of the disk (to $\sim 1 \, \Reff$ or $\sim 4$~kpc; Fig.~\ref{fig: obs vs model disp}). The cold molecular gas has a dispersion of $\sigma_{0,\,\mathrm{mol.}} \sim 37.3^{+2.5}_{-2.8} \ \mathrm{km/s}$, and the ionized gas has $\sigma_{0,\,\mathrm{ion.}} \sim 64.3^{+5.0}_{-5.1} \ \mathrm{km/s}$. 
Both velocity dispersions match well with the 
$\sigma_{0,\,\mathrm{ion.}}$ and $\sigma_{0,\,\mathrm{mol.}}$ evolution trends derived from non-lensed massive SFGs (\citealt{Ubler2019}). 

We derive the stellar and cold gas mass maps and the corresponding Toomre $Q$ map in J0901, finding a strong cold gas $Q$-peak ($Q \gtrsim 3$) within the central kilo-parsec of J0901 and a highly-unstable cold gas ring at radii $\sim$2--4~kpc ($Q \sim 0.3$). 
Together, the dense central peak in stellar mass surface density inside the cold gas ring structure are suggestive of an inside-out quenching scenario (e.g., \citealt{Martig2009,Genzel2014a}). These features observed in J0901 are also in qualitative agreement with structures identified in post-wet compaction phases in recent high-resolution numerical simulations of high-$z$ turbulent SFGs (\citealt{Dekel2020b}). It will be important in future work to investigate the frequency of such signatures with high-resolution observations of larger samples of massive $z \sim 2$ SFGs.

\vspace{3ex}

\acknowledgments

We thank the anonymous referee for the very helpful review.
This paper makes use of the following ALMA data:
\incode{ADS/JAO.ALMA#2016.1.00406.S}, 
\incode{ADS/JAO.ALMA#2013.1.00952.S}.
ALMA is a partnership of ESO (representing its member states), NSF (USA) and NINS (Japan), together with NRC (Canada), MOST and ASIAA (Taiwan), and KASI (Republic of Korea), in cooperation with the Republic of Chile. The Joint ALMA Observatory is operated by ESO, AUI/NRAO and NAOJ. 
Based in part on observations collected at the European Southern Observatory under ESO programmes 092.A-0082(A), 093.A-0110(A), and 094.A-0568(A). 
This research was supported by the Excellence Cluster ORIGINS which is funded by the Deutsche Forschungsgemeinschaft (DFG, German Research Foundation) under Germany's Excellence Strategy -EX-2094-390783311.
H\"{U} gratefully acknowledges support by the Isaac Newton Trust and by the Kavli Foundation through a Newton-Kavli Junior Fellowship.

\facilities{%
ALMA,
SINFONI/VLT
}

\software{%
Astropy \citep{astropy:2013, astropy:2018, astropy:2022}, 
CASA \citep{CASA}, 
DysmalPy \citep{Price2021},
Emcee \citep{emcee}, 
FAST \citep{Kriek2009,Krumholz2018b}, 
Galfit \citep{Peng2002,Peng2010}, 
Glafic \citep{Oguri2010, Oguri2010asclsoft}, 
TinyTim \citep{TinyTim}, 
Photutils \citep{photutils}, 
Scipy \citep{scipy}, 
PyMC3 \citep{pymc3}
}

\newpage

\appendix

\counterwithin{figure}{section}

\section{J0901 data products gallery}
\label{appx: all products}

\begin{figure*}[hptb]
\centering%
\includegraphics[width=0.75\textwidth]{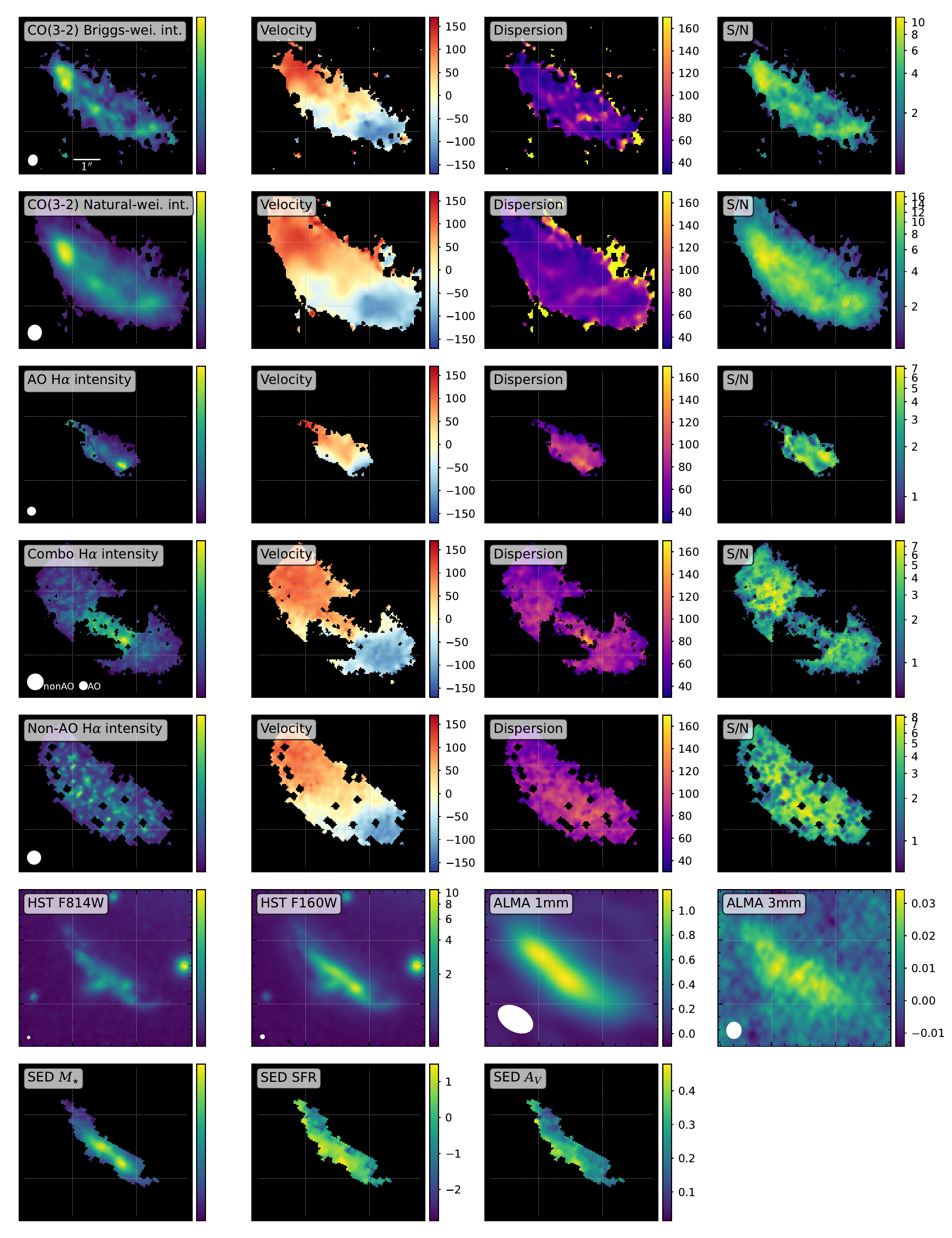}
\caption{%
Image-plane data of the J0901 SE arc. The first two rows are the maps of the ALMA CO(3--2) line integrated intensity, velocity, velocity dispersion and S/N of the integrated intensity (\textit{1st row}: Briggs-weighting; \textit{2nd row}: natural-weighting). 
The 3rd to 5th rows are the \Halpha{} data products from the VLT SINFONI AO, ``combo'' and non-AO data cubes (Sect.~\ref{subsec: SINFONI}; Appendix~\ref{appx: combo}), respectively. 
For \Halpha{} products we only show the narrow-line (non-outflow) component from our pixel-by-pixel spectral line fitting (Appendix~\ref{appx: line fitting}). 
The last two rows show the 
\textit{HST} F814W and F160W images, ALMA 1mm and 3mm dust continuum images, and the pixel-by-pixel SED-fitted stellar mass, SFR and $A_V$ maps (see our SED fitting in Appendix~\ref{appx: SED fitting}). 
The fields of view are the same in all panels ($8'' \times 7''$) and north is up. 
\label{fig: image plane all images}
}
\end{figure*}

\begin{figure*}[hptb]
\centering%
\includegraphics[width=0.75\textwidth]{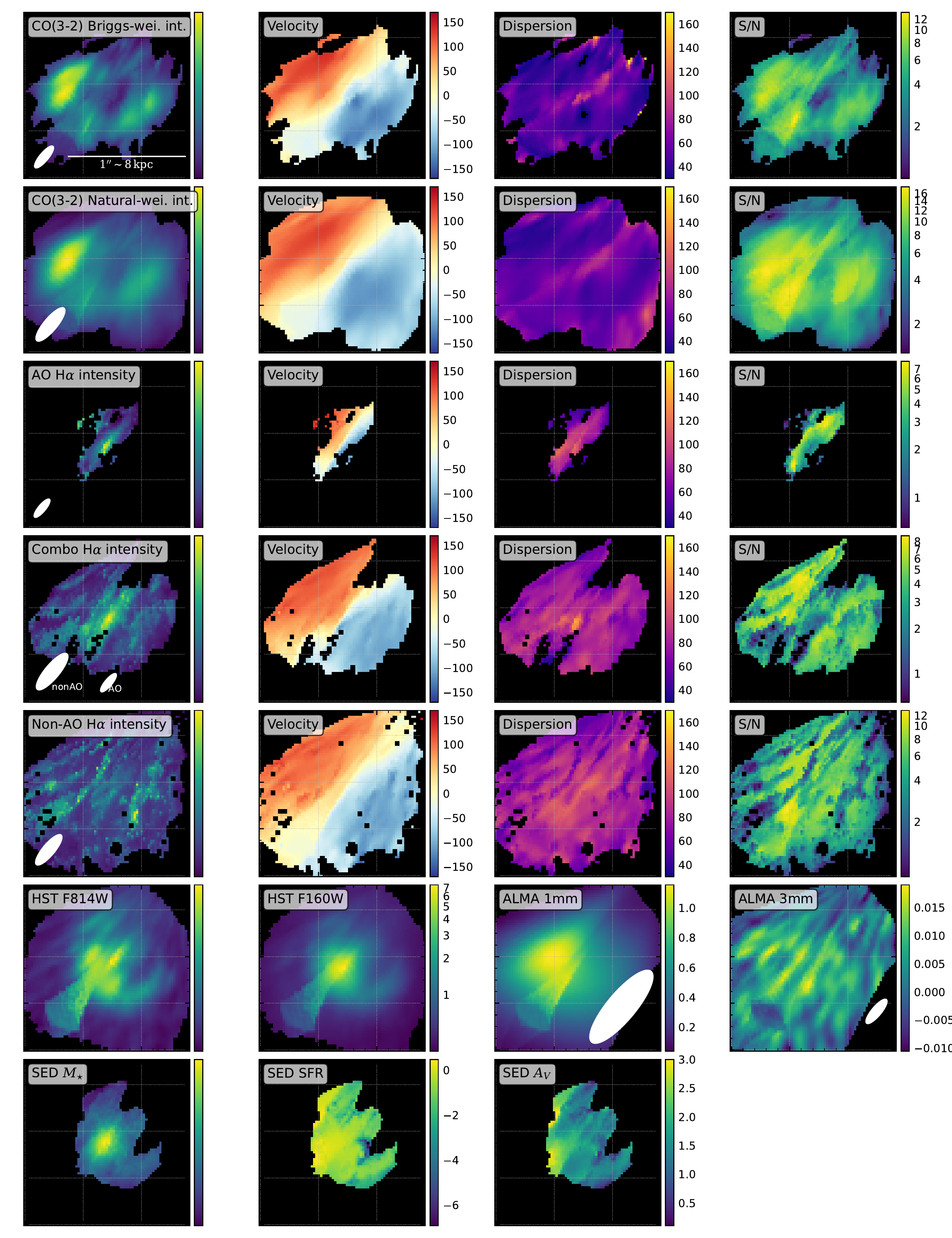}
\caption{%
Source-plane maps corresponding to the panels in Fig.~\ref{fig: image plane images}. 
The line maps are re-extracted from the delensed data cubes (instead of direct delensing of the 2D images in Fig.~\ref{fig: image plane images}). 
All panels have the same field of view of $1.5'' \times 1.5''$ and north is up. 
\label{fig: source plane all images}
}
\end{figure*}

We show a gallery of all our J0901 CO and \Halpha{} products, along with the \textit{HST} F814W, F160W, ALMA archival 1mm dust continuum (beam~$\sim1.45'' \times 0.91''$ at PA$\sim60^{\circ}$; project code: 2013.1.00952.S), ALMA 3mm dust continuum observed together with our CO(3--2) data, and SED-fitted stellar mass, SFR and $A_V$ maps in Figs.~\ref{fig: image plane all images} (image plane) and \ref{fig: source plane all images} (source plane).

\section{Additional data processing and analyses}
\label{appx: analysis}

\subsection{Astrometry correction, PSF matching, and foreground subtraction for HST and SINFONI data}
\label{appx: astrometry}

The astrometric calibration of the \textit{HST} imaging is based on 9 stars identified in the GAIA DR3 database\,\footnote{\url{https://gea.esac.esa.int/archive/}} with proper motion information. \incode{Galfit} (\citealt{Peng2002,Peng2010}) is used to accurately determine the source positions in the images. 

The astrometry corrections for the SINFONI AO and non-AO data are based on the alignment between the extracted $K$-band continuum and the \textit{HST} $H$-band continuum, as the small SINFONI field of view does not contain any bright star.

We matched the \textit{HST} ACS/WFC and WFC3/IR images to a common PSF before performing the spatially-resolved SED fitting. 
We built PSFs using the TinyTim software (\citealt{TinyTim}) and made convolution kernels using the Python \incode{photutils} package (\incode{photutils.psf.create_matching_kernel} function). Then the \incode{astropy} and \incode{reproj} packages are used for convolution and reprojection to a common pixel scale. 
We examined the radial profiles of our PSFs using several unsaturated stars in our \textit{HST} images as well as the CANDELS \textit{HST} images, finding good agreements yet note that sometimes the peak pixel of real stars is 20--30\% lower than that of our ideal PSFs, which is likely because of a sub-pixel sampling/smearing issue. The choice of a perfect PSF is not a key issue in our analysis and not obviously altering our SED fitting derived stellar mass and/or other properties.

\begin{figure}[htb]
\centering
\includegraphics[width=\textwidth]{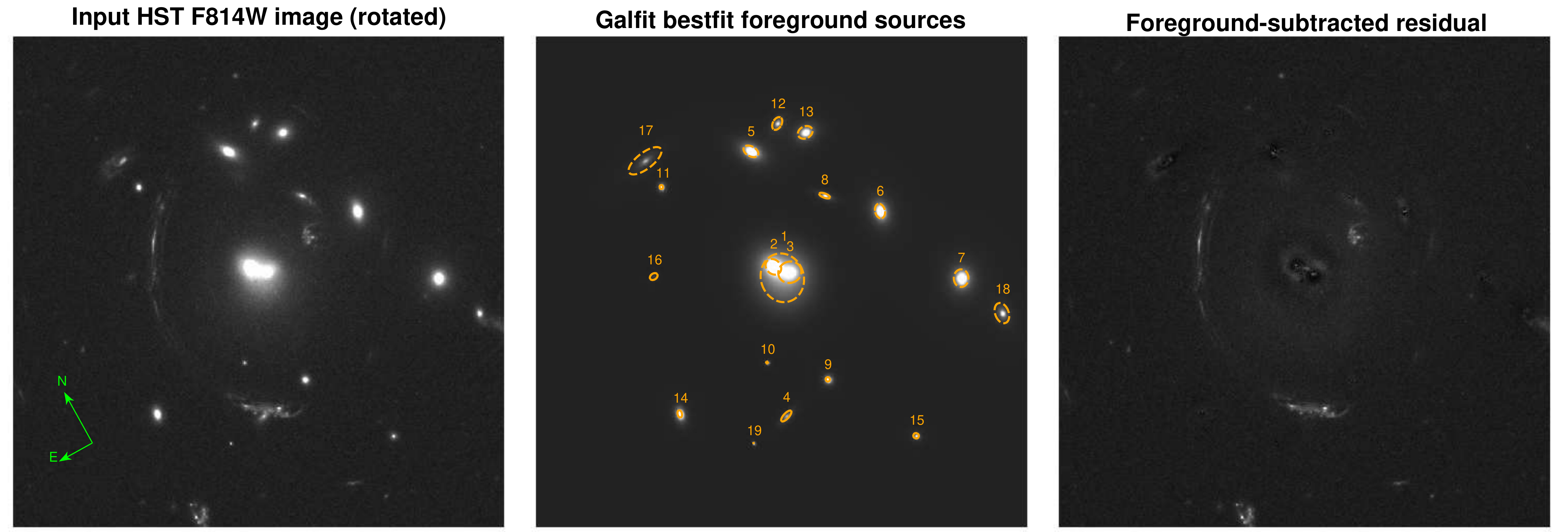}
\vspace{2ex}
\caption{%
Illustration of our \incode{Galfit} source fitting in the Gaia-DR3 astrometry corrected \textit{HST} images for the foreground lens galaxy cluster member locations. The left panel is the astrometry corrected \textit{HST} image. The middle panel is our best-fit \incode{Galfit} models convolved with the PSF, with ellipses indicating each fitted source. The orange ellipse is the foreground galaxy that blends with the southeast arc of J0901, and we treated it carefully by iteratively fitting and fixing its photometric parameters. 
The right panel is the residual image where the foreground galaxies' emission are subtracted and J0901 emission is clearly visible. 
}
\label{fig: galfit}
\end{figure}

In order to reconstruct the source-plane \textit{HST} maps of J0901, galaxies in the foreground lensing cluster need to be subtracted. 
This is done by first running \incode{Galfit} to fit \Sersic{} profiles simultaneously for 19 foreground galaxies, plus 14 sources which are manually added to represent the J0901 emission for a better deblending. 
For the foreground emission we included an extended component at the lensing cluster center, representing a diffuse cluster light which is needed for a good fit. 
Then we fixed the foreground galaxies' photometric parameters to their best-fits and run \incode{Galfit} again with only the foreground galaxies. This produces a foreground-only model image and a residual image where only J0901 emission remains. 
The resulting images are presented in Fig~\ref{fig: galfit}.

\vspace{1ex}

\subsection{Combining SINFONI AO and non-AO data sets}
\label{appx: combo}

As mentioned in Sect.~\ref{subsec: SINFONI}, we combine the SINFONI AO and non-AO data into one ``combo'' data cube for our kinematic fitting. 
We stitch the AO and non-AO cubes so that AO cube pixels fill the inner part and non-AO cube pixels fill the outer part. 
To improve the S/N, we did a two-pixel-FWHM Gaussian smoothing to the AO data and an 1.5-pixel-FWHM Gaussian smoothing to the non-AO data before combining (with pixel size regridded to $0.05''$ in advance). 
The transition radius for the stitching is determined empirically and has no major effect as we tested. 
The stitched cube therefore has two PSFs and inhomogeneous noise, but represents a total on-source integration time of about 19~hours. 
The AO PSF was adopted for kinematic modeling of the stitched combo cube.  It is adequate for the central regions of the galaxy, where the observed velocity and dispersion variations are strongest; for the outer disk regions, it is smaller than the actual resolution but for the case of J0901, it has little impact because in these regions the velocity curve and the velocity dispersion are fairly flat.

To further understand the effect of fitting the combo or individual data sets, we performed independent kinematic fitting for the AO, non-AO and ``combo'' data sets, then compared their results in Appendix~\ref{appx: kinematic fitting for all}. 
We find no significant inconsistency given the fitting uncertainties (mostly limited by the area we probed which is slightly beyond $1 \, \Reff$). 
However, the combo data set gives the full information of the rotation curve with the best inner spatial resolution, therefore is taken as our fiducial \Halpha{} data set throughout the paper.

\vspace{1ex}

\subsection{Lens modeling}
\label{appx: lensing}

J0901 is lensed by the gravitational field of a low-$z$ massive galaxy cluster. The brightest central galaxy (BCG) is a massive elliptical galaxy confirmed at $z=0.34612\pm0.00019$ with SDSS DR7 spectroscopy (\citealt{Diehl2009}). 
As shown in Fig.~\ref{fig: galfit}, we adopt 16 cluster member galaxies visible in the \textit{HST} data and with a similar color.
The entire J0901 is doubly-lensed into the southeast (SE) and west (W) arcs. 
Additionally, the majority of J0901 except for its southeast part in the source plane is inside the caustics of the major lens and thus is quadruply-lensed, forming the northeast (NE) arc in the image plane (see also \citealt{Tagore2014}). 
Moreover, there is a galaxy lens very close to the southeast (SE) arc, hereafter ``the southern perturber'', which significantly distorts part of the SE arc as seen in the \textit{HST} image and the CO velocity field. The distorted regions are multiply-lensed around the perturber, creating two apparent nuclei (\citealt{Sharon2019,Davies2020a}) and a twisted velocity field therein. 
Coincidentally, there is a higher-redshift galaxy, nicknamed ``\textit{Sith}'' (\citealt{Tagore2014}) also lensed by the cluster and is seen as four faint images. 
(It's redshift is about $z = 3.23^{+0.13}_{-0.11}$ from our MCMC lens modeling, consistent with the determined value of $z\sim3.1$ by independent lens modeling in \citealt{Davies2020a}). 
All these complexities require a careful modeling of a large number ($\sim 50$) of varying parameters for the dominant (massive, and/or close to the lensed images) lens galaxies and the cluster's dark matter halo. 

\begin{figure}[htb]
\centering%
\includegraphics[width=\textwidth]{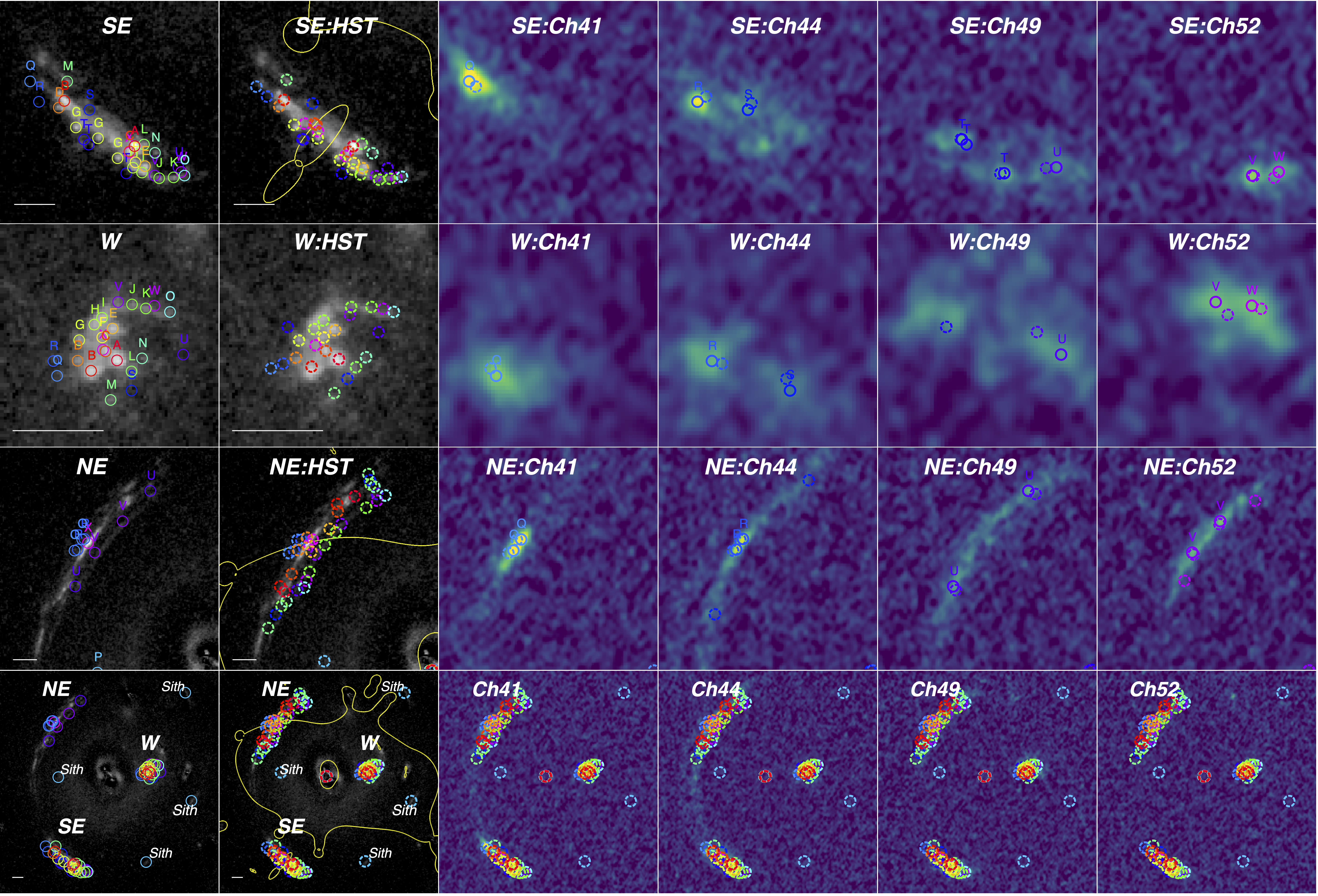}
\vspace{1ex}
\caption{%
The manually-marked knot lensed positions in the SE, W and NE arcs of our lensed $z=2.259$ galaxy J0901 and another lensed $z\sim3.1$ galaxy ``Sith''. 
\textit{Left two columns} show the same \textit{HST} F814W image (astrometry-corrected with Gaia DR3) but with different knot symbols: first column shows input knots (solid circles) and second column shows fitted ones (dashed circles) with critical lines (yellow line). 
\textit{Right four columns} show the ALMA CO(3--2) channel maps, where both input (solid) and output (dashed) knot positions are shown. 
From top to bottom, the field of view is zoomed to the SE, W, NE arcs and the full area, respectively. 
Knots \incode{A}--\incode{O} are marked based on the bright spots in the \textit{HST} image. The knot \incode{P} represents the higher-$z$ galaxy ``Sith''. 
knots \incode{Q}--\incode{W} are identified from the ALMA channel maps which is spatially complementary to the \textit{HST} data. 
Knot \incode{X} marks the most magnified position corresponding to the NE arc's brightest spot. 
The horizontal white line at the bottom left of each panel indicates 1$''$. 
}
\label{fig: lens modeling}
\end{figure}

We use the astrometry-corrected, foreground-subtracted \textit{HST} F814W image data, together with the new high-resolution ALMA CO(3--2) data cube, plus our \incode{Galfit}-fitted galaxy positions and magnitudes as the starting point for the lens modeling. 
We visually examine the \textit{HST} data and each channel map ($\sim$22~km/s) of the ALMA CO data cube to define a collection of ``knots'' and their positions in the image plane. 
We are able to define 36 bright knots in the HST data and 26 in the ALMA CO channel maps (peak $\mathrm{S/N} \sim 10 \text{--} 20$). These knots correspond to 16 and 7 compact stellar and CO emission in the source plane, respectively, plus \textit{Sith}. 
For each knot image a positional uncertainty is assigned based on the peak pixel's $\mathrm{S/N}$ and the resolution of the data, and is used during our lens model fitting. 
In Fig.~\ref{fig: lens modeling}, we show the knot positions in the \textit{HST} and ALMA channel maps. 

The combination of \textit{HST} and high-resolution ALMA channel map is the key improvement in this work. The $\sim 0.36''$ ALMA data not only provide locations of knots which are invisible in the \textit{HST} data and are at large galactocentric radii, due to heavy dust attention or too few stars, but also high spectral resolution which unambiguously separates knots in the velocity space, even including some in the highly-distorted NE arc. 

With the visually-identified knots and lens galaxies' locations and magnitudes, we performed lens modeling using the {\sc Glafic} software (\citealt{Oguri2010,Oguri2010asclsoft}). 
We first run the direct least-$\chi^2$ fitting using {\sc Glafic}. It is computationally efficient but is sensitive to the initial guess of parameters and may be trapped into local $\chi^2$ minima. For example, the brightest spot in the NE arc should correspond to a significantly-magnified ($>10$) region in J0901 (but is not the brightest spot in SE and W images) and the critical line should cross the NE arc near this position. This puts a strong limit in the lens halo mass and shape. We tried adopting a very large mass as the initial guess, finding that the $\chi^2$ minimization does not always converge given the large number of free parameters. 

\begin{figure}[htbp]
\centering%
\includegraphics[width=\linewidth]{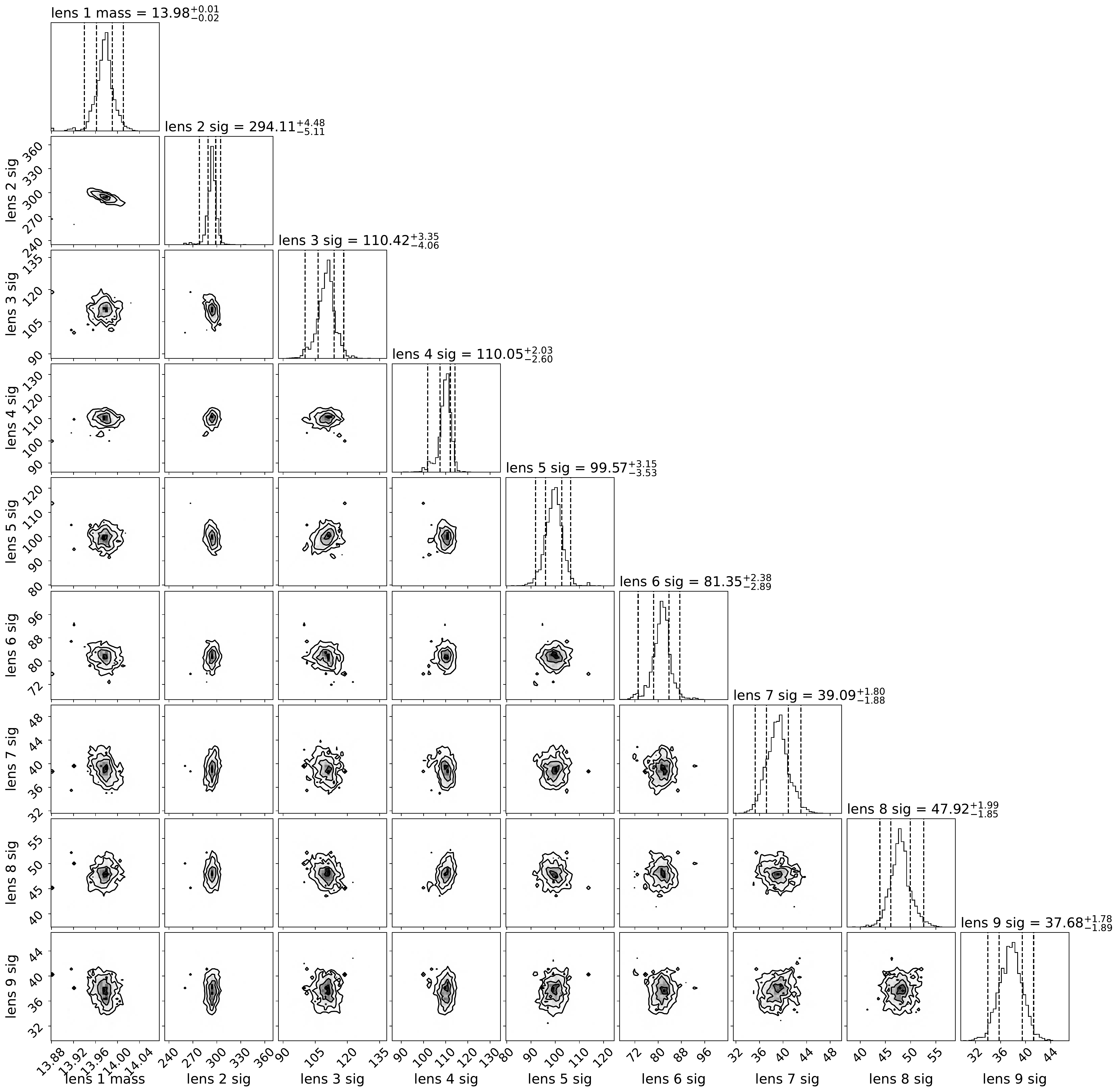}
\vspace{2ex}
\caption{
The posterior distributions of the lens cluster halo mass and cluster member galaxies' dispersions (equivalent to masses) from our MCMC fitting. 
Each panel shows the co-posterior distribution of each two parameter pair as labeled in $x$ and $y$ axes. 
Prior boundaries are $\pm0.1$~dex for these mass parameters. 
}
\label{fig: lensing MCMC corner plot}
\end{figure}

Then, we performed a complementary MCMC-based fitting. We use the Python {\sc emcee} package and run {\sc Glafic} in each MCMC iteration to sample the high-dimensional parameter space. This method is much more time-consuming but can effectively reveal parameter degeneracy and assess uncertainties. We performed the MCMC fitting with $\sim 100$ random walkers (twice the number of free parameters) and $10^3$ iterations. During each iteration, the MCMC sampler runs {\sc Glafic} with fixed parameters which are controlled by the sampler itself, then a likelihood is computed based on the corresponding offsets between each pair of input (source-plane) and output (image-plane) knots. If no corresponding lensed image is found for a knot, then we ignore its likelihood. 
To justify this approach, we tried other likelihood computation methods, for example, setting a very low likelihood in the case of missing an image-plane knot, which, however, often leads to no convergence. 
In Fig.~\ref{fig: lensing MCMC corner plot}, we present the tight PDFs of the lens mass parameters from our MCMC fitting.

\begin{figure}[htb]
\centering%
\includegraphics[width=0.95\textwidth]{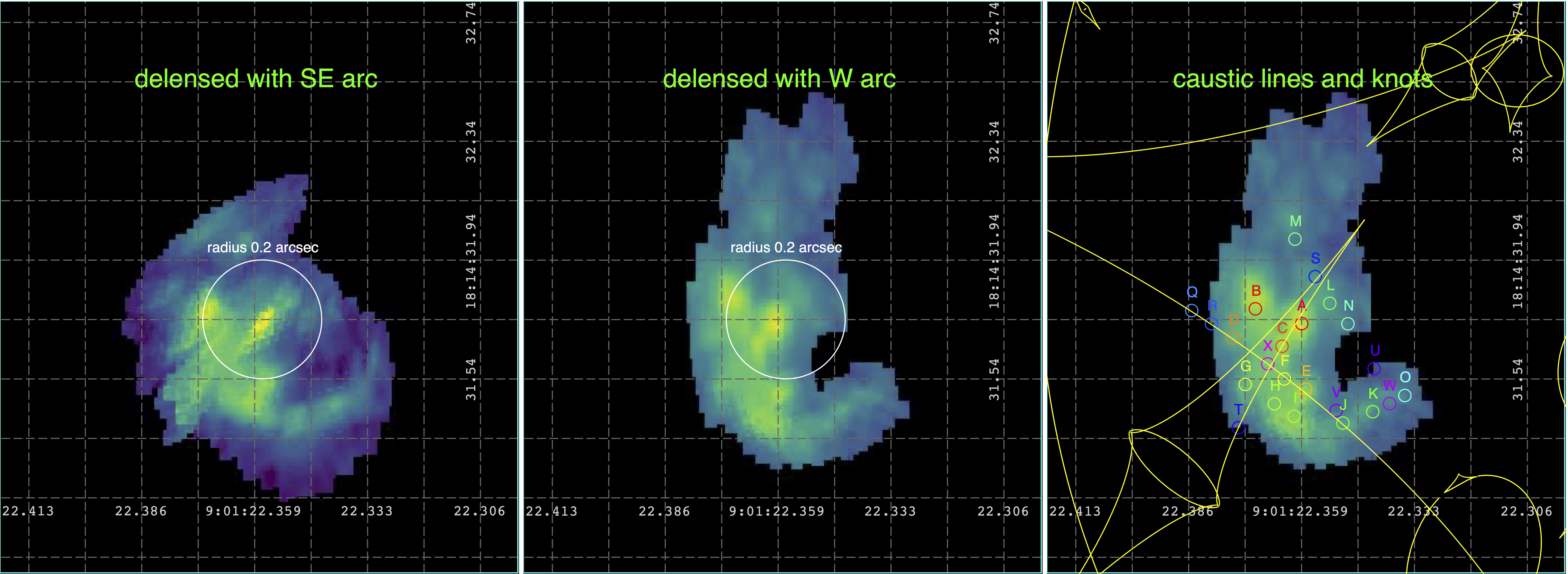}\\
\includegraphics[width=0.95\textwidth]{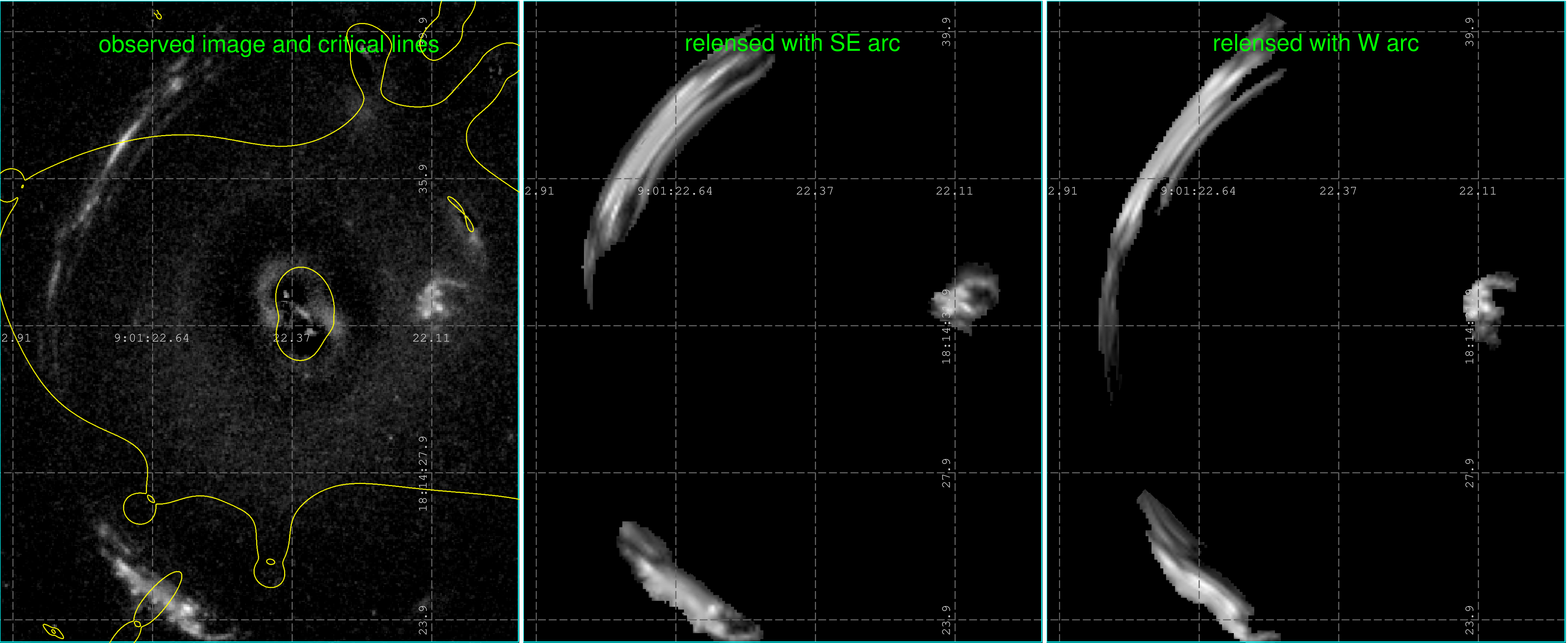}
\vspace{3.5ex}
\caption{%
\textit{Upper}: Source-plane \textit{HST} F814W images delensed from the SE and W arcs, in the first and second panels, respectively. The third panel is identical to the second one except for overlaying the identified knots (same as in Fig.~\ref{fig: lens modeling}) and caustic lines in the source plane. 
\textit{Lower}: Comparison of the observed and SE and W-arc relensed images, from left to right, respectively. The observed image is identical to the \textit{HST} image shown in Figs.~\ref{fig: galfit} and \ref{fig: lens modeling}. The relensed images are constructed by the method laid out in Appendix~\ref{appx: lensing}. Critical lines of our lens model is overlaid in the first lower panel.
}
\label{fig: delensing relensing}
\end{figure}

Our lens modeling could be further improved by: increasing the angular resolution for more accurate knot positions especially in the ALMA CO data, obtaining spectroscopic information for all lens galaxies, and confirming the redshift of \textit{Sith}.
Based on the current data, we verified that the uncertainty of the lens model to our kinematic study is limited to about 20\%. This is estimated by generating a few more testing lens models using parameters within the 2-sigma MCMC confidence level, then deriving delensed images and performing radial analysis as presented in Sect.~\ref{subsec: mass distribution}. 
In Fig.~\ref{fig: delensing relensing}, we further show the source-plane maps delensed with the SE and W arcs, respectively. 
Minor differences seen in the comparison are because the magnification factors are different in the two arcs. The same image-plane PSF corresponds to different source-plane resolutions in the two arcs, therefore their delensed maps naturally have minor differences. The bright spots and the faint spiral-arm like feature at south in the source plane do correspond to each other in Fig.~\ref{fig: delensing relensing}. The lower panels of Fig.~\ref{fig: delensing relensing} are mapped from the source-plane mesh grids containing the image brightness (top row) onto the image-plane mesh grids. Spatial smearing of the relensed images are caused by differential PSF across the source plane and the finite resolution of the mesh grid. The overall qualitative agreement nonetheless demonstrates our lens model is robust.

\begin{figure*}[htb]
\centering%
\begin{interactive}{animation}{{Plot_PSF_lensing_animation_00}.mp4}
\includegraphics[width=0.8\textwidth]{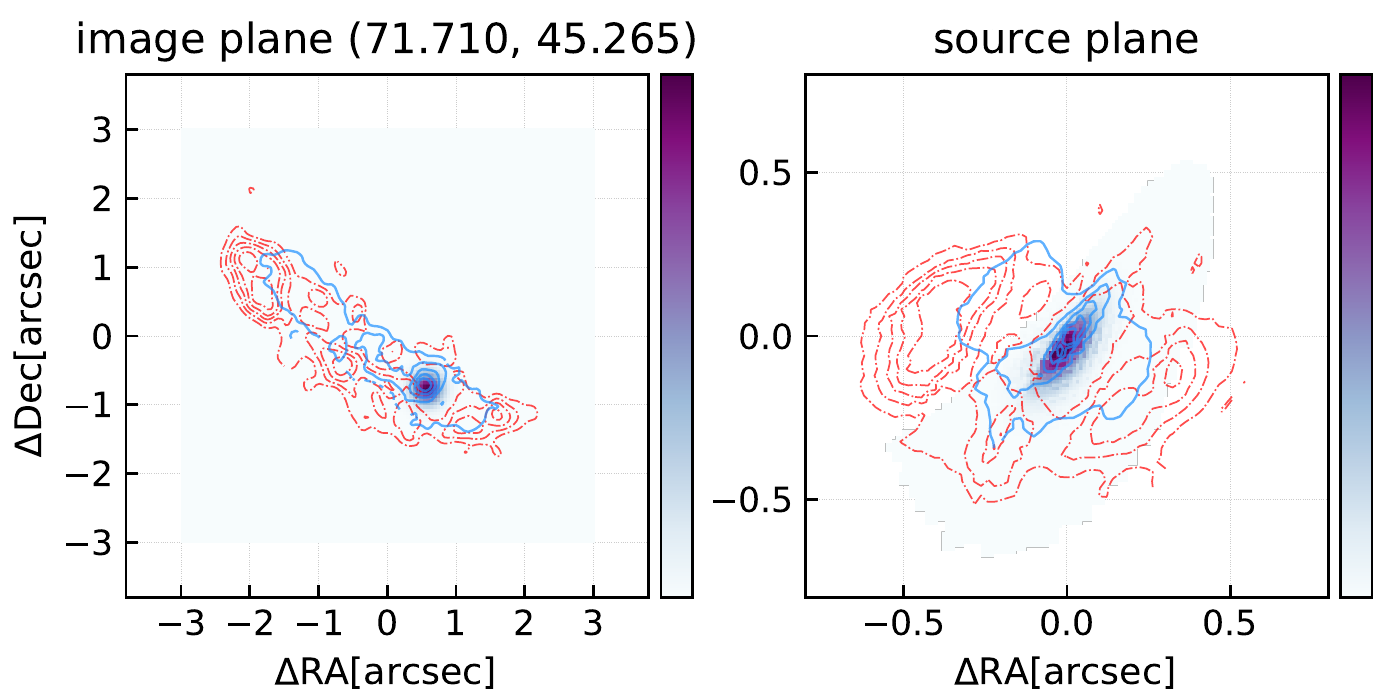}
\end{interactive}
\vspace{2ex}
\caption{%
Illustration of an image-plane PSF with a FWHM size of 0.36$''$ at an example location (left panel) and its delensed spatial distribution in the source plane (right panel). 
The red and blue contours overlaid are from the Briggs-weighting CO and combo-\Halpha{} intensity maps in Figs.~\ref{fig: image plane images} and \ref{fig: source plane images}. 
This figure is available as an animation online, where the deformation of a moving image-plane PSF (left panel) in the source plane (right panel) is illustrated. The movement of the image-plane PSF (left panel) is from top left to bottom right within the emission contour. The inline static figure is one frame of the animation with the image-plane PSF locating at the \Halpha{} intensity peak.
\label{fig: source plane psf}
}
\end{figure*}

\vspace{1ex}

\subsection{Delensing with mesh grid}
\label{appx: delensing}

Our delensing is based on the adaptive mesh grid file produced by {\sc Glafic}, which maps each grid cell as rectangles or irregular 4-vertex polygons between the image and source plane (either can be rectangle and the other irregular). 
We set the image-plane grid to be rectangle, with cell size varying from 1/4 to 2 pixels depending on the distance to critical lines. 
Each image-plane cell is then mapped to an irregular 4-vertex polygon in the source plane according to the adaptive mesh. 
We then calculate bi-linear interpolation in each irregular source-plane cell to obtain the delensed 2D image. 
In the case of data cube, we perform the delensing channel-by-channel. 
We choose a pixel size of $0.02''$ in the source plane, sufficient for sampling the source-plane PSF shape.

In order to visualize the variation of the source-plane PSF, we generated an array of image-plane 2D PSF profiles and delensed them. 
In Fig.~\ref{fig: source plane psf}, we illustrate how the moving $\sim 0.36''$ PSF in the image plane is delensed into the source plane. 
It is elongated along the northeast-southwest direction and the minor axis FWHM is about $0.07''$, or 560~pc, which coincidentally aligns with the kinematic major axis of J0901, thus providing a high spatial resolution for the J0901 rotation curve. 
Similarly we analyzed the source-plane PSF away from the center, finding a minor axis FWHM of about $0.14''$, or 1.1~kpc, and with a slightly rotated position angle.

\vspace{1ex}

\subsection{SED fitting}
\label{appx: SED fitting}

After delensing the foreground-subtracted, PSF-matched \textit{HST} images, we perform SED fitting to the five-band photometry pixel-by-pixel, using the FAST software (\citealt{Kriek2009,FAST}). It fits composite stellar population SEDs to the photometric data with star formation history (SFH), attenuation and filter response taken into consideration. 
As widely used for massive SFGs at high-$z$ (e.g., \citealt{Wuyts2011a}), we adopt solar metallicity, $\tau$-declining SFH, and \citet{Calzetti2000} attenuation law. 
Despite that the five photometric bands probe rest-frame UV to optical but not near-IR, the stellar mass is usually the most robustly constrained parameter from the SED fitting, compared to other parameters like age, attenuation and star formation history parameters (see, e.g., \citealt{Bell2000, Wuyts2012, Lang2014}).

\vspace{1ex}

\subsection{CO and \Halpha's line map creation}
\label{appx: line fitting}

We perform pixel-by-pixel spectral line fitting to the continuum-subtracted CO and \Halpha{} data cubes using a MCMC-based, custom, flexible multi-component and multi-constraint fitting procedure.

For the SINFONI data, we simultaneously fit a narrow and broad component to the \Halpha{} line and \Ntwo{} doublet to account for the star formation-dominated and outflow-dominated emission. 
This two-component fitting is necessary because the outflow emission is strong especially in the nuclear regions of J0901, related to the presence of the AGN (\citealt{Genzel2014b,Davies2020a}). 
In practice, the implementation of the outflow is that
for each main line component, we add an 1D Gaussian whose line width is parametrized by a variable as an increment to the main line component's width. 
This procedure makes sure that the outflow component is always broader than the main line. We further constrain the outflow line center to be within $\pm 500 \,\kms$ from the main line, so that it will not try to fit unphysically broad noise features. 
To avoid overfitting the data, especially in low S/N pixels, 
we tie the line centers and widths of the \Ntwo{} doublet to those of \Halpha{} for each of the narrow and broad emission.
The MCMC fitting is implemented with the \incode{pymc3} package, with flat line amplitude, velocity and dispersion priors. 
The uncertainty of each free or tied parameters is determined from the MC sampling, for which we used 3000 samplings.

For the CO data cube, a single 1D Gaussian profile is adequate to fit each pixel's spectrum as there is no detectable broad outflow component. 
We tested the impact of fitting a broad component, finding that the difference between the velocity dispersion of narrow component and that of the single-Gaussian fit is generally $<20\%$, and the outflow component broader than the dominant narrow component has an insignificant contribution (median amplitude ratio of the broad and narrow components is $\sim 0.07$) with large uncertainty.

\begin{figure*}[hptb]
\centering%
\includegraphics[width=0.92\textwidth]{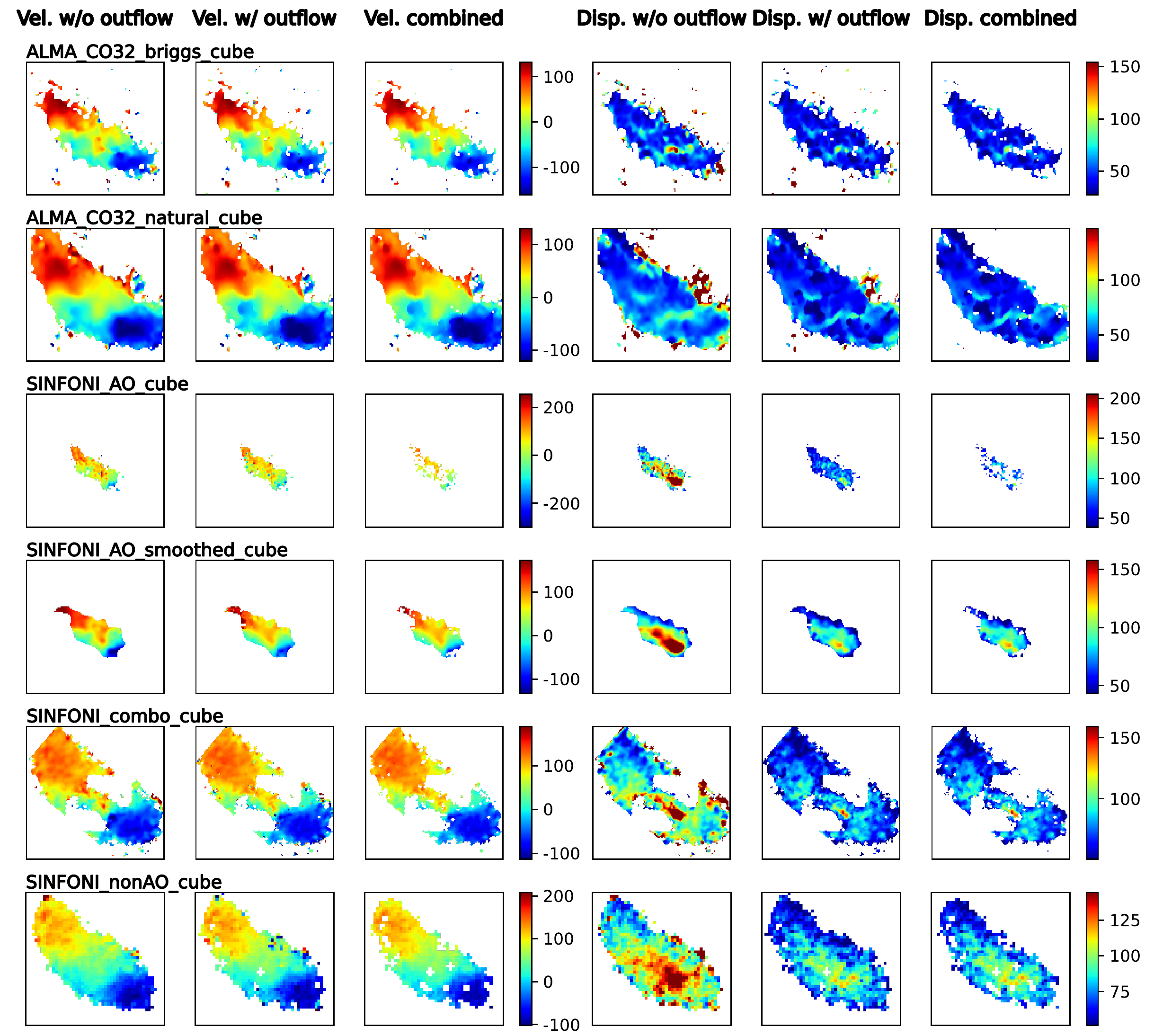}
\vspace{0.5ex}
\caption{%
Pixel-by-pixel spectral line fitting in the image-plane with and without outflow for all our data sets (one row for each data set). 
The left three columns are velocity maps, and right three columns are velocity dispersion maps. 
The first (fourth) column shows the velocity (dispersion) from the fitting without outflow, i.e., a single Gaussian line for CO or three Gaussian components for \Halpha{}+\Ntwo{}. 
The second (fifth) column shows the velocity (dispersion) from the fitting with outflow, i.e., two Gaussian components for CO and six Gaussian components for \Halpha{}+\Ntwo{} main lines and broad outflow components (showing the narrow component). 
The third (sixth) column shows the consistent pixels between the fitting with and without outflow, for \Halpha{} this is used for our later kinematic study (see in Appendix~\ref{appx: line fitting}). 
\label{fig: image plane outflow}
}
\end{figure*}

In Fig.~\ref{fig: image plane outflow}, we compare the fitting results for CO and \Halpha{} in the image plane, with and without outflow, respectively. 
The effect of including an outflow component is very minor for the velocity dispersion of CO as aforementioned, and is significant for that of \Halpha{}.
To ensure most reliable outflow removal for \Halpha{}, we select only pixels whose fitted main line velocities and velocity dispersions are consistent between the fitting with and without outflow (the ``combined'' panels in Fig.~\ref{fig: image plane outflow}, i.e., with velocity and velocity dispersion S/N~$>1$ and velocities agreeing within 50~km/s ($\sim$1.5$\times$ the LSF Gaussian sigma), for our kinematic study. 
This approach significantly reduces the apparent \Halpha{} dispersion in the nuclear area from $>$200~km/s to about 100~km/s (uncorrected for LSF), whereas the dispersions at the largest radii are mostly unaffected.

\newpage

\section{Image-plane kinematic fitting plots}
\label{appx: image-plane kinematic fitting plots}

\begin{figure*}[htb]
\centering%
\includegraphics[width=0.95\textwidth]{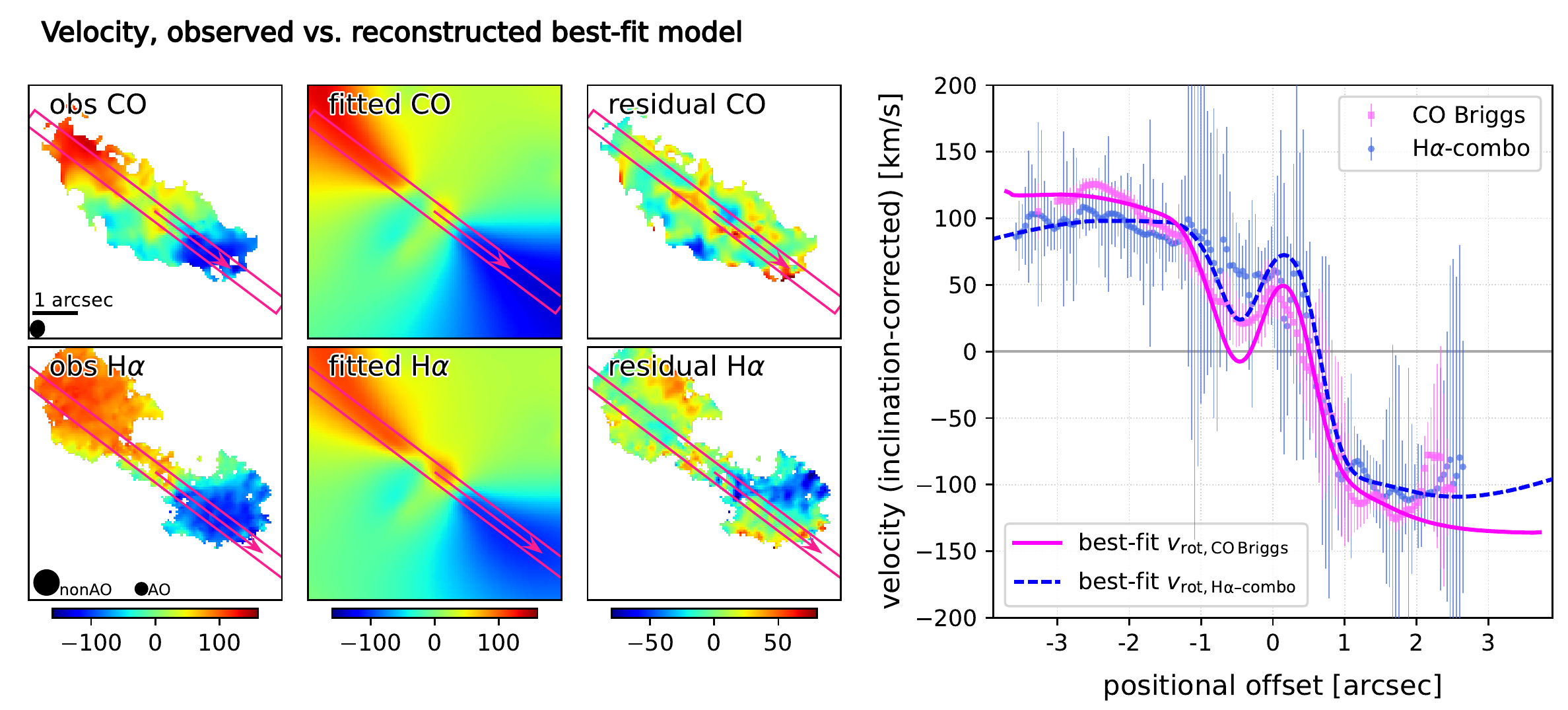}
\caption{%
Similar to Fig.~\ref{fig: obs vs model vel}, but showing the velocity maps and 1D profiles in the image plane. 
See Fig.~\ref{fig: obs vs model vel} caption for the details. 
}
\label{fig: obs vs model for velocity in image plane}
\end{figure*}

\begin{figure*}[htb]
\centering%
\includegraphics[width=0.95\textwidth]{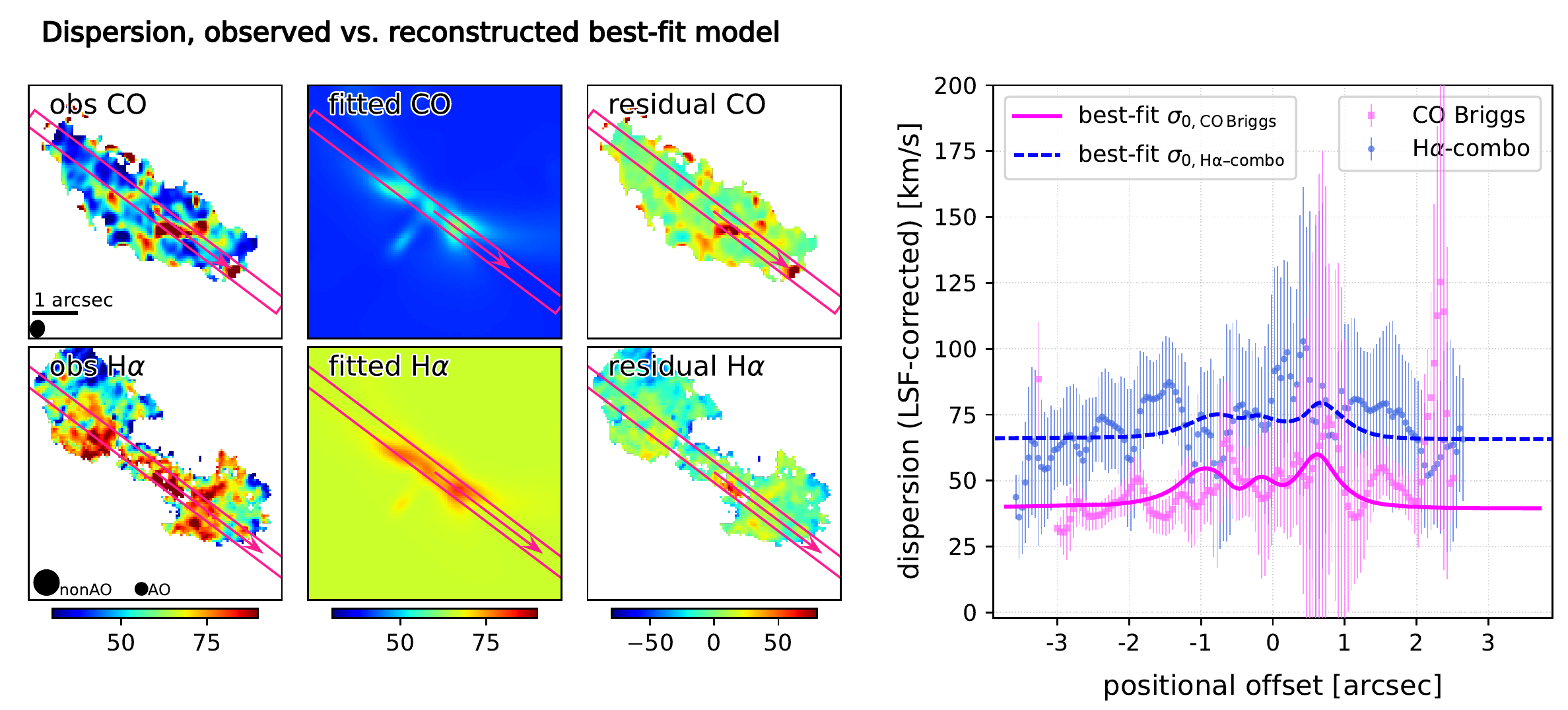}
\caption{%
Similar to Fig.~\ref{fig: obs vs model disp}, but showing the velocity dispersions in the image plane. 
See Fig.~\ref{fig: obs vs model disp} caption for the details.
}
\label{fig: obs vs model for dispersion in image plane}
\end{figure*}

Figs.~\ref{fig: obs vs model for velocity in image plane} and \ref{fig: obs vs model for dispersion in image plane} show the direct comparison of the observed data and our best-fit model in the image plane (Sect.~\ref{subsec: kinematic fitting}). 
We use our new code {\sc DysmalPy+Lensing} to directly fit the observed velocity and velocity dispersion along a pseudo slit in the image plane. The best-fit model has been convolved with the PSF and LSF when comparing to the observed data. Residual maps are shown in the third columns in Figs.~\ref{fig: obs vs model for velocity in image plane} and \ref{fig: obs vs model for dispersion in image plane}. The right-most panels show the velocity and velocity dispersion profiles extracted along the pseudo slit. As the comparison is in the image plane, the 1D profiles show twisted shapes because of the lensing.

\newpage

\section{Comparing kinematic fitting with different data sets, in different data spaces and lensing planes}
\label{appx: kinematic fitting for all}

\begin{figure*}[htbp]
\centering%
\includegraphics[width=0.492\textwidth, trim=6mm 0 0 0, clip]{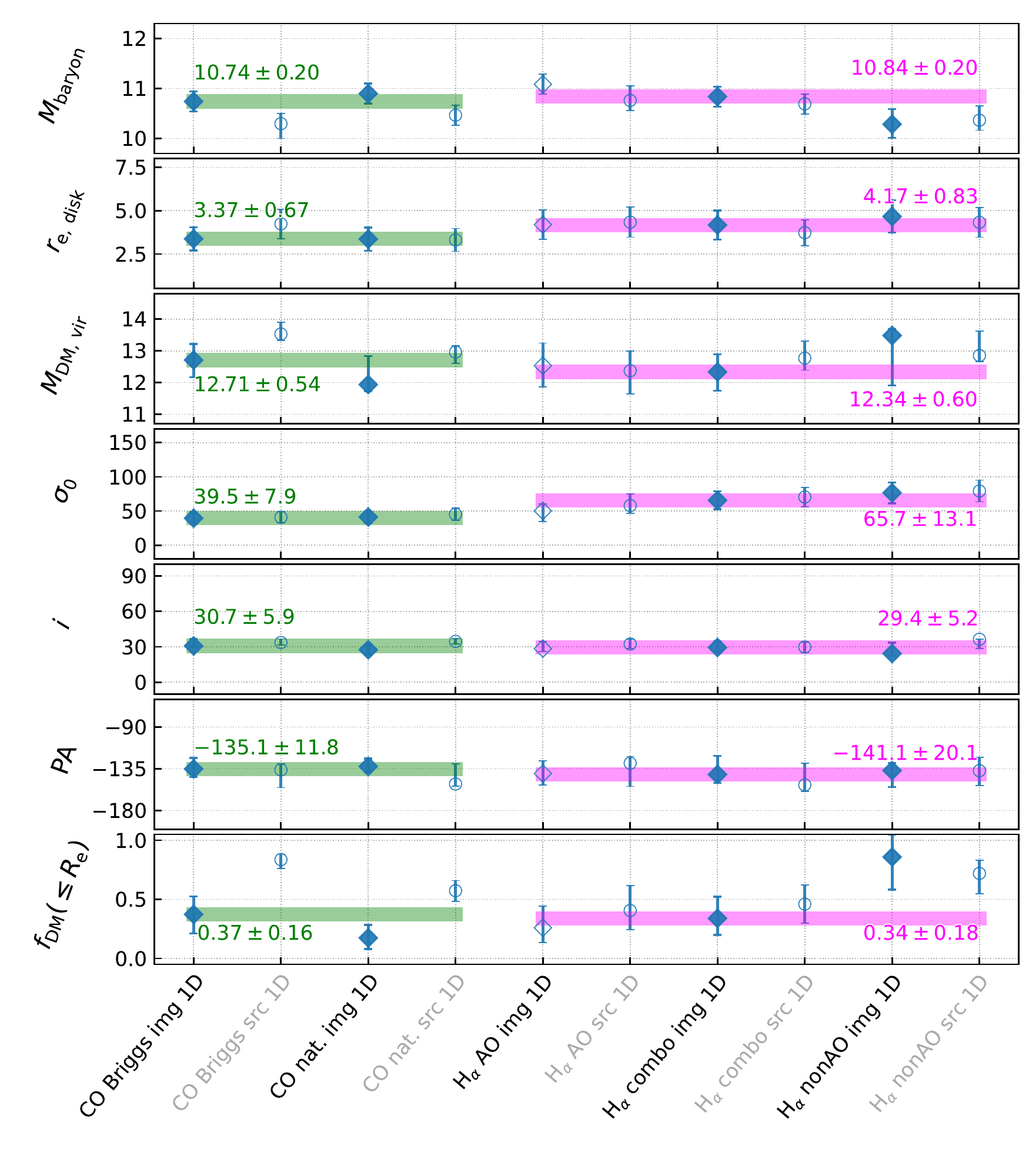}
\includegraphics[width=0.492\textwidth, trim=6mm 0 0 0, clip]{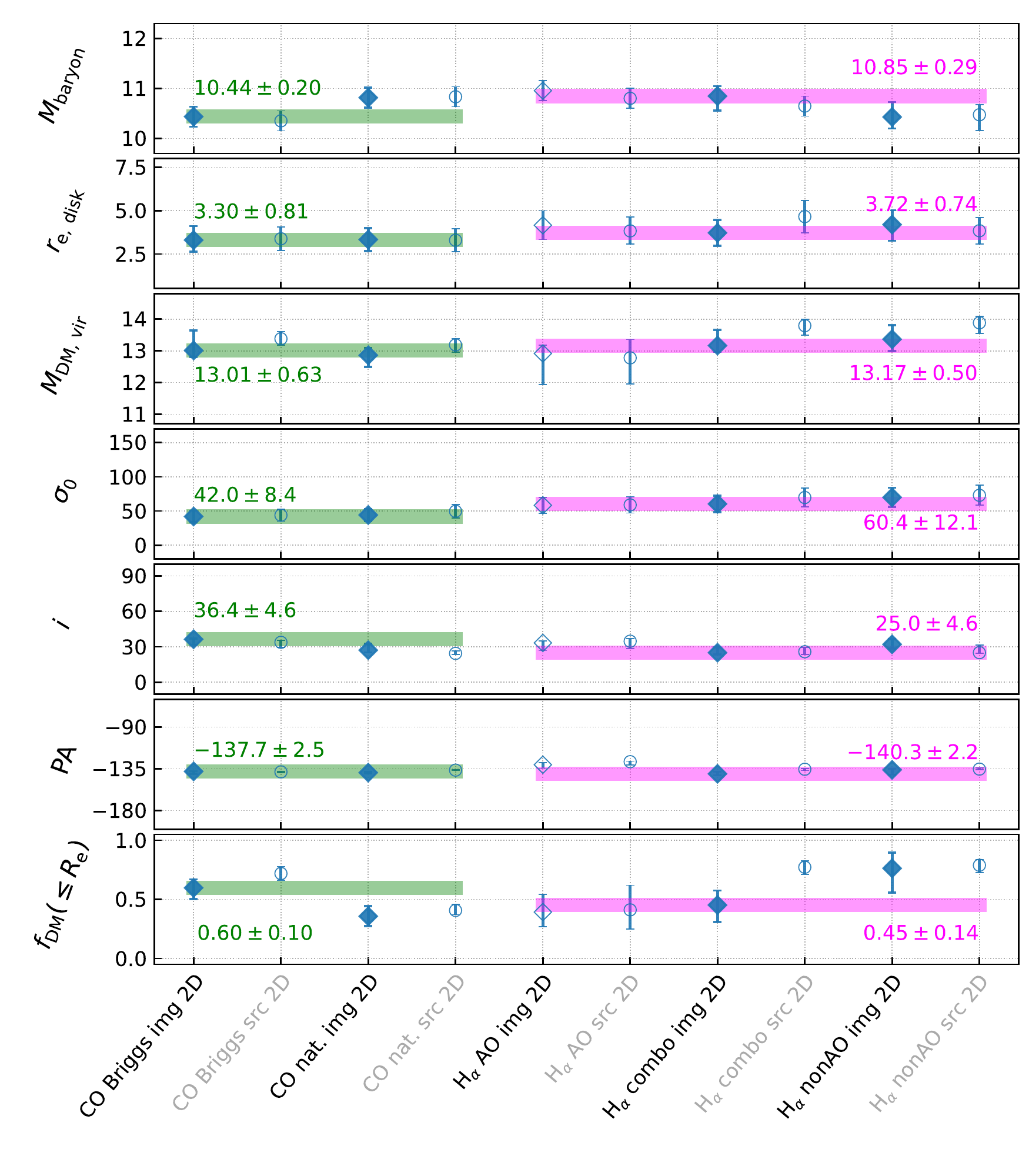}
\vspace{-0.5ex}
\caption{%
Comparison of our best-fit parameters from our \textsc{DysmalPy}(\textsc{+Lensing}) MCMC kinematic fittings to all our data sets. 
\textit{Left} panels are 1D data-space fittings and \textit{right} panels are those fitted in 2D data space. 
As labeled in the $x$-axis, the data sets are ALMA Briggs-weighting CO (beam~$\sim0.55''$), natural-weighting CO (beam~$\sim0.36''$), SINFONI AO (smoothed to PSF~$\sim0.3''$), combo (PSF~$\sim0.3''$--$0.6''$), and non-AO (PSF~$\sim0.5''$) \Halpha{} cubes. 
Solid symbols represent the fitting in the image plane (img), and open symbols are those in the source plane (src) which we argued in Appendix~\ref{appx: kinematic fitting for all} that they are unreliable due to the heterogeneous PSF. 
The panels from top to bottom are six fitted and one derivative parameters: the total disk+bulge baryon mass ($\Mbar$), disk \Sersic{} profile's effective radius ($r_{\mathrm{e,\,disk}}$), dark matter halo's virial mass ($M_{\mathrm{DM,\,vir}}$), disk's intrinsic dispersion ($\sig0$), inclination ($i$; 0 for face-on) and position angle (PA). 
Error bar indicates the uncertainty derived from our MCMC kinematic fitting (Sect.~\ref{subsec: kinematic fitting}). 
In each panel, we highlight the Briggs-weighting CO data set best-fits with a green horizontal line and text, and the \Halpha{} combo best-fits with a magenta line and text. 
The 2D fittings are highly affected by the features off the major axis and can deviate from the 1D fittings, but in general agree within the uncertainties. 
\label{fig: kinematics for all}
}
\end{figure*}

We tested kinematic fitting in the image- and source-plane, in 1D and 2D data spaces, and with five different data sets: 
\textit{i)} Briggs- and 
\textit{ii)} natural-weighting CO, 
\textit{iii)} AO, 
\textit{iv)} ``combo'', and 
\textit{v)} non-AO \Halpha{} data cubes. 
Fig.~\ref{fig: kinematics for all} compares the kinematic fitting results of all data products, fitted in either 1D or 2D data space, and in either image or source plane. 

Firstly, the $\Mbar$ (i.e., $M_{\mathrm{disk}} + M_{\mathrm{bulge}}$), $\sig0$, $\Reff$ and geometric parameters are tightly constrained with $\sim 0.2$~dex (or $\sim 20$\%) uncertainties. 
Only $\Mvir$ has a large uncertainty of $\sim 0.6$~dex (see Table~\ref{Table1} and Fig.~\ref{fig: kinematics for all}). 
The CO and the combo \Halpha{} results have the smallest error bars because of they have both best spatial coverage and resolution. The AO data alone without covering the outer rotation curve leads to a too high $\Mbar$ and too low $\Mvir$ as expected, whereas the non-AO data alone leads to large uncertainties. 

Secondly, fitting in the source plane without considering the varying PSF deflected shapes leads to incorrect results, e.g., sometimes leading to large discrepancies especially in $\Mvir$ and $\Mbar$ ($\sim 0.5 \text{--} 0.8$~dex). 
Our tests demonstrate that implementing the lensing transformation and fitting in the image-plane is very necessary. 

Thirdly, fitting in the 1D and 2D data space leads to good agreement in $\Mbar$, disk/bulge radii, $\sig0$ and geometric parameters, but the 2D data-space fittings result in systematically larger $\Mvir$ values for J0901 (although still within uncertainty). 
This discrepancy may be related to the precision of lensing deflection calculation in pixel grids, and the limited angular resolution and potential non-circular motions in 2D. The 1D data-space fitting focuses on the major kinematic axis profile and is therefore better for determining the rotation curve (see also discussions in \citealt{Genzel2017,Genzel2020,Price2021}). 

Given these facts, we therefore focused on the fitting results in the 1D data space from the Briggs-weighting CO and combo-\Halpha{} data sets which can recover the rotation curve the best. 
The key fitting results are listed in Table~\ref{Table1}.

\newpage

\FloatBarrier


\input{Article_J0901_main.bbl}
\bibliographystyle{aasjournal}

\end{document}

%% file: Input_usepackage.tex
\usepackage{graphicx}
\usepackage{float}
\usepackage{lipsum}
\usepackage{amsfonts,amsmath,amssymb}
\usepackage{natbib}
\usepackage{url}
\usepackage{xcolor}
\usepackage{hyperref}
\usepackage{parskip}
\usepackage[numbered]{bookmark}
\usepackage{calculator}
\usepackage{placeins} 
\let\splitbox\relax
\usepackage{adjustbox} 
\usepackage{enumitem} 
\usepackage{xifthen}
\usepackage{chngcntr} 

\usepackage{bm}

\usepackage{tabularx}

\usepackage{tikz}

\hypersetup{
    colorlinks,
    linkcolor={blue},
    citecolor={blue!80!black},
    urlcolor={magenta!80!black},
}

\usepackage{listings} 
\usepackage{color}
\definecolor{lightgray}{gray}{0.9}

\usepackage{lmodern}
\usepackage{slantsc}

\usepackage{multirow}

\usepackage{verbatimbox}

%% file: Input_newcommand.tex
\newcommand{\TODO}[2][]{%
\ifthenelse{\isempty{#1}}%
{\textbf{\textcolor{red}{TODO: #2}}}%
{\textbf{\textcolor{red}{TODO (#1) #2}}}%
}

\def\Sersic{S\'{e}rsic}

\def\gtsima{$\; \buildrel > \over \sim \;$}
\def\ltsima{$\; \buildrel < \over \sim \;$}
\def\prosima{$\; \buildrel \propto \over \sim \;$}
\def\gsim{\lower.5ex\hbox{\gtsima}}
\def\lsim{\lower.5ex\hbox{\ltsima}}
\def\simgt{\lower.5ex\hbox{\gtsima}}
\def\simlt{\lower.5ex\hbox{\ltsima}}
\def\simpr{\lower.5ex\hbox{\prosima}}

\def\beq{\begin{equation}}
\def\eeq{\end{equation}}
\def\8mu{8\,$\mu{\rm m}$}
\def\16mu{16\,$\mu{\rm m}$}
\def\24mu{24\,$\mu{\rm m}$}
\def\70mu{70\,$\mu{\rm m}$}

\def\alphaCO{\alpha_{\mathrm{CO}}}

\def\nH2{n_{\mathrm{H}_2}}
\def\NH2_{N_{\mathrm{H}_2}}

\def\lgNH2_{\mathrm{log}_{10}\,(N_{\mathrm{H}_2} / \mathrm{cm^{-2}})}

\def\HIrom{\mathrm{H}\textnormal{\textup{\uppercase\expandafter{\romannumeral 1}}}}
\def\HIIrom{\mathrm{H}\textnormal{\textup{\uppercase\expandafter{\romannumeral 2}}}}

\def\Msun{\mathrm{M}_{\odot}}
\def\Msyr{\mathrm{M}_{\odot}\,\mathrm{yr}^{-1}}

\def\Kkmspc2{\mathrm{K\,km\,s^{-1}\,pc^{2}}}
\def\kms{\mathrm{km\,s^{-1}}}
\def\SFR{\mathrm{SFR}}

\def\Mstar{M_{\star}}

\def\Mgas{M_{\mathrm{gas}}}

\def\Mvir{M_{\mathrm{DM,\,vir}}}
\def\Mbar{M_{\mathrm{baryon}}}

\def\Reff{R_{\mathrm{e}}}
\def\Rhalf{R_{\mathrm{half}}}

\def\fH2_{f_{\mathrm{H_2}}}
\def\MH2_{M_{\mathrm{H_2}}}
\def\fDM{f_{\mathrm{DM}}}
\def\Sigbar{\Sigma_{\mathrm{baryon}}}

\def\kms{\mathrm{km/s}}

\def\michi2{\textsc{MiChi2}}

\def\NIIHA{[N\,\textsc{ii}]/H$\alpha$}
\def\Ntwo{[N\,\textsc{ii}]}
\def\Halpha{H$\alpha$}
\def\OIIIHB{[O\,\textsc{iii}]/H$\beta$}

\def\Nfive{[N\,\textsc{v}]}
\def\Ctwo{[C\,\textsc{ii}]}

\def\vrot{v_{\mathrm{rot}}}

\def\fgas{f_{\mathrm{gas}}}
\def\SFR{\mathrm{SFR}}
\def\kpc{\mathrm{kpc}}
\def\sig0{\sigma_{0}}

\newcommand{\J}[2][]{%
	\ifthenelse{\isempty{#1}}%
	{\begingroup%
		\def\JlowerN{}%
		\ADD{#2}{-1}{\JlowerN}%
		J=#2{\to}\JlowerN%
		\endgroup%
	}%
	{J=#1{\to}#2}%
}

\newcommand{\rom}[1]{{{\uppercase\expandafter{\romannumeral #1}}}}
\newcommand{\romup}[1]{{\textup{\uppercase\expandafter{\romannumeral #1}}}}

\newcommand{\incode}[1]{{\raggedright\lstinline|#1|}}
\newcommand{\incodep}[1]{{\raggedright\lstinline|"#1"|}}

\defcitealias{Liudz2015}{L15}
\defcitealias{DL07}{DL07}
\defcitealias{V20}{V20}

\renewcommand{\figurename}{Fig.}

\usepackage{xifthen}
\usepackage{contour} 
\usepackage{twoopt} 
\usepackage[normalem]{ulem}
\newcommandtwoopt{\MYREVISED}[3][][]{%
\ifthenelse{\isempty{#3}}%
{%
\ifthenelse{\isempty{#2}}%
{
\ifthenelse{\equal{#1}{\myrevisiontag}}{%
\bm{\textcolor{black}{#1}}%
}{}%
}%
{
\ifthenelse{\equal{#1}{\myrevisiontag}}{%
\textcolor{red!30!black}{\sout{#2}}%
}{}%
}%
}%
{%
\ifthenelse{\isempty{#2}}%
{
\ifthenelse{\equal{#1}{\myrevisiontag}}{%
\bm{\textcolor{black}{#3}}%
}{#3}%
}%
{
\ifthenelse{\equal{#1}{\myrevisiontag}}{%
\textcolor{red!30!black}{\sout{#2}}%
\bm{\textcolor{black}{#3}}%
}{#3}%
}%
}%
}

%% file: Input_authors.tex
\correspondingauthor{Daizhong Liu}
\email{dzliu@mpe.mpg.de}

\author[0000-0001-9773-7479]{Daizhong Liu}
\affiliation{Max-Planck-Institut f\"ur Extraterrestrische Physik (MPE), Giessenbachstr. 1, D-85748 Garching, Germany}

\author[0000-0003-4264-3381]{N.~M. F\"{o}rster Schreiber}
\affiliation{Max-Planck-Institut f\"ur Extraterrestrische Physik (MPE), Giessenbachstr. 1, D-85748 Garching, Germany}

\author[0000-0002-2767-9653]{R. Genzel}
\affiliation{Max-Planck-Institut f\"ur Extraterrestrische Physik (MPE), Giessenbachstr. 1, D-85748 Garching, Germany}

\author[0000-0003-0291-9582]{D. Lutz}
\affiliation{Max-Planck-Institut f\"ur Extraterrestrische Physik (MPE), Giessenbachstr. 1, D-85748 Garching, Germany}

\author[0000-0002-0108-4176]{S.~H. Price}
\affiliation{Max-Planck-Institut f\"ur Extraterrestrische Physik (MPE), Giessenbachstr. 1, D-85748 Garching, Germany}
\affiliation{Department of Physics and Astronomy and PITT PACC, University of Pittsburgh, Pittsburgh, PA 15260, USA}

\author[0000-0001-7457-4371]{L. L. Lee}
\affiliation{Max-Planck-Institut f\"ur Extraterrestrische Physik (MPE), Giessenbachstr. 1, D-85748 Garching, Germany}

\author[0000-0002-7892-396X]{Andrew J. Baker}
\affiliation{Department of Physics and Astronomy, University of the Western Cape, Robert Sobukwe Road, Bellville 7535, South Africa} 
\affiliation{Department of Physics and Astronomy, Rutgers, the State University of New Jersey, 136 Frelinghuysen Road, Piscataway, NJ 08854-8019, USA}

\author[0000-0001-6879-9822]{A. Burkert}
\affiliation{Max-Planck-Institut f\"ur Extraterrestrische Physik (MPE), Giessenbachstr. 1, D-85748 Garching, Germany}

\author[0000-0002-4343-0479]{R.~T. Coogan}
\affiliation{Max-Planck-Institut f\"ur Extraterrestrische Physik (MPE), Giessenbachstr. 1, D-85748 Garching, Germany}

\author[0000-0003-4949-7217]{R.~I. Davies}
\affiliation{Max-Planck-Institut f\"ur Extraterrestrische Physik (MPE), Giessenbachstr. 1, D-85748 Garching, Germany}

\author[0000-0002-3324-4824]{R.~L. Davies}
\affiliation{Centre for Astrophysics and Supercomputing, Swinburne Univ. of Technology, P.O. Box 218, Hawthorn, VIC 3122, Australia}
\affiliation{ARC Centre of Excellence for All Sky Astrophysics in 3 Dimensions (ASTRO 3D), Australia} 

\author[0000-0002-2775-0595]{R. Herrera-Camus}
\affiliation{Max-Planck-Institut f\"ur Extraterrestrische Physik (MPE), Giessenbachstr. 1, D-85748 Garching, Germany}
\affiliation{Departamento de Astronom\'{i}a, Universidad de Concepci\'{o}n, Barrio Universitario, Concepci\'{o}n, Chile}

\author[0000-0002-2993-1576]{Tadayuki Kodama}
\affiliation{Astronomical Institute, Tohoku University, 6-3, Aramaki, Aoba, Sendai, Miyagi, 980-8578, Japan}

\author[0000-0002-2419-3068]{Minju~M. Lee}
\affiliation{Max-Planck-Institut f\"ur Extraterrestrische Physik (MPE), Giessenbachstr. 1, D-85748 Garching, Germany}

\author[0000-0003-1785-1357]{A. Nestor}
\affiliation{School of Physics and Astronomy, Tel Aviv University, Ramat Aviv 69978, Israel}

\author[0000-0002-1428-1558]{C. Pulsoni}
\affiliation{Max-Planck-Institut f\"ur Extraterrestrische Physik (MPE), Giessenbachstr. 1, D-85748 Garching, Germany}

\author[0000-0002-7093-7355]{A. Renzini}
\affiliation{INAF - Osservatorio Astronomico di Padova, Vicolo dell'Osservatorio 5, I-35122 Padova, Italy}

\author[0000-0002-6250-5608]{Chelsea E. Sharon}
\affiliation{Yale-NUS College, 16 College Ave West 01-220, 138527, Singapore}

\author[0000-0002-2125-4670]{T.~T. Shimizu}
\affiliation{Max-Planck-Institut f\"ur Extraterrestrische Physik (MPE), Giessenbachstr. 1, D-85748 Garching, Germany}

\author[0000-0002-1485-9401]{L.~J. Tacconi}
\affiliation{Max-Planck-Institut f\"ur Extraterrestrische Physik (MPE), Giessenbachstr. 1, D-85748 Garching, Germany}

\author[0000-0001-9728-8909]{Ken-ichi Tadaki}
\affiliation{National Astronomical Observatory of Japan, 2-21-1 Osawa, Mitaka, Tokyo 181-8588, Japan}

\author[0000-0003-4891-0794]{H. \"{U}bler}
\affiliation{Cavendish Laboratory, University of Cambridge, 19 JJ Thomson Avenue, Cambridge CB3 0HE, UK}
\affiliation{Kavli Institute for Cosmology, University of Cambridge, Madingley Road, Cambridge CB3 0HA, UK}

%% file: Input_table_1.tex
\begin{table}[htbp]
    \centering
    \caption{J0901 Properties}
    \begin{tabular}{l l}
    \hline
    \hline
    Name & SDSS~J090122.37+181432.3 \\
    R.A. Dec. (J2000) & 09h01m22.59s 18d14m24.20s \\
    Redshift & 2.259 \\
    $\log (M_{\star} \,/\, \Msun)$ \tablenotemark{a} & $\sim 11.2$ \\
    $\mathrm{SFR} \,/\, (\Msyr)$ \tablenotemark{a} & $\sim 200$ \\
    $\Delta \mathrm{MS} \,/\, \mathrm{dex} $ \tablenotemark{b} & $-0.05$ \\
    \hline
    \textit{(Photometry)} & \\
    $\log (M_{\star}({\mathrm{phot.}}) \,/\, \Msun)$ & $11.04 \pm 0.3$ \\
    $\log (M_{\mathrm{mol,\,gas}}({\mathrm{phot.}}) \,/\, \Msun)$ & $10.88 \pm 0.3$ \\
    $\log (M_{\mathrm{baryon}}({\mathrm{phot.}}) \,/\, \Msun)$ & $11.26 \pm 0.3$ \\
    $\Reff ({\mathrm{phot.}}) \,/\, \kpc$ & $3.85$ \\
    $\Rhalf ({\mathrm{phot.}}) \,/\, \kpc$ & $3.23$ \\
    \hline
    \textit{(Kinematics)} & \\
    Inclination & $30.7_{-0.8}^{+5.9}$ \\
    Position angle & $-135.1_{-8.9}^{+11.8}$ \\
    $v_{\mathrm{rot}, \, \Reff} \,/\, (\mathrm{km \, s^{-1}})$ & $240.7_{-20.3}^{+31.8}$ \\
    $\sigma_{0, \, \mathrm{cold\,gas}} \,/\, (\mathrm{km \, s^{-1}})$ & $37.3_{-2.8}^{+2.5}$ \\
    $\sigma_{0, \, \mathrm{ionized\,gas}} \,/\, (\mathrm{km \, s^{-1}})$ & $64.3_{-5.1}^{+5.0}$ \\
    $\log M_{\mathrm{DM}} (\mathrm{virial}) \,/\, \Msun$ & $12.64_{-0.51}^{+0.48}$ \\
    $\log M_{\mathrm{baryon}} ({\mathrm{kin.}}) \,/\, \Msun$ & $10.72_{-0.14}^{+0.14}$ \\
    CO $f_{\mathrm{DM}} (\Reff)$ & $0.44_{-0.16}^{+0.15}$ \\ 
    H$\alpha$ $f_{\mathrm{DM}} (\Reff)$ & $0.36_{-0.14}^{+0.18}$ \\
    \hline
    \hline
    \end{tabular}
    \begin{minipage}[t]{0.95\linewidth}
        \tablenotetext{a}{From \citet{Davies2020a}. }
        \tablenotetext{b}{Using \citet{Speagle2014} MS. }
    \end{minipage}
    \label{Table1}
\end{table}

%% file: Input_fig_5.tex
\begin{figure*}[ht]
\centering%
\includegraphics[width=0.92\textwidth]{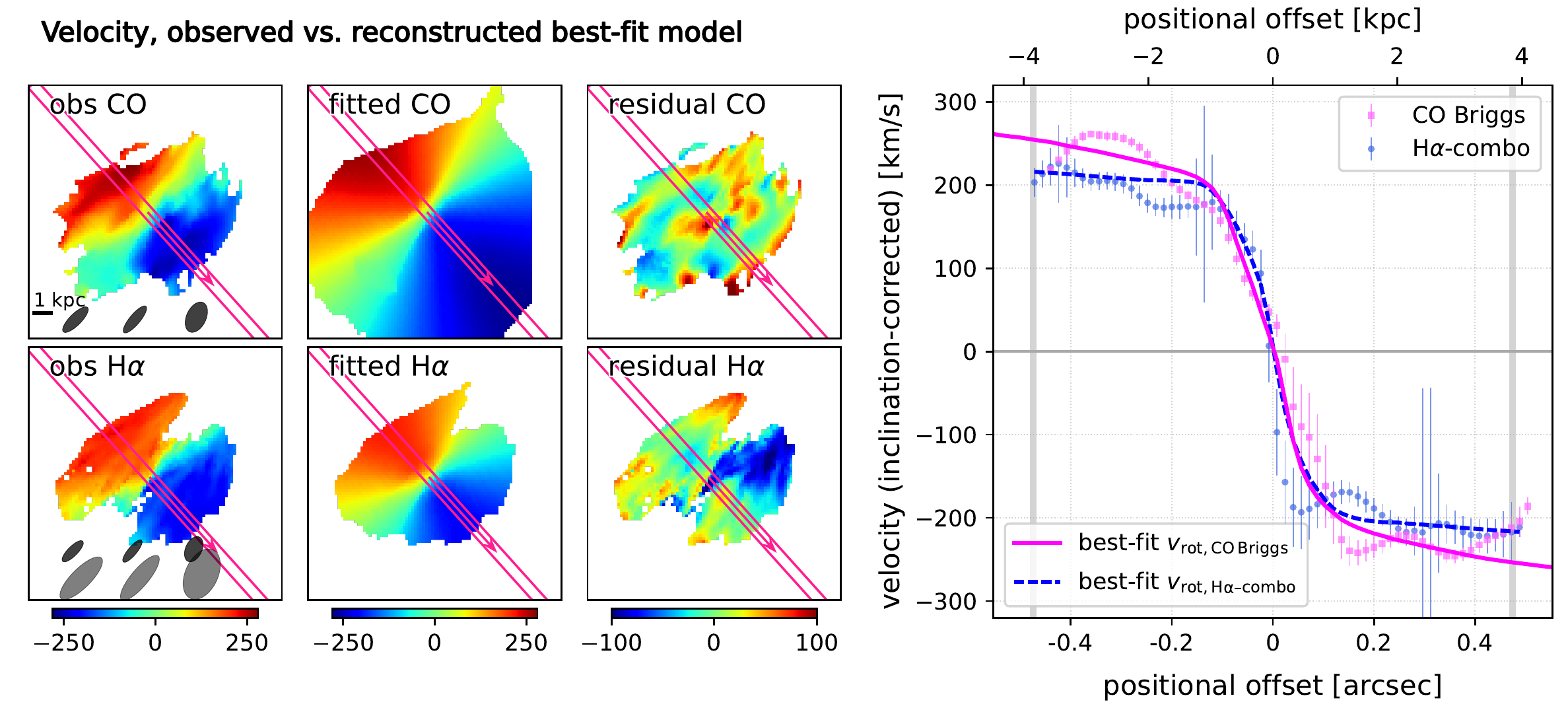}\\
\vspace{1ex}
\caption{%
Comparison of the observed (delensed) and best-fit model reconstructed 2D velocity maps and 1D pseudo-slit extraction profiles. 
\textit{Left panels} are the observed and delensed Briggs-weighting CO (upper) and combo-\Halpha{} (lower) velocity maps. PSFs are shown at the bottom as in Fig.~\ref{fig: mass distribution} (see caption therein). For the combo-\Halpha{}, smaller PSFs correspond to the AO data and larger ones to the non-AO data. 
\textit{Middle-left panels} are the reconstructed and delensed velocity maps based on our best-fit kinematic models. 
\textit{Middle-right panels} show the residual maps.
\textit{The rightmost panel} shows the extracted 1D velocity profiles in a pseudo-slit along the kinematic major axis shown as a magenta rectangle in the left panels. The magenta arrow inside the slit rectangle starts at positional offset 0 and points towards positive offsets. Error bars in the rightmost panel are the uncertainties derived from our line fitting. For \Halpha{}, both a narrow-line and a broad-line outflow components are fitted, and only the narrow component is shown here. 
\label{fig: obs vs model vel}
}
\end{figure*}

%% file: Input_fig_6.tex
\begin{figure*}[ht]
\centering%
\includegraphics[width=0.92\textwidth]{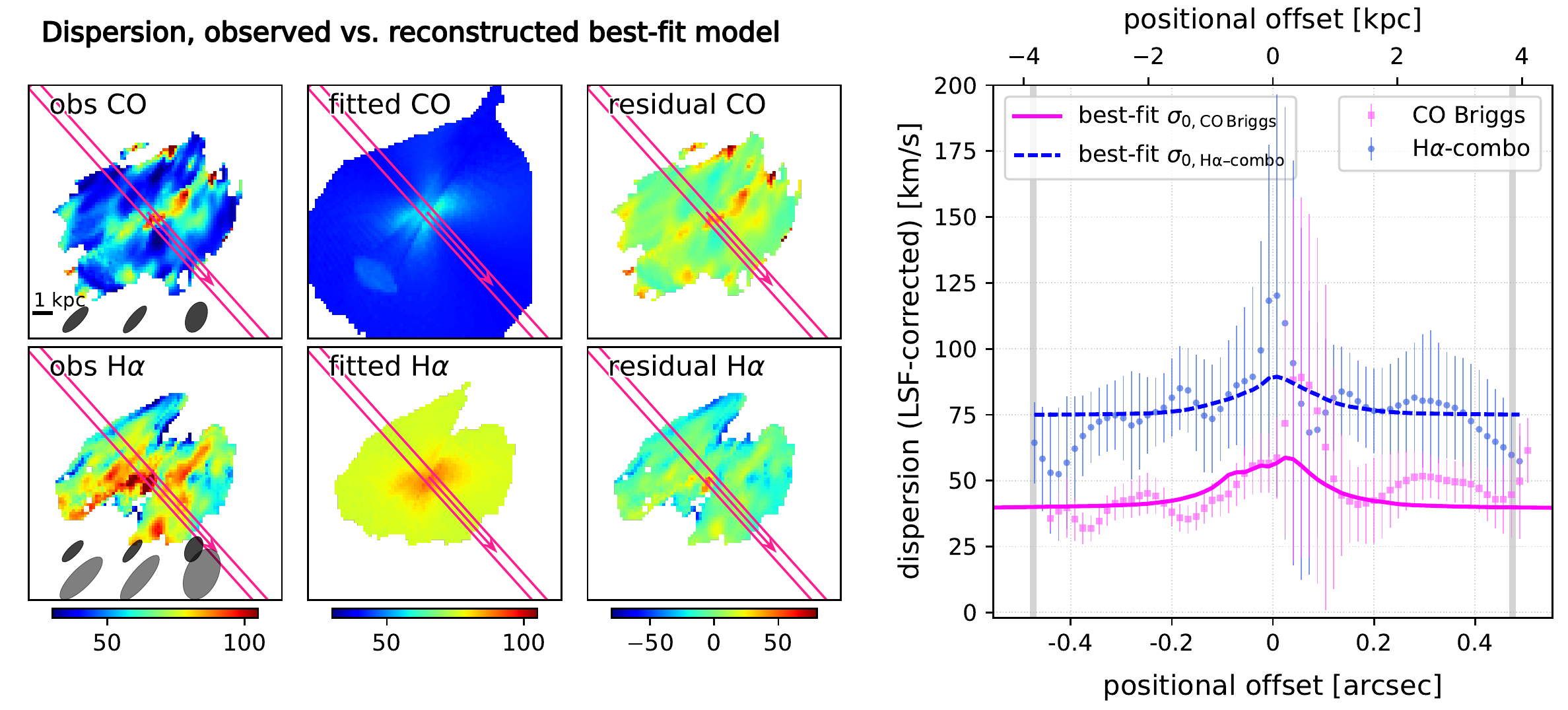}\\
\vspace{1ex}
\caption{%
Similar to Fig.~\ref{fig: obs vs model vel}, showing the velocity dispersion 2D maps and 1D pseudo-slit extraction profiles. See Fig.~\ref{fig: obs vs model vel} caption. 
\label{fig: obs vs model disp}
}
\end{figure*}

%% file: Input_fig_7.tex
\begin{figure}[ht]
\centering%
\includegraphics[width=0.95\linewidth]{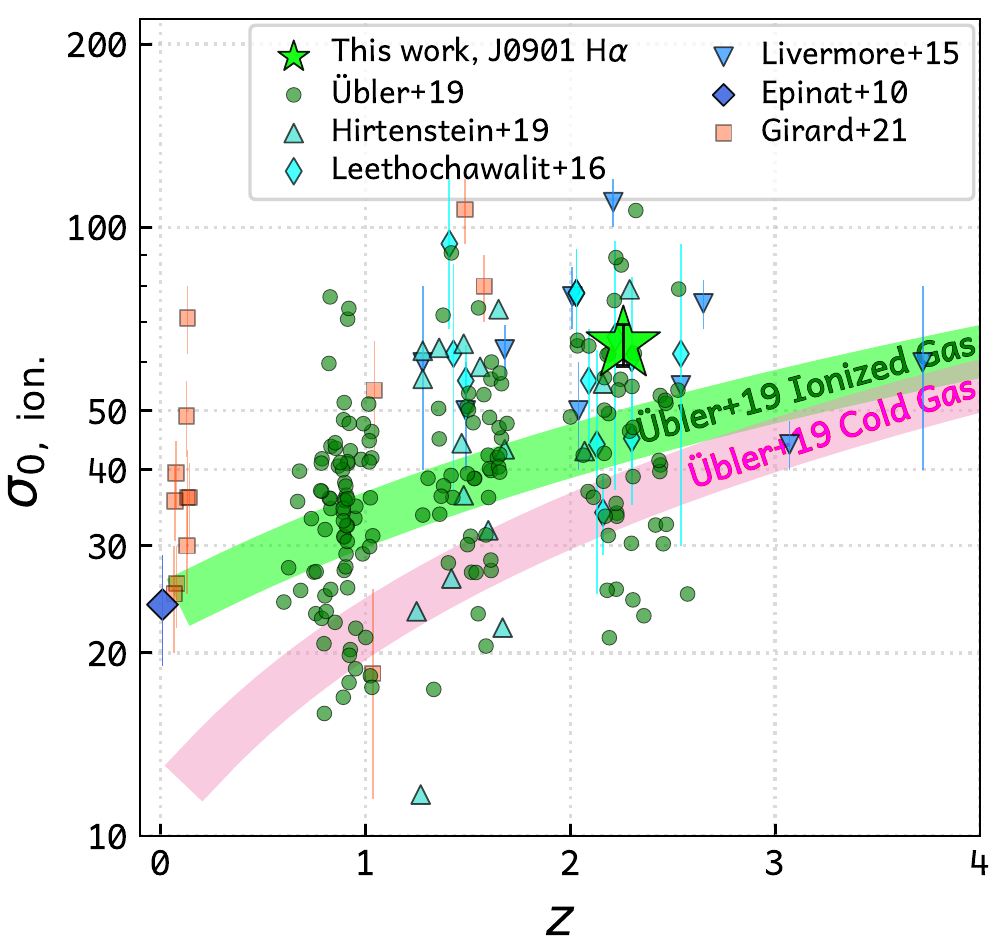}
\includegraphics[width=0.95\linewidth]{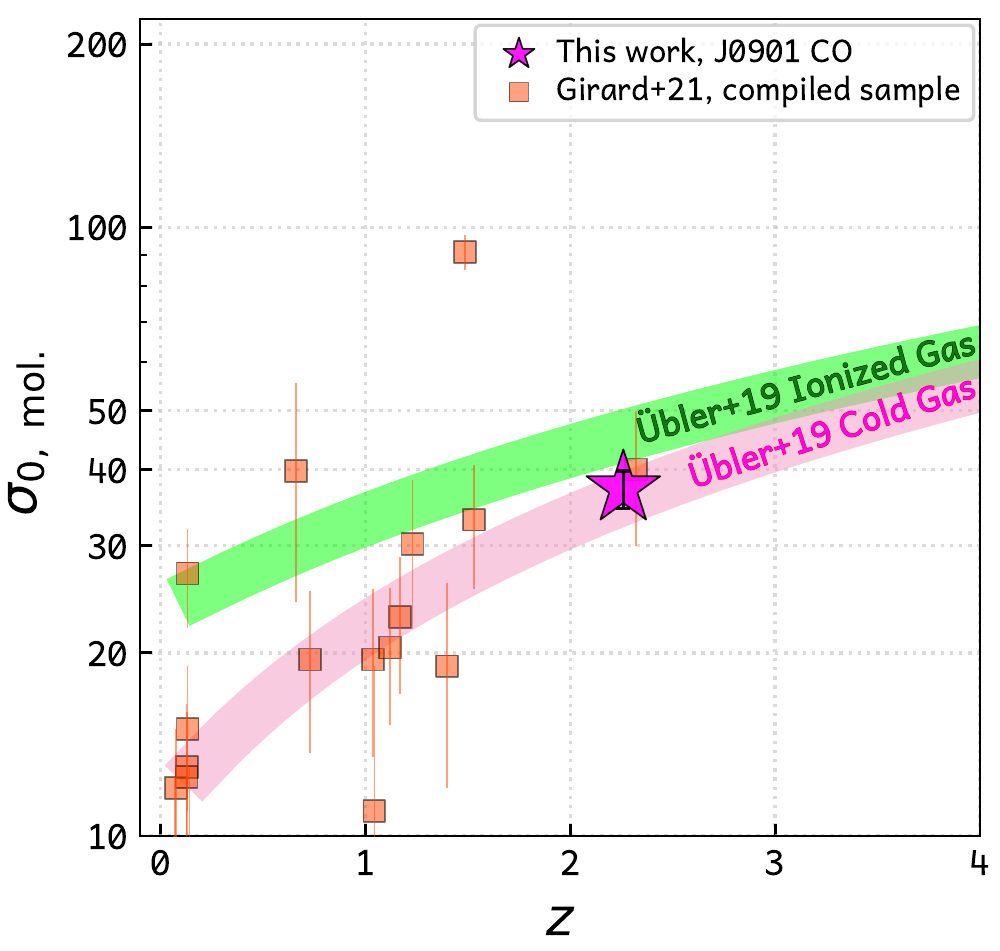}\\
\vspace{1ex}
\caption{%
Ionized (\textit{upper panel}) and cold gas (\textit{lower panel}) velocity dispersions in the J0901 disk compared to other massive $z\sim 1$--3 SFGs (\citealt{Livermore2015, Leethochawalit2016, Hirtenstein2019, Ubler2019, Girard2021}) and the $z\sim0$ GHASP survey (\citealt{Epinat2008,Epinat2010}). 
The shaded bands are the \citet{Ubler2019} empirical evolution trends for the dispersions of ionized gas (green):
$\sigma_{0,\,\mathrm{ion.}} = 23.3 + 9.8 \; z$, 
and cold gas (magenta): 
$\sigma_{0,\,\mathrm{mol.}} = 10.9 + 11.0 \; z$. 
\label{fig: disk dispersion}
}
\end{figure}

%% file: Input_fig_8.tex
\begin{figure}[ht]
\centering%
\includegraphics[width=0.9\linewidth]{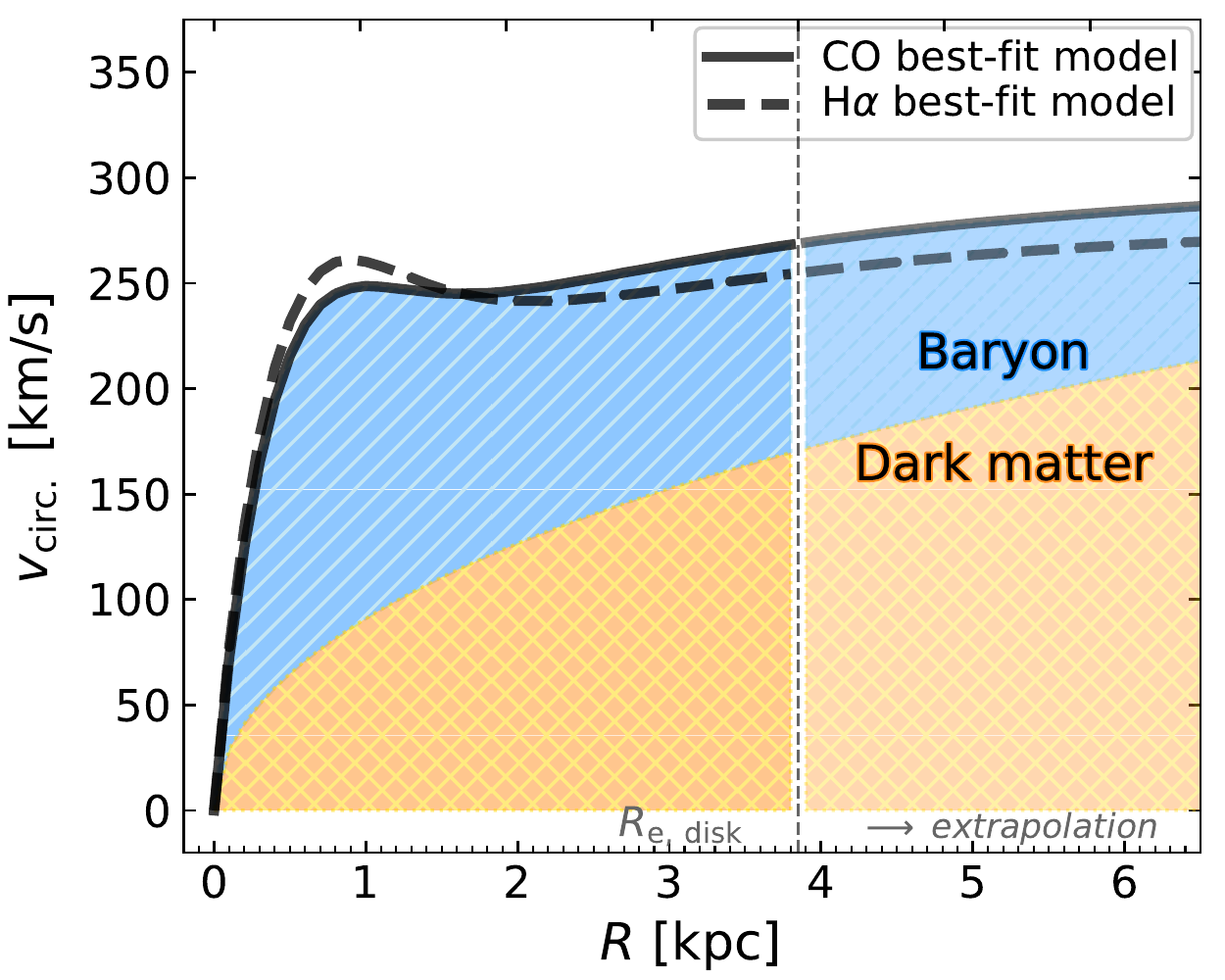}\\
\vspace{1ex}
\caption{%
Best-fit models' intrinsic circular velocity, corrected for inclination and without beam smearing and asymmetric drift. 
The baryonic and dark matter contributions to the circular velocity are shown as the blue and yellow shadings, respectively, and stacked on each other. The total circular velocity profiles are shown as the black lines (solid for CO- and dashed for \Halpha{}-based kinematics). 
The dashed vertical line indicates the $\Reff$ of J0901. 
The fading beyond $\Reff$ indicates regions with little data, where the model curves are extrapolated. 
\label{fig: intrinsic rotation curve}
}
\end{figure}

%% file: Input_fig_9.tex
\begin{figure}[ht]
\centering%
\includegraphics[width=0.9\linewidth]{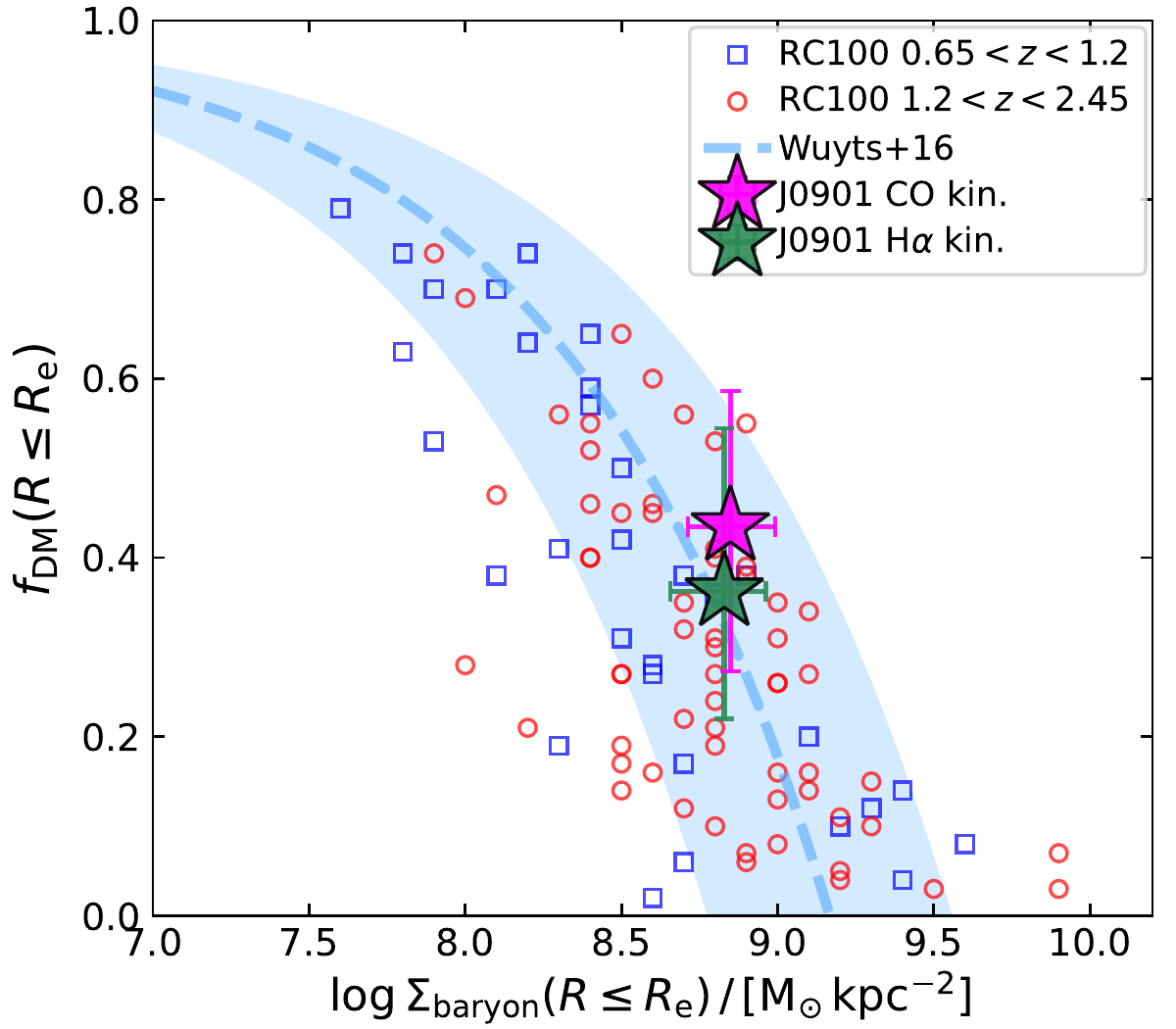}\\
\vspace{1ex}
\caption{%
Dark matter fraction $\fDM(<\Reff)$ versus baryon surface density $\Sigbar(<\Reff)$ within the baryonic disk effective radius $\Reff$. 
Small open symbols are the latest IFU studies of 100 massive $z=0.65$--$2.45$ galaxies from \citep[][RC100; extending the work of \citealt{Genzel2020} RC41]{Nestor2022}. 
The RC100 galaxies are divided in two equal-cosmic-interval redshift bins, $0.65 < z < 1.2$ and $1.2 < z < 2.45$. 
The dashed line indicates the \citet{Wuyts2016} empirical fit: $\log (1 - y) = {(-0.34 + 0.51 \, (x - 8.5))}$, and the blue shaded area indicates a $\pm 0.2$~dex scatter. 
The J0901 kinematically-fitted $\fDM$ versus $\Sigbar$ (and 16- and 84-th percentiles) are shown as the magenta and green solid stars (and error bars) for the CO and \Halpha{} data sets, respectively. 
\label{fig: fDM Sigbar}
}
\end{figure}

%% file: Input_fig_10.tex
\begin{figure}[ht]
\centering%
\includegraphics[width=0.95\linewidth]{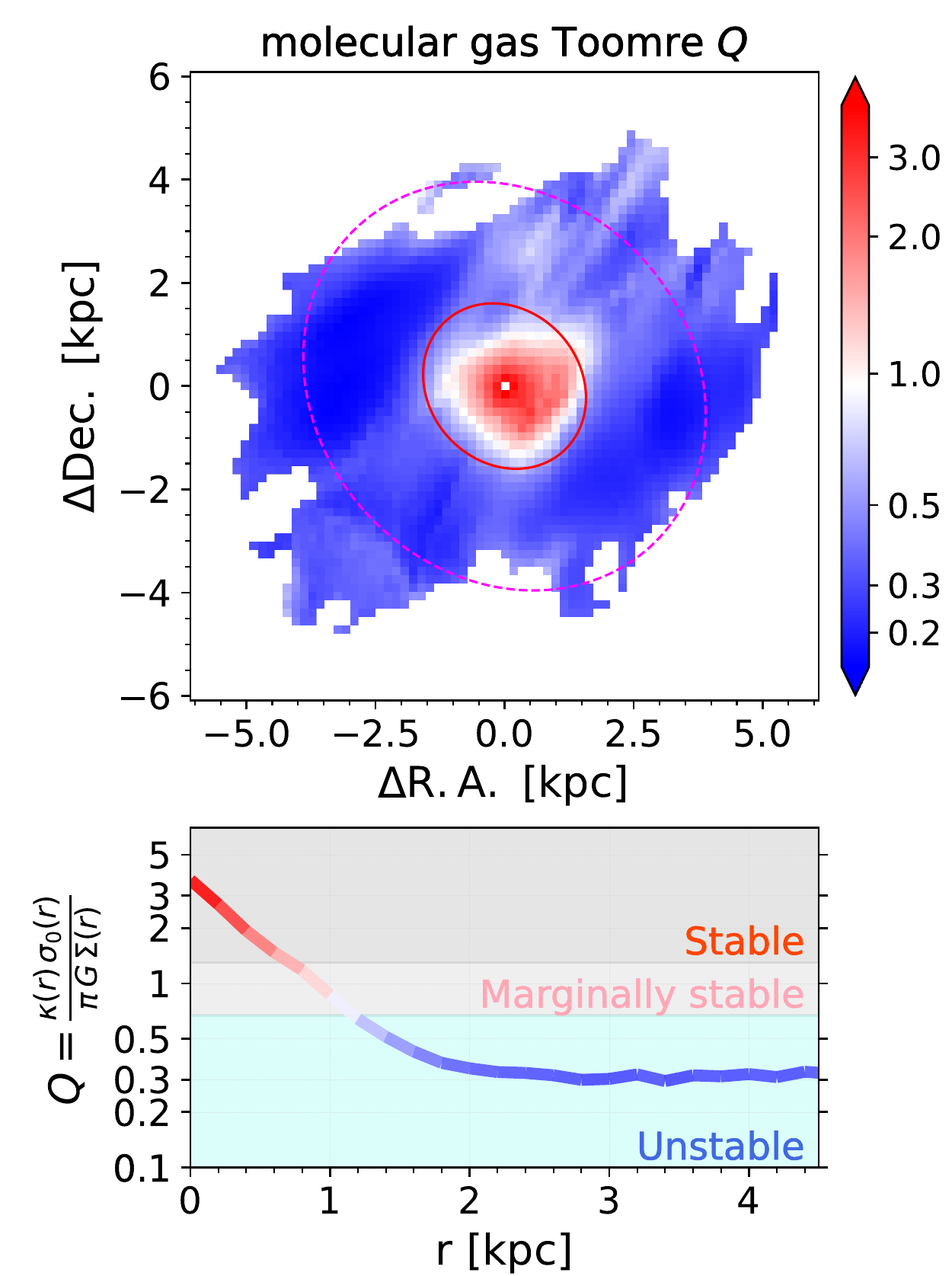}\\
\vspace{1ex}
\caption{%
\textit{Upper panel}: spatial distribution of the Toomre $Q$ for the molecular gas in J0901, computed following \citet[][Eq.~2]{Genzel2014a}, $Q_{\mathrm{gas}} = \frac{\kappa (r) \sigma_{0}(r)}{\pi G \Sigma_{\mathrm{gas}}(r)}$,
using our molecular gas surface density map (Fig.~\ref{fig: mass distribution}) and the best-fit rotation curve (Fig.~\ref{fig: intrinsic rotation curve}). 
The two ellipses are the same as in Fig.~\ref{fig: mass distribution}. 
\textit{Lower panel}: the azimuthally-averaged radial profile of the molecular gas $Q$. The lower and upper marginally (un)stable $Q_{\mathrm{crit}}$ are 0.67 and 1.3, respectively (for a thick disk with gas and stars; \citealt{Genzel2014a}). 
\label{fig: Toomre Q}
}
\end{figure}

%% file: Article_J0901_main.bbl
\begin{thebibliography}{}
\expandafter\ifx\csname natexlab\endcsname\relax\def\natexlab#1{#1}\fi
\providecommand{\url}[1]{\href{#1}{#1}}
\providecommand{\dodoi}[1]{doi:~\href{http://doi.org/#1}{\nolinkurl{#1}}}
\providecommand{\doeprint}[1]{\href{http://ascl.net/#1}{\nolinkurl{http://ascl.net/#1}}}
\providecommand{\doarXiv}[1]{\href{https://arxiv.org/abs/#1}{\nolinkurl{https://arxiv.org/abs/#1}}}

\bibitem[{{Astropy Collaboration} {et~al.}(2013){Astropy Collaboration},
  {Robitaille}, {Tollerud}, {Greenfield}, {Droettboom}, {Bray}, {Aldcroft},
  {Davis}, {Ginsburg}, {Price-Whelan}, {Kerzendorf}, {Conley}, {Crighton},
  {Barbary}, {Muna}, {Ferguson}, {Grollier}, {Parikh}, {Nair}, {Unther},
  {Deil}, {Woillez}, {Conseil}, {Kramer}, {Turner}, {Singer}, {Fox}, {Weaver},
  {Zabalza}, {Edwards}, {Azalee Bostroem}, {Burke}, {Casey}, {Crawford},
  {Dencheva}, {Ely}, {Jenness}, {Labrie}, {Lim}, {Pierfederici}, {Pontzen},
  {Ptak}, {Refsdal}, {Servillat}, \& {Streicher}}]{astropy:2013}
{Astropy Collaboration}, {Robitaille}, T.~P., {Tollerud}, E.~J., {et~al.} 2013,
  \aap, 558, A33, \dodoi{10.1051/0004-6361/201322068}

\bibitem[{{Astropy Collaboration} {et~al.}(2018){Astropy Collaboration},
  {Price-Whelan}, {Sip{\H{o}}cz}, {G{\"u}nther}, {Lim}, {Crawford}, {Conseil},
  {Shupe}, {Craig}, {Dencheva}, {Ginsburg}, {VanderPlas}, {Bradley},
  {P{\'e}rez-Su{\'a}rez}, {de Val-Borro}, {Aldcroft}, {Cruz}, {Robitaille},
  {Tollerud}, {Ardelean}, {Babej}, {Bach}, {Bachetti}, {Bakanov}, {Bamford},
  {Barentsen}, {Barmby}, {Baumbach}, {Berry}, {Biscani}, {Boquien}, {Bostroem},
  {Bouma}, {Brammer}, {Bray}, {Breytenbach}, {Buddelmeijer}, {Burke},
  {Calderone}, {Cano Rodr{\'\i}guez}, {Cara}, {Cardoso}, {Cheedella}, {Copin},
  {Corrales}, {Crichton}, {D'Avella}, {Deil}, {Depagne}, {Dietrich}, {Donath},
  {Droettboom}, {Earl}, {Erben}, {Fabbro}, {Ferreira}, {Finethy}, {Fox},
  {Garrison}, {Gibbons}, {Goldstein}, {Gommers}, {Greco}, {Greenfield},
  {Groener}, {Grollier}, {Hagen}, {Hirst}, {Homeier}, {Horton}, {Hosseinzadeh},
  {Hu}, {Hunkeler}, {Ivezi{\'c}}, {Jain}, {Jenness}, {Kanarek}, {Kendrew},
  {Kern}, {Kerzendorf}, {Khvalko}, {King}, {Kirkby}, {Kulkarni}, {Kumar},
  {Lee}, {Lenz}, {Littlefair}, {Ma}, {Macleod}, {Mastropietro}, {McCully},
  {Montagnac}, {Morris}, {Mueller}, {Mumford}, {Muna}, {Murphy}, {Nelson},
  {Nguyen}, {Ninan}, {N{\"o}the}, {Ogaz}, {Oh}, {Parejko}, {Parley}, {Pascual},
  {Patil}, {Patil}, {Plunkett}, {Prochaska}, {Rastogi}, {Reddy Janga},
  {Sabater}, {Sakurikar}, {Seifert}, {Sherbert}, {Sherwood-Taylor}, {Shih},
  {Sick}, {Silbiger}, {Singanamalla}, {Singer}, {Sladen}, {Sooley},
  {Sornarajah}, {Streicher}, {Teuben}, {Thomas}, {Tremblay}, {Turner},
  {Terr{\'o}n}, {van Kerkwijk}, {de la Vega}, {Watkins}, {Weaver}, {Whitmore},
  {Woillez}, {Zabalza}, \& {Astropy Contributors}}]{astropy:2018}
{Astropy Collaboration}, {Price-Whelan}, A.~M., {Sip{\H{o}}cz}, B.~M., {et~al.}
  2018, \aj, 156, 123, \dodoi{10.3847/1538-3881/aabc4f}

\bibitem[{{Astropy Collaboration} {et~al.}(2022){Astropy Collaboration},
  {Price-Whelan}, {Lim}, {Earl}, {Starkman}, {Bradley}, {Shupe}, {Patil},
  {Corrales}, {Brasseur}, {N{"o}the}, {Donath}, {Tollerud}, {Morris},
  {Ginsburg}, {Vaher}, {Weaver}, {Tocknell}, {Jamieson}, {van Kerkwijk},
  {Robitaille}, {Merry}, {Bachetti}, {G{"u}nther}, {Aldcroft},
  {Alvarado-Montes}, {Archibald}, {B{'o}di}, {Bapat}, {Barentsen}, {Baz{'a}n},
  {Biswas}, {Boquien}, {Burke}, {Cara}, {Cara}, {Conroy}, {Conseil}, {Craig},
  {Cross}, {Cruz}, {D'Eugenio}, {Dencheva}, {Devillepoix}, {Dietrich},
  {Eigenbrot}, {Erben}, {Ferreira}, {Foreman-Mackey}, {Fox}, {Freij}, {Garg},
  {Geda}, {Glattly}, {Gondhalekar}, {Gordon}, {Grant}, {Greenfield}, {Groener},
  {Guest}, {Gurovich}, {Handberg}, {Hart}, {Hatfield-Dodds}, {Homeier},
  {Hosseinzadeh}, {Jenness}, {Jones}, {Joseph}, {Kalmbach}, {Karamehmetoglu},
  {Ka{l}uszy{'n}ski}, {Kelley}, {Kern}, {Kerzendorf}, {Koch}, {Kulumani},
  {Lee}, {Ly}, {Ma}, {MacBride}, {Maljaars}, {Muna}, {Murphy}, {Norman},
  {O'Steen}, {Oman}, {Pacifici}, {Pascual}, {Pascual-Granado}, {Patil},
  {Perren}, {Pickering}, {Rastogi}, {Roulston}, {Ryan}, {Rykoff}, {Sabater},
  {Sakurikar}, {Salgado}, {Sanghi}, {Saunders}, {Savchenko}, {Schwardt},
  {Seifert-Eckert}, {Shih}, {Jain}, {Shukla}, {Sick}, {Simpson},
  {Singanamalla}, {Singer}, {Singhal}, {Sinha}, {Sip{H{o}}cz}, {Spitler},
  {Stansby}, {Streicher}, {{{S}}umak}, {Swinbank}, {Taranu}, {Tewary},
  {Tremblay}, {Val-Borro}, {Van Kooten}, {Vasovi{'c}}, {Verma}, {de Miranda
  Cardoso}, {Williams}, {Wilson}, {Winkel}, {Wood-Vasey}, {Xue}, {Yoachim},
  {Zhang}, {Zonca}, \& {Astropy Project Contributors}}]{astropy:2022}
{Astropy Collaboration}, {Price-Whelan}, A.~M., {Lim}, P.~L., {et~al.} 2022,
  apj, 935, 167, \dodoi{10.3847/1538-4357/ac7c74}

\bibitem[{{Barro} {et~al.}(2017){Barro}, {Kriek}, {P{\'e}rez-Gonz{\'a}lez},
  {Diaz-Santos}, {Price}, {Rujopakarn}, {Pandya}, {Koo}, {Faber}, {Dekel},
  {Primack}, \& {Kocevski}}]{Barro2017}
{Barro}, G., {Kriek}, M., {P{\'e}rez-Gonz{\'a}lez}, P.~G., {et~al.} 2017,
  \apjl, 851, L40, \dodoi{10.3847/2041-8213/aa9f0d}

\bibitem[{{Bell} \& {de Jong}(2000)}]{Bell2000}
{Bell}, E.~F., \& {de Jong}, R.~S. 2000, \mnras, 312, 497,
  \dodoi{10.1046/j.1365-8711.2000.03138.x}

\bibitem[{{Bolatto} {et~al.}(2015){Bolatto}, {Warren}, {Leroy}, {Tacconi},
  {Bouch{\'e}}, {F{\"o}rster Schreiber}, {Genzel}, {Cooper}, {Fisher},
  {Combes}, {Garc{\'\i}a-Burillo}, {Burkert}, {Bournaud}, {Weiss}, {Saintonge},
  {Wuyts}, \& {Sternberg}}]{Bolatto2015}
{Bolatto}, A.~D., {Warren}, S.~R., {Leroy}, A.~K., {et~al.} 2015, \apj, 809,
  175, \dodoi{10.1088/0004-637X/809/2/175}

\bibitem[{{Bournaud} {et~al.}(2007){Bournaud}, {Elmegreen}, \&
  {Elmegreen}}]{Bournaud2007}
{Bournaud}, F., {Elmegreen}, B.~G., \& {Elmegreen}, D.~M. 2007, \apj, 670, 237,
  \dodoi{10.1086/522077}

\bibitem[{{Bradley} {et~al.}(2020){Bradley}, {Sip{\H{o}}cz}, {Robitaille},
  {Tollerud}, {Vin{\'\i}cius}, {Deil}, {Barbary}, {Wilson}, {Busko},
  {G{\"u}nther}, {Cara}, {Conseil}, {Bostroem}, {Droettboom}, {Bray}, {Andersen
  Bratholm}, {Lim}, {Barentsen}, {Craig}, {Pascual}, {Perren}, {Greco},
  {Donath}, {de Val-Borro}, {Kerzendorf}, {Bach}, {Weaver}, {D'Eugenio},
  {Souchereau}, \& {Ferreira}}]{photutils}
{Bradley}, L., {Sip{\H{o}}cz}, B., {Robitaille}, T., {et~al.} 2020,
  {astropy/photutils: 1.0.0}, 1.0.0, Zenodo,  Zenodo,
  \dodoi{10.5281/zenodo.4044744}

\bibitem[{{Bullock} {et~al.}(2001){Bullock}, {Kolatt}, {Sigad}, {Somerville},
  {Kravtsov}, {Klypin}, {Primack}, \& {Dekel}}]{Bullock2001}
{Bullock}, J.~S., {Kolatt}, T.~S., {Sigad}, Y., {et~al.} 2001, \mnras, 321,
  559, \dodoi{10.1046/j.1365-8711.2001.04068.x}

\bibitem[{{Burkert} {et~al.}(2010){Burkert}, {Genzel}, {Bouch{\'e}}, {Cresci},
  {Khochfar}, {Sommer-Larsen}, {Sternberg}, {Naab}, {F{\"o}rster Schreiber},
  {Tacconi}, {Shapiro}, {Hicks}, {Lutz}, {Davies}, {Buschkamp}, \&
  {Genel}}]{Burkert2010}
{Burkert}, A., {Genzel}, R., {Bouch{\'e}}, N., {et~al.} 2010, \apj, 725, 2324,
  \dodoi{10.1088/0004-637X/725/2/2324}

\bibitem[{{Burkert} {et~al.}(2016){Burkert}, {F{\"o}rster Schreiber}, {Genzel},
  {Lang}, {Tacconi}, {Wisnioski}, {Wuyts}, {Bandara}, {Beifiori}, {Bender},
  {Brammer}, {Chan}, {Davies}, {Dekel}, {Fabricius}, {Fossati}, {Kulkarni},
  {Lutz}, {Mendel}, {Momcheva}, {Nelson}, {Naab}, {Renzini}, {Saglia},
  {Sharples}, {Sternberg}, {Wilman}, \& {Wuyts}}]{Burkert2016}
{Burkert}, A., {F{\"o}rster Schreiber}, N.~M., {Genzel}, R., {et~al.} 2016,
  \apj, 826, 214, \dodoi{10.3847/0004-637X/826/2/214}

\bibitem[{{Calistro Rivera} {et~al.}(2018){Calistro Rivera}, {Hodge}, {Smail},
  {Swinbank}, {Weiss}, {Wardlow}, {Walter}, {Rybak}, {Chen}, {Brandt},
  {Coppin}, {da Cunha}, {Dannerbauer}, {Greve}, {Karim}, {Knudsen},
  {Schinnerer}, {Simpson}, {Venemans}, \& {van der Werf}}]{CalistroRivera2018}
{Calistro Rivera}, G., {Hodge}, J.~A., {Smail}, I., {et~al.} 2018, \apj, 863,
  56, \dodoi{10.3847/1538-4357/aacffa}

\bibitem[{{Calzetti} {et~al.}(2000){Calzetti}, {Armus}, {Bohlin}, {Kinney},
  {Koornneef}, \& {Storchi-Bergmann}}]{Calzetti2000}
{Calzetti}, D., {Armus}, L., {Bohlin}, R.~C., {et~al.} 2000, \apj, 533, 682,
  \dodoi{10.1086/308692}

\bibitem[{{Cappellari} {et~al.}(2012){Cappellari}, {McDermid}, {Alatalo},
  {Blitz}, {Bois}, {Bournaud}, {Bureau}, {Crocker}, {Davies}, {Davis}, {de
  Zeeuw}, {Duc}, {Emsellem}, {Khochfar}, {Krajnovi{\'c}}, {Kuntschner},
  {Lablanche}, {Morganti}, {Naab}, {Oosterloo}, {Sarzi}, {Scott}, {Serra},
  {Weijmans}, \& {Young}}]{Cappellari2012}
{Cappellari}, M., {McDermid}, R.~M., {Alatalo}, K., {et~al.} 2012, \nat, 484,
  485, \dodoi{10.1038/nature10972}

\bibitem[{{Ceverino} {et~al.}(2010){Ceverino}, {Dekel}, \&
  {Bournaud}}]{Ceverino2010}
{Ceverino}, D., {Dekel}, A., \& {Bournaud}, F. 2010, \mnras, 404, 2151,
  \dodoi{10.1111/j.1365-2966.2010.16433.x}

\bibitem[{{Chabrier}(2003)}]{Chabrier2003}
{Chabrier}, G. 2003, \pasp, 115, 763, \dodoi{10.1086/376392}

\bibitem[{{Cresci} {et~al.}(2009){Cresci}, {Hicks}, {Genzel}, {F{\"o}rster
  Schreiber}, {Davies}, {Bouch{\'e}}, {Buschkamp}, {Genel}, {Shapiro},
  {Tacconi}, {Sommer-Larsen}, {Burkert}, {Eisenhauer}, {Gerhard}, {Lutz},
  {Naab}, {Sternberg}, {Cimatti}, {Daddi}, {Erb}, {Kurk}, {Lilly}, {Renzini},
  {Shapley}, {Steidel}, \& {Caputi}}]{Cresci2009}
{Cresci}, G., {Hicks}, E.~K.~S., {Genzel}, R., {et~al.} 2009, \apj, 697, 115,
  \dodoi{10.1088/0004-637X/697/1/115}

\bibitem[{{Danovich} {et~al.}(2015){Danovich}, {Dekel}, {Hahn}, {Ceverino}, \&
  {Primack}}]{Danovich2015}
{Danovich}, M., {Dekel}, A., {Hahn}, O., {Ceverino}, D., \& {Primack}, J. 2015,
  \mnras, 449, 2087, \dodoi{10.1093/mnras/stv270}

\bibitem[{{Davidzon} {et~al.}(2017){Davidzon}, {Ilbert}, {Laigle}, {Coupon},
  {McCracken}, {Delvecchio}, {Masters}, {Capak}, {Hsieh}, {Le F{\`e}vre},
  {Tresse}, {Bethermin}, {Chang}, {Faisst}, {Le Floc'h}, {Steinhardt}, {Toft},
  {Aussel}, {Dubois}, {Hasinger}, {Salvato}, {Sanders}, {Scoville}, \&
  {Silverman}}]{Davidzon2017}
{Davidzon}, I., {Ilbert}, O., {Laigle}, C., {et~al.} 2017, \aap, 605, A70,
  \dodoi{10.1051/0004-6361/201730419}

\bibitem[{{Davies} {et~al.}(2011){Davies}, {F{\"o}rster Schreiber}, {Cresci},
  {Genzel}, {Bouch{\'e}}, {Burkert}, {Buschkamp}, {Genel}, {Hicks}, {Kurk},
  {Lutz}, {Newman}, {Shapiro}, {Sternberg}, {Tacconi}, \& {Wuyts}}]{Davies2011}
{Davies}, R., {F{\"o}rster Schreiber}, N.~M., {Cresci}, G., {et~al.} 2011,
  \apj, 741, 69, \dodoi{10.1088/0004-637X/741/2/69}

\bibitem[{{Davies} {et~al.}(2020){Davies}, {F{\"o}rster Schreiber}, {Lutz},
  {Genzel}, {Belli}, {Shimizu}, {Contursi}, {Davies}, {Herrera-Camus}, {Lee},
  {Naab}, {Price}, {Renzini}, {Schruba}, {Sternberg}, {Tacconi}, {{\"U}bler},
  {Wisnioski}, \& {Wuyts}}]{Davies2020a}
{Davies}, R.~L., {F{\"o}rster Schreiber}, N.~M., {Lutz}, D., {et~al.} 2020,
  \apj, 894, 28, \dodoi{10.3847/1538-4357/ab86ad}

\bibitem[{{Dekel} \& {Birnboim}(2006)}]{Dekel2006}
{Dekel}, A., \& {Birnboim}, Y. 2006, \mnras, 368, 2,
  \dodoi{10.1111/j.1365-2966.2006.10145.x}

\bibitem[{{Dekel} \& {Burkert}(2014)}]{Dekel2014}
{Dekel}, A., \& {Burkert}, A. 2014, \mnras, 438, 1870,
  \dodoi{10.1093/mnras/stt2331}

\bibitem[{{Dekel} {et~al.}(2003){Dekel}, {Devor}, \& {Hetzroni}}]{Dekel2003b}
{Dekel}, A., {Devor}, J., \& {Hetzroni}, G. 2003, \mnras, 341, 326,
  \dodoi{10.1046/j.1365-8711.2003.06432.x}

\bibitem[{{Dekel} {et~al.}(2009{\natexlab{a}}){Dekel}, {Sari}, \&
  {Ceverino}}]{Dekel2009b}
{Dekel}, A., {Sari}, R., \& {Ceverino}, D. 2009{\natexlab{a}}, \apj, 703, 785,
  \dodoi{10.1088/0004-637X/703/1/785}

\bibitem[{{Dekel} {et~al.}(2009{\natexlab{b}}){Dekel}, {Birnboim}, {Engel},
  {Freundlich}, {Goerdt}, {Mumcuoglu}, {Neistein}, {Pichon}, {Teyssier}, \&
  {Zinger}}]{Dekel2009a}
{Dekel}, A., {Birnboim}, Y., {Engel}, G., {et~al.} 2009{\natexlab{b}}, \nat,
  457, 451, \dodoi{10.1038/nature07648}

\bibitem[{{Dekel} {et~al.}(2020){Dekel}, {Lapiner}, {Ginzburg}, {Freundlich},
  {Jiang}, {Finish}, {Kretschmer}, {Lin}, {Ceverino}, {Primack}, {Giavalisco},
  \& {Ji}}]{Dekel2020b}
{Dekel}, A., {Lapiner}, S., {Ginzburg}, O., {et~al.} 2020, \mnras, 496, 5372,
  \dodoi{10.1093/mnras/staa1713}

\bibitem[{{Dekel} {et~al.}(2021){Dekel}, {Freundlich}, {Jiang}, {Lapiner},
  {Burkert}, {Ceverino}, {Du}, {Genzel}, \& {Primack}}]{Dekel2021}
{Dekel}, A., {Freundlich}, J., {Jiang}, F., {et~al.} 2021, \mnras, 508, 999,
  \dodoi{10.1093/mnras/stab2416}

\bibitem[{{Diehl} {et~al.}(2009){Diehl}, {Allam}, {Annis}, {Buckley-Geer},
  {Frieman}, {Kubik}, {Kubo}, {Lin}, {Tucker}, \& {West}}]{Diehl2009}
{Diehl}, H.~T., {Allam}, S.~S., {Annis}, J., {et~al.} 2009, \apj, 707, 686,
  \dodoi{10.1088/0004-637X/707/1/686}

\bibitem[{{Dutton} \& {Macci{\`o}}(2014)}]{Dutton2014}
{Dutton}, A.~A., \& {Macci{\`o}}, A.~V. 2014, \mnras, 441, 3359,
  \dodoi{10.1093/mnras/stu742}

\bibitem[{{El-Zant} {et~al.}(2001){El-Zant}, {Shlosman}, \&
  {Hoffman}}]{ElZant2001}
{El-Zant}, A., {Shlosman}, I., \& {Hoffman}, Y. 2001, \apj, 560, 636,
  \dodoi{10.1086/322516}

\bibitem[{{Elagali} {et~al.}(2018){Elagali}, {Lagos}, {Wong}, {Staveley-Smith},
  {Trayford}, {Schaller}, {Yuan}, \& {Abadi}}]{Elagali2018}
{Elagali}, A., {Lagos}, C. D.~P., {Wong}, O.~I., {et~al.} 2018, \mnras, 481,
  2951, \dodoi{10.1093/mnras/sty2462}

\bibitem[{{Elmegreen} \& {Elmegreen}(2006)}]{Elmegreen2006rings}
{Elmegreen}, D.~M., \& {Elmegreen}, B.~G. 2006, \apj, 651, 676,
  \dodoi{10.1086/507863}

\bibitem[{{Epinat} {et~al.}(2010){Epinat}, {Amram}, {Balkowski}, \&
  {Marcelin}}]{Epinat2010}
{Epinat}, B., {Amram}, P., {Balkowski}, C., \& {Marcelin}, M. 2010, \mnras,
  401, 2113, \dodoi{10.1111/j.1365-2966.2009.15688.x}

\bibitem[{{Epinat} {et~al.}(2008){Epinat}, {Amram}, \& {Marcelin}}]{Epinat2008}
{Epinat}, B., {Amram}, P., \& {Marcelin}, M. 2008, \mnras, 390, 466,
  \dodoi{10.1111/j.1365-2966.2008.13796.x}

\bibitem[{{Escala} \& {Larson}(2008)}]{Escala2008}
{Escala}, A., \& {Larson}, R.~B. 2008, \apjl, 685, L31, \dodoi{10.1086/592271}

\bibitem[{{Fadely} {et~al.}(2010){Fadely}, {Allam}, {Baker}, {Lin}, {Lutz},
  {Shapley}, {Shin}, {Allyn Smith}, {Strauss}, \& {Tucker}}]{Fadely2010}
{Fadely}, R., {Allam}, S.~S., {Baker}, A.~J., {et~al.} 2010, \apj, 723, 729,
  \dodoi{10.1088/0004-637X/723/1/729}

\bibitem[{{Foreman-Mackey} {et~al.}(2013){Foreman-Mackey}, {Hogg}, {Lang}, \&
  {Goodman}}]{emcee}
{Foreman-Mackey}, D., {Hogg}, D.~W., {Lang}, D., \& {Goodman}, J. 2013, \pasp,
  125, 306, \dodoi{10.1086/670067}

\bibitem[{{F{\"o}rster Schreiber} \& {Wuyts}(2020)}]{ForsterSchreiber2020}
{F{\"o}rster Schreiber}, N.~M., \& {Wuyts}, S. 2020, \araa, 58, 661,
  \dodoi{10.1146/annurev-astro-032620-021910}

\bibitem[{{F{\"o}rster Schreiber} {et~al.}(2006){F{\"o}rster Schreiber},
  {Genzel}, {Lehnert}, {Bouch{\'e}}, {Verma}, {Erb}, {Shapley}, {Steidel},
  {Davies}, {Lutz}, {Nesvadba}, {Tacconi}, {Eisenhauer}, {Abuter}, {Gilbert},
  {Gillessen}, \& {Sternberg}}]{ForsterSchreiber2006}
{F{\"o}rster Schreiber}, N.~M., {Genzel}, R., {Lehnert}, M.~D., {et~al.} 2006,
  \apj, 645, 1062, \dodoi{10.1086/504403}

\bibitem[{{F{\"o}rster Schreiber} {et~al.}(2009){F{\"o}rster Schreiber},
  {Genzel}, {Bouch{\'e}}, {Cresci}, {Davies}, {Buschkamp}, {Shapiro},
  {Tacconi}, {Hicks}, {Genel}, {Shapley}, {Erb}, {Steidel}, {Lutz},
  {Eisenhauer}, {Gillessen}, {Sternberg}, {Renzini}, {Cimatti}, {Daddi},
  {Kurk}, {Lilly}, {Kong}, {Lehnert}, {Nesvadba}, {Verma}, {McCracken},
  {Arimoto}, {Mignoli}, \& {Onodera}}]{ForsterSchreiber2009}
{F{\"o}rster Schreiber}, N.~M., {Genzel}, R., {Bouch{\'e}}, N., {et~al.} 2009,
  \apj, 706, 1364, \dodoi{10.1088/0004-637X/706/2/1364}

\bibitem[{{F{\"o}rster Schreiber} {et~al.}(2014){F{\"o}rster Schreiber},
  {Genzel}, {Newman}, {Kurk}, {Lutz}, {Tacconi}, {Wuyts}, {Bandara}, {Burkert},
  {Buschkamp}, {Carollo}, {Cresci}, {Daddi}, {Davies}, {Eisenhauer}, {Hicks},
  {Lang}, {Lilly}, {Mainieri}, {Mancini}, {Naab}, {Peng}, {Renzini}, {Rosario},
  {Shapiro Griffin}, {Shapley}, {Sternberg}, {Tacchella}, {Vergani},
  {Wisnioski}, {Wuyts}, \& {Zamorani}}]{ForsterSchreiber2014}
{F{\"o}rster Schreiber}, N.~M., {Genzel}, R., {Newman}, S.~F., {et~al.} 2014,
  \apj, 787, 38, \dodoi{10.1088/0004-637X/787/1/38}

\bibitem[{{F{\"o}rster Schreiber} {et~al.}(2018){F{\"o}rster Schreiber},
  {Renzini}, {Mancini}, {Genzel}, {Bouch{\'e}}, {Cresci}, {Hicks}, {Lilly},
  {Peng}, {Burkert}, {Carollo}, {Cimatti}, {Daddi}, {Davies}, {Genel}, {Kurk},
  {Lang}, {Lutz}, {Mainieri}, {McCracken}, {Mignoli}, {Naab}, {Oesch},
  {Pozzetti}, {Scodeggio}, {Shapiro Griffin}, {Shapley}, {Sternberg},
  {Tacchella}, {Tacconi}, {Wuyts}, \& {Zamorani}}]{ForsterSchreiber2018}
{F{\"o}rster Schreiber}, N.~M., {Renzini}, A., {Mancini}, C., {et~al.} 2018,
  \apjs, 238, 21, \dodoi{10.3847/1538-4365/aadd49}

\bibitem[{{Genel} {et~al.}(2008){Genel}, {Genzel}, {Bouch{\'e}}, {Sternberg},
  {Naab}, {F{\"o}rster Schreiber}, {Shapiro}, {Tacconi}, {Lutz}, {Cresci},
  {Buschkamp}, {Davies}, \& {Hicks}}]{Genel2008}
{Genel}, S., {Genzel}, R., {Bouch{\'e}}, N., {et~al.} 2008, \apj, 688, 789,
  \dodoi{10.1086/592241}

\bibitem[{{Genzel} {et~al.}(2006){Genzel}, {Tacconi}, {Eisenhauer},
  {F{\"o}rster Schreiber}, {Cimatti}, {Daddi}, {Bouch{\'e}}, {Davies},
  {Lehnert}, {Lutz}, {Nesvadba}, {Verma}, {Abuter}, {Shapiro}, {Sternberg},
  {Renzini}, {Kong}, {Arimoto}, \& {Mignoli}}]{Genzel2006}
{Genzel}, R., {Tacconi}, L.~J., {Eisenhauer}, F., {et~al.} 2006, \nat, 442,
  786, \dodoi{10.1038/nature05052}

\bibitem[{{Genzel} {et~al.}(2008){Genzel}, {Burkert}, {Bouch{\'e}}, {Cresci},
  {F{\"o}rster Schreiber}, {Shapley}, {Shapiro}, {Tacconi}, {Buschkamp},
  {Cimatti}, {Daddi}, {Davies}, {Eisenhauer}, {Erb}, {Genel}, {Gerhard},
  {Hicks}, {Lutz}, {Naab}, {Ott}, {Rabien}, {Renzini}, {Steidel}, {Sternberg},
  \& {Lilly}}]{Genzel2008}
{Genzel}, R., {Burkert}, A., {Bouch{\'e}}, N., {et~al.} 2008, \apj, 687, 59,
  \dodoi{10.1086/591840}

\bibitem[{{Genzel} {et~al.}(2011){Genzel}, {Newman}, {Jones}, {F{\"o}rster
  Schreiber}, {Shapiro}, {Genel}, {Lilly}, {Renzini}, {Tacconi}, {Bouch{\'e}},
  {Burkert}, {Cresci}, {Buschkamp}, {Carollo}, {Ceverino}, {Davies}, {Dekel},
  {Eisenhauer}, {Hicks}, {Kurk}, {Lutz}, {Mancini}, {Naab}, {Peng},
  {Sternberg}, {Vergani}, \& {Zamorani}}]{Genzel2011}
{Genzel}, R., {Newman}, S., {Jones}, T., {et~al.} 2011, \apj, 733, 101,
  \dodoi{10.1088/0004-637X/733/2/101}

\bibitem[{{Genzel} {et~al.}(2013){Genzel}, {Tacconi}, {Kurk}, {Wuyts},
  {Combes}, {Freundlich}, {Bolatto}, {Cooper}, {Neri}, {Nordon}, {Bournaud},
  {Burkert}, {Comerford}, {Cox}, {Davis}, {F{\"o}rster Schreiber},
  {Garc{\'\i}a-Burillo}, {Gracia-Carpio}, {Lutz}, {Naab}, {Newman},
  {Saintonge}, {Shapiro Griffin}, {Shapley}, {Sternberg}, \&
  {Weiner}}]{Genzel2013}
{Genzel}, R., {Tacconi}, L.~J., {Kurk}, J., {et~al.} 2013, \apj, 773, 68,
  \dodoi{10.1088/0004-637X/773/1/68}

\bibitem[{{Genzel} {et~al.}(2014{\natexlab{a}}){Genzel}, {F{\"o}rster
  Schreiber}, {Lang}, {Tacchella}, {Tacconi}, {Wuyts}, {Bandara}, {Burkert},
  {Buschkamp}, {Carollo}, {Cresci}, {Davies}, {Eisenhauer}, {Hicks}, {Kurk},
  {Lilly}, {Lutz}, {Mancini}, {Naab}, {Newman}, {Peng}, {Renzini}, {Shapiro
  Griffin}, {Sternberg}, {Vergani}, {Wisnioski}, {Wuyts}, \&
  {Zamorani}}]{Genzel2014a}
{Genzel}, R., {F{\"o}rster Schreiber}, N.~M., {Lang}, P., {et~al.}
  2014{\natexlab{a}}, \apj, 785, 75, \dodoi{10.1088/0004-637X/785/1/75}

\bibitem[{{Genzel} {et~al.}(2014{\natexlab{b}}){Genzel}, {F{\"o}rster
  Schreiber}, {Rosario}, {Lang}, {Lutz}, {Wisnioski}, {Wuyts}, {Wuyts},
  {Bandara}, {Bender}, {Berta}, {Kurk}, {Mendel}, {Tacconi}, {Wilman},
  {Beifiori}, {Brammer}, {Burkert}, {Buschkamp}, {Chan}, {Carollo}, {Davies},
  {Eisenhauer}, {Fabricius}, {Fossati}, {Kriek}, {Kulkarni}, {Lilly},
  {Mancini}, {Momcheva}, {Naab}, {Nelson}, {Renzini}, {Saglia}, {Sharples},
  {Sternberg}, {Tacchella}, \& {van Dokkum}}]{Genzel2014b}
{Genzel}, R., {F{\"o}rster Schreiber}, N.~M., {Rosario}, D., {et~al.}
  2014{\natexlab{b}}, \apj, 796, 7, \dodoi{10.1088/0004-637X/796/1/7}

\bibitem[{{Genzel} {et~al.}(2015){Genzel}, {Tacconi}, {Lutz}, {Saintonge},
  {Berta}, {Magnelli}, {Combes}, {Garc{\'{\i}}a-Burillo}, {Neri}, {Bolatto},
  {Contini}, {Lilly}, {Boissier}, {Boone}, {Bouch{\'e}}, {Bournaud}, {Burkert},
  {Carollo}, {Colina}, {Cooper}, {Cox}, {Feruglio}, {F{\"o}rster Schreiber},
  {Freundlich}, {Gracia-Carpio}, {Juneau}, {Kovac}, {Lippa}, {Naab}, {Salome},
  {Renzini}, {Sternberg}, {Walter}, {Weiner}, {Weiss}, \& {Wuyts}}]{Genzel2015}
{Genzel}, R., {Tacconi}, L.~J., {Lutz}, D., {et~al.} 2015, \apj, 800, 20,
  \dodoi{10.1088/0004-637X/800/1/20}

\bibitem[{{Genzel} {et~al.}(2017){Genzel}, {F{\"o}rster Schreiber},
  {{\"U}bler}, {Lang}, {Naab}, {Bender}, {Tacconi}, {Wisnioski}, {Wuyts},
  {Alexander}, {Beifiori}, {Belli}, {Brammer}, {Burkert}, {Carollo}, {Chan},
  {Davies}, {Fossati}, {Galametz}, {Genel}, {Gerhard}, {Lutz}, {Mendel},
  {Momcheva}, {Nelson}, {Renzini}, {Saglia}, {Sternberg}, {Tacchella},
  {Tadaki}, \& {Wilman}}]{Genzel2017}
{Genzel}, R., {F{\"o}rster Schreiber}, N.~M., {{\"U}bler}, H., {et~al.} 2017,
  \nat, 543, 397, \dodoi{10.1038/nature21685}

\bibitem[{{Genzel} {et~al.}(2020){Genzel}, {Price}, {{\"U}bler}, {F{\"o}rster
  Schreiber}, {Shimizu}, {Tacconi}, {Bender}, {Burkert}, {Contursi}, {Coogan},
  {Davies}, {Davies}, {Dekel}, {Herrera-Camus}, {Lee}, {Lutz}, {Naab}, {Neri},
  {Nestor}, {Renzini}, {Saglia}, {Schuster}, {Sternberg}, {Wisnioski}, \&
  {Wuyts}}]{Genzel2020}
{Genzel}, R., {Price}, S.~H., {{\"U}bler}, H., {et~al.} 2020, \apj, 902, 98,
  \dodoi{10.3847/1538-4357/abb0ea}

\bibitem[{{Girard} {et~al.}(2019){Girard}, {Dessauges-Zavadsky}, {Combes},
  {Chisholm}, {Patr{\'\i}cio}, {Richard}, \& {Schaerer}}]{Girard2019}
{Girard}, M., {Dessauges-Zavadsky}, M., {Combes}, F., {et~al.} 2019, \aap, 631,
  A91, \dodoi{10.1051/0004-6361/201935896}

\bibitem[{{Girard} {et~al.}(2018){Girard}, {Dessauges-Zavadsky}, {Schaerer},
  {Cirasuolo}, {Turner}, {Cava}, {Rodr{\'\i}guez-Mu{\~n}oz}, {Richard}, \&
  {P{\'e}rez-Gonz{\'a}lez}}]{Girard2018}
{Girard}, M., {Dessauges-Zavadsky}, M., {Schaerer}, D., {et~al.} 2018, \aap,
  613, A72, \dodoi{10.1051/0004-6361/201731988}

\bibitem[{{Girard} {et~al.}(2021){Girard}, {Fisher}, {Bolatto}, {Abraham},
  {Bassett}, {Glazebrook}, {Herrera-Camus}, {Jim{\'e}nez}, {Lenki{\'c}}, \&
  {Obreschkow}}]{Girard2021}
{Girard}, M., {Fisher}, D.~B., {Bolatto}, A.~D., {et~al.} 2021, \apj, 909, 12,
  \dodoi{10.3847/1538-4357/abd5b9}

\bibitem[{{Goldreich} \& {Lynden-Bell}(1965)}]{Goldreich1965a}
{Goldreich}, P., \& {Lynden-Bell}, D. 1965, \mnras, 130, 97,
  \dodoi{10.1093/mnras/130.2.97}

\bibitem[{{Hainline} {et~al.}(2009){Hainline}, {Shapley}, {Kornei}, {Pettini},
  {Buckley-Geer}, {Allam}, \& {Tucker}}]{Hainline2009}
{Hainline}, K.~N., {Shapley}, A.~E., {Kornei}, K.~A., {et~al.} 2009, \apj, 701,
  52, \dodoi{10.1088/0004-637X/701/1/52}

\bibitem[{{Herrera-Camus} {et~al.}(2019){Herrera-Camus}, {Tacconi}, {Genzel},
  {F{\"o}rster Schreiber}, {Lutz}, {Bolatto}, {Wuyts}, {Renzini}, {Lilly},
  {Belli}, {{\"U}bler}, {Shimizu}, {Davies}, {Sturm}, {Combes}, {Freundlich},
  {Garc{\'\i}a-Burillo}, {Cox}, {Burkert}, {Naab}, {Colina}, {Saintonge},
  {Cooper}, {Feruglio}, \& {Weiss}}]{HerreraCamus2019}
{Herrera-Camus}, R., {Tacconi}, L., {Genzel}, R., {et~al.} 2019, \apj, 871, 37,
  \dodoi{10.3847/1538-4357/aaf6a7}

\bibitem[{{Hirtenstein} {et~al.}(2019){Hirtenstein}, {Jones}, {Wang}, {Wetzel},
  {El-Badry}, {Hoag}, {Treu}, {Brada{\v{c}}}, \& {Morishita}}]{Hirtenstein2019}
{Hirtenstein}, J., {Jones}, T., {Wang}, X., {et~al.} 2019, \apj, 880, 54,
  \dodoi{10.3847/1538-4357/ab113e}

\bibitem[{{Hopkins}(2018)}]{Hopkins2018Review}
{Hopkins}, A.~M. 2018, \pasa, 35, e039, \dodoi{10.1017/pasa.2018.29}

\bibitem[{{Johnson} {et~al.}(2018){Johnson}, {Harrison}, {Swinbank}, {Tiley},
  {Stott}, {Bower}, {Smail}, {Bunker}, {Sobral}, {Turner}, {Best}, {Bureau},
  {Cirasuolo}, {Jarvis}, {Magdis}, {Sharples}, {Bland-Hawthorn}, {Catinella},
  {Cortese}, {Croom}, {Federrath}, {Glazebrook}, {Sweet}, {Bryant}, {Goodwin},
  {Konstantopoulos}, {Lawrence}, {Medling}, {Owers}, \&
  {Richards}}]{Johnson2018}
{Johnson}, H.~L., {Harrison}, C.~M., {Swinbank}, A.~M., {et~al.} 2018, \mnras,
  474, 5076, \dodoi{10.1093/mnras/stx3016}

\bibitem[{{Jones} {et~al.}(2010){Jones}, {Swinbank}, {Ellis}, {Richard}, \&
  {Stark}}]{Jones2010}
{Jones}, T.~A., {Swinbank}, A.~M., {Ellis}, R.~S., {Richard}, J., \& {Stark},
  D.~P. 2010, \mnras, 404, 1247, \dodoi{10.1111/j.1365-2966.2010.16378.x}

\bibitem[{{Kaasinen} {et~al.}(2020){Kaasinen}, {Walter}, {Novak}, {Neeleman},
  {Smail}, {Boogaard}, {Cunha}, {Weiss}, {Liu}, {Decarli}, {Popping},
  {Diaz-Santos}, {Cort{\'e}s}, {Aravena}, {Werf}, {Riechers}, {Inami}, {Hodge},
  {Rix}, \& {Cox}}]{Kaasinen2020}
{Kaasinen}, M., {Walter}, F., {Novak}, M., {et~al.} 2020, \apj, 899, 37,
  \dodoi{10.3847/1538-4357/aba438}

\bibitem[{{Kassin} {et~al.}(2012){Kassin}, {Weiner}, {Faber}, {Gardner},
  {Willmer}, {Coil}, {Cooper}, {Devriendt}, {Dutton}, {Guhathakurta}, {Koo},
  {Metevier}, {Noeske}, \& {Primack}}]{Kassin2012}
{Kassin}, S.~A., {Weiner}, B.~J., {Faber}, S.~M., {et~al.} 2012, \apj, 758,
  106, \dodoi{10.1088/0004-637X/758/2/106}

\bibitem[{{Kriek} {et~al.}(2009){Kriek}, {van Dokkum}, {Labb{\'e}}, {Franx},
  {Illingworth}, {Marchesini}, \& {Quadri}}]{Kriek2009}
{Kriek}, M., {van Dokkum}, P.~G., {Labb{\'e}}, I., {et~al.} 2009, \apj, 700,
  221, \dodoi{10.1088/0004-637X/700/1/221}

\bibitem[{{Kriek} {et~al.}(2018){Kriek}, {van Dokkum}, {Labb{\'e}}, {Franx},
  {Illingworth}, {Marchesini}, {Quadri}, {Aird}, {Coil}, \&
  {Georgakakis}}]{FAST}
---. 2018, {FAST: Fitting and Assessment of Synthetic Templates}.
\newblock \doeprint{1803.008}

\bibitem[{{Krist} {et~al.}(2011){Krist}, {Hook}, \& {Stoehr}}]{TinyTim}
{Krist}, J.~E., {Hook}, R.~N., \& {Stoehr}, F. 2011, in Society of
  Photo-Optical Instrumentation Engineers (SPIE) Conference Series, Vol. 8127,
  Optical Modeling and Performance Predictions V, ed. M.~A. {Kahan}, 81270J,
  \dodoi{10.1117/12.892762}

\bibitem[{{Krumholz} \& {Burkert}(2010)}]{Krumholz2010c}
{Krumholz}, M., \& {Burkert}, A. 2010, \apj, 724, 895,
  \dodoi{10.1088/0004-637X/724/2/895}

\bibitem[{{Krumholz} \& {Burkhart}(2016)}]{Krumholz2016a}
{Krumholz}, M.~R., \& {Burkhart}, B. 2016, \mnras, 458, 1671,
  \dodoi{10.1093/mnras/stw434}

\bibitem[{{Krumholz} {et~al.}(2018){Krumholz}, {Burkhart}, {Forbes}, \&
  {Crocker}}]{Krumholz2018b}
{Krumholz}, M.~R., {Burkhart}, B., {Forbes}, J.~C., \& {Crocker}, R.~M. 2018,
  \mnras, 477, 2716, \dodoi{10.1093/mnras/sty852}

\bibitem[{{Lang} {et~al.}(2014){Lang}, {Wuyts}, {Somerville}, {F{\"o}rster
  Schreiber}, {Genzel}, {Bell}, {Brammer}, {Dekel}, {Faber}, {Ferguson},
  {Grogin}, {Kocevski}, {Koekemoer}, {Lutz}, {McGrath}, {Momcheva}, {Nelson},
  {Primack}, {Rosario}, {Skelton}, {Tacconi}, {van Dokkum}, \&
  {Whitaker}}]{Lang2014}
{Lang}, P., {Wuyts}, S., {Somerville}, R.~S., {et~al.} 2014, \apj, 788, 11,
  \dodoi{10.1088/0004-637X/788/1/11}

\bibitem[{{Lang} {et~al.}(2017){Lang}, {F{\"o}rster Schreiber}, {Genzel},
  {Wuyts}, {Wisnioski}, {Beifiori}, {Belli}, {Bender}, {Brammer}, {Burkert},
  {Chan}, {Davies}, {Fossati}, {Galametz}, {Kulkarni}, {Lutz}, {Mendel},
  {Momcheva}, {Naab}, {Nelson}, {Saglia}, {Seitz}, {Tacchella}, {Tacconi},
  {Tadaki}, {{\"U}bler}, {van Dokkum}, \& {Wilman}}]{Lang2017}
{Lang}, P., {F{\"o}rster Schreiber}, N.~M., {Genzel}, R., {et~al.} 2017, \apj,
  840, 92, \dodoi{10.3847/1538-4357/aa6d82}

\bibitem[{{Law} {et~al.}(2009){Law}, {Steidel}, {Erb}, {Larkin}, {Pettini},
  {Shapley}, \& {Wright}}]{Law2009}
{Law}, D.~R., {Steidel}, C.~C., {Erb}, D.~K., {et~al.} 2009, \apj, 697, 2057,
  \dodoi{10.1088/0004-637X/697/2/2057}

\bibitem[{{Leethochawalit} {et~al.}(2016){Leethochawalit}, {Jones}, {Ellis},
  {Stark}, {Richard}, {Zitrin}, \& {Auger}}]{Leethochawalit2016}
{Leethochawalit}, N., {Jones}, T.~A., {Ellis}, R.~S., {et~al.} 2016, \apj, 820,
  84, \dodoi{10.3847/0004-637X/820/2/84}

\bibitem[{{Liu} {et~al.}(2019){Liu}, {Schinnerer}, {Groves}, {Magnelli},
  {Lang}, {Leslie}, {Jim{\'e}nez-Andrade}, {Riechers}, {Popping}, {Magdis},
  {Daddi}, {Sargent}, {Gao}, {Fudamoto}, {Oesch}, \& {Bertoldi}}]{Liudz2019b}
{Liu}, D., {Schinnerer}, E., {Groves}, B., {et~al.} 2019, \apj, 887, 235,
  \dodoi{10.3847/1538-4357/ab578d}

\bibitem[{{Livermore} {et~al.}(2015){Livermore}, {Jones}, {Richard}, {Bower},
  {Swinbank}, {Yuan}, {Edge}, {Ellis}, {Kewley}, {Smail}, {Coppin}, \&
  {Ebeling}}]{Livermore2015}
{Livermore}, R.~C., {Jones}, T.~A., {Richard}, J., {et~al.} 2015, \mnras, 450,
  1812, \dodoi{10.1093/mnras/stv686}

\bibitem[{{Ludlow} {et~al.}(2014){Ludlow}, {Navarro}, {Angulo},
  {Boylan-Kolchin}, {Springel}, {Frenk}, \& {White}}]{Ludlow2014}
{Ludlow}, A.~D., {Navarro}, J.~F., {Angulo}, R.~E., {et~al.} 2014, \mnras, 441,
  378, \dodoi{10.1093/mnras/stu483}

\bibitem[{{Martig} {et~al.}(2009){Martig}, {Bournaud}, {Teyssier}, \&
  {Dekel}}]{Martig2009}
{Martig}, M., {Bournaud}, F., {Teyssier}, R., \& {Dekel}, A. 2009, \apj, 707,
  250, \dodoi{10.1088/0004-637X/707/1/250}

\bibitem[{{Martizzi} {et~al.}(2012){Martizzi}, {Teyssier}, {Moore}, \&
  {Wentz}}]{Martizzi2012}
{Martizzi}, D., {Teyssier}, R., {Moore}, B., \& {Wentz}, T. 2012, \mnras, 422,
  3081, \dodoi{10.1111/j.1365-2966.2012.20879.x}

\bibitem[{{Molina} {et~al.}(2019){Molina}, {Ibar}, {Smail}, {Swinbank},
  {Villard}, {Escala}, {Sobral}, \& {Hughes}}]{Molina2019}
{Molina}, J., {Ibar}, E., {Smail}, I., {et~al.} 2019, \mnras, 487, 4856,
  \dodoi{10.1093/mnras/stz1643}

\bibitem[{{Moster} {et~al.}(2018){Moster}, {Naab}, \& {White}}]{Moster2018}
{Moster}, B.~P., {Naab}, T., \& {White}, S. D.~M. 2018, \mnras, 477, 1822,
  \dodoi{10.1093/mnras/sty655}

\bibitem[{{Moster} {et~al.}(2020){Moster}, {Naab}, \& {White}}]{Moster2020}
---. 2020, \mnras, 499, 4748, \dodoi{10.1093/mnras/staa3019}

\bibitem[{{Navarro} {et~al.}(1996){Navarro}, {Frenk}, \& {White}}]{NFW}
{Navarro}, J.~F., {Frenk}, C.~S., \& {White}, S. D.~M. 1996, \apj, 462, 563,
  \dodoi{10.1086/177173}

\bibitem[{{Nestor Shachar} {et~al.}(2022){Nestor Shachar}, {Price},
  {F{\"o}rster Schreiber}, {Genzel}, {Shimizu}, {Tacconi}, {{\"U}bler},
  {Burkert}, {Davies}, {Deke}, {Herrera-Camus}, {Lee}, {Liu}, {Lutz}, {Naab},
  {Neri}, {Renzini}, {Saglia}, {Schuster}, {Sternberg}, {Wisnioski}, \&
  {Wuyts}}]{Nestor2022}
{Nestor Shachar}, A., {Price}, S.~H., {F{\"o}rster Schreiber}, N.~M., {et~al.}
  2022, arXiv e-prints, arXiv:2209.12199.
\newblock \doarXiv{2209.12199}

\bibitem[{{Ogiya} \& {Nagai}(2022)}]{Ogiya2022}
{Ogiya}, G., \& {Nagai}, D. 2022, \mnras, 514, 555,
  \dodoi{10.1093/mnras/stac1311}

\bibitem[{{Oguri}(2010{\natexlab{a}})}]{Oguri2010}
{Oguri}, M. 2010{\natexlab{a}}, \pasj, 62, 1017, \dodoi{10.1093/pasj/62.4.1017}

\bibitem[{{Oguri}(2010{\natexlab{b}})}]{Oguri2010asclsoft}
---. 2010{\natexlab{b}}, {glafic: Software Package for Analyzing Gravitational
  Lensing}.
\newblock \doeprint{1010.012}

\bibitem[{{Patr{\'\i}cio} {et~al.}(2018){Patr{\'\i}cio}, {Richard}, {Carton},
  {Contini}, {Epinat}, {Brinchmann}, {Schmidt}, {Krajnovi{\'c}}, {Bouch{\'e}},
  {Weilbacher}, {Pell{\'o}}, {Caruana}, {Maseda}, {Finley}, {Bauer},
  {Martinez}, {Mahler}, {Lagattuta}, {Cl{\'e}ment}, {Soucail}, \&
  {Wisotzki}}]{Patricio2018}
{Patr{\'\i}cio}, V., {Richard}, J., {Carton}, D., {et~al.} 2018, \mnras, 477,
  18, \dodoi{10.1093/mnras/sty555}

\bibitem[{{Peirani} {et~al.}(2017){Peirani}, {Dubois}, {Volonteri},
  {Devriendt}, {Bundy}, {Silk}, {Pichon}, {Kaviraj}, {Gavazzi}, \&
  {Habouzit}}]{Peirani2017}
{Peirani}, S., {Dubois}, Y., {Volonteri}, M., {et~al.} 2017, \mnras, 472, 2153,
  \dodoi{10.1093/mnras/stx2099}

\bibitem[{{Peng} {et~al.}(2002){Peng}, {Ho}, {Impey}, \& {Rix}}]{Peng2002}
{Peng}, C.~Y., {Ho}, L.~C., {Impey}, C.~D., \& {Rix}, H.-W. 2002, \aj, 124,
  266, \dodoi{10.1086/340952}

\bibitem[{{Peng} {et~al.}(2010){Peng}, {Ho}, {Impey}, \& {Rix}}]{Peng2010}
---. 2010, \aj, 139, 2097, \dodoi{10.1088/0004-6256/139/6/2097}

\bibitem[{{Peng} \& {Renzini}(2020)}]{PengYJ2020}
{Peng}, Y.-j., \& {Renzini}, A. 2020, \mnras, 491, L51,
  \dodoi{10.1093/mnrasl/slz163}

\bibitem[{{Price} {et~al.}(2021){Price}, {Shimizu}, {Genzel}, {{\"U}bler},
  {F{\"o}rster Schreiber}, {Tacconi}, {Davies}, {Coogan}, {Lutz}, {Wuyts},
  {Wisnioski}, {Nestor}, {Sternberg}, {Burkert}, {Bender}, {Contursi},
  {Davies}, {Herrera-Camus}, {Lee}, {Naab}, {Neri}, {Renzini}, {Saglia},
  {Schruba}, \& {Schuster}}]{Price2021}
{Price}, S.~H., {Shimizu}, T.~T., {Genzel}, R., {et~al.} 2021, \apj, 922, 143,
  \dodoi{10.3847/1538-4357/ac22ad}

\bibitem[{{Privon} {et~al.}(2017){Privon}, {Aalto}, {Falstad}, {Muller},
  {Gonz{\'a}lez-Alfonso}, {Sliwa}, {Treister}, {Costagliola}, {Armus}, {Evans},
  {Garcia-Burillo}, {Izumi}, {Sakamoto}, {van der Werf}, \& {Chu}}]{Privon2017}
{Privon}, G.~C., {Aalto}, S., {Falstad}, N., {et~al.} 2017, \apj, 835, 213,
  \dodoi{10.3847/1538-4357/835/2/213}

\bibitem[{{Rhoads} {et~al.}(2014){Rhoads}, {Malhotra}, {Allam}, {Carilli},
  {Combes}, {Finkelstein}, {Finkelstein}, {Frye}, {Gerin}, {Guillard},
  {Nesvadba}, {Rigby}, {Spaans}, \& {Strauss}}]{Rhoads2014}
{Rhoads}, J.~E., {Malhotra}, S., {Allam}, S., {et~al.} 2014, \apj, 787, 8,
  \dodoi{10.1088/0004-637X/787/1/8}

\bibitem[{{Rizzo} {et~al.}(2021){Rizzo}, {Vegetti}, {Fraternali}, {Stacey}, \&
  {Powell}}]{Rizzo2021}
{Rizzo}, F., {Vegetti}, S., {Fraternali}, F., {Stacey}, H.~R., \& {Powell}, D.
  2021, \mnras, 507, 3952, \dodoi{10.1093/mnras/stab2295}

\bibitem[{{Rizzo} {et~al.}(2020){Rizzo}, {Vegetti}, {Powell}, {Fraternali},
  {McKean}, {Stacey}, \& {White}}]{Rizzo2020}
{Rizzo}, F., {Vegetti}, S., {Powell}, D., {et~al.} 2020, \nat, 584, 201,
  \dodoi{10.1038/s41586-020-2572-6}

\bibitem[{{Rybak} {et~al.}(2019){Rybak}, {Calistro Rivera}, {Hodge}, {Smail},
  {Walter}, {van der Werf}, {da Cunha}, {Chen}, {Dannerbauer}, {Ivison},
  {Karim}, {Simpson}, {Swinbank}, \& {Wardlow}}]{Rybak2019}
{Rybak}, M., {Calistro Rivera}, G., {Hodge}, J.~A., {et~al.} 2019, \apj, 876,
  112, \dodoi{10.3847/1538-4357/ab0e0f}

\bibitem[{{Saintonge} {et~al.}(2013){Saintonge}, {Lutz}, {Genzel}, {Magnelli},
  {Nordon}, {Tacconi}, {Baker}, {Bandara}, {Berta}, {F{\"o}rster Schreiber},
  {Poglitsch}, {Sturm}, {Wuyts}, \& {Wuyts}}]{Saintonge2013}
{Saintonge}, A., {Lutz}, D., {Genzel}, R., {et~al.} 2013, \apj, 778, 2,
  \dodoi{10.1088/0004-637X/778/1/2}

\bibitem[{Salvatier {et~al.}(2016)Salvatier, Wiecki, \& Fonnesbeck}]{pymc3}
Salvatier, J., Wiecki, T.~V., \& Fonnesbeck, C. 2016, PeerJ Comput. Sci., 2,
  e55, \dodoi{10.7717/peerj-cs.55}

\bibitem[{{Sharma} {et~al.}(2018){Sharma}, {Richard}, {Yuan}, {Gupta},
  {Kewley}, {Patr{\'\i}cio}, {Leethochawalit}, \& {Jones}}]{Sharma2018}
{Sharma}, S., {Richard}, J., {Yuan}, T., {et~al.} 2018, \mnras, 481, 1427,
  \dodoi{10.1093/mnras/sty2352}

\bibitem[{{Sharon} {et~al.}(2019){Sharon}, {Tagore}, {Baker}, {Rivera},
  {Keeton}, {Lutz}, {Genzel}, {Wilner}, {Hicks}, {Allam}, \&
  {Tucker}}]{Sharon2019}
{Sharon}, C.~E., {Tagore}, A.~S., {Baker}, A.~J., {et~al.} 2019, \apj, 879, 52,
  \dodoi{10.3847/1538-4357/ab22b9}

\bibitem[{{Simons} {et~al.}(2017){Simons}, {Kassin}, {Weiner}, {Faber},
  {Trump}, {Heckman}, {Koo}, {Pacifici}, {Primack}, {Snyder}, \& {de la
  Vega}}]{Simons2017}
{Simons}, R.~C., {Kassin}, S.~A., {Weiner}, B.~J., {et~al.} 2017, \apj, 843,
  46, \dodoi{10.3847/1538-4357/aa740c}

\bibitem[{{Speagle} {et~al.}(2014){Speagle}, {Steinhardt}, {Capak}, \&
  {Silverman}}]{Speagle2014}
{Speagle}, J.~S., {Steinhardt}, C.~L., {Capak}, P.~L., \& {Silverman}, J.~D.
  2014, \apjs, 214, 15, \dodoi{10.1088/0067-0049/214/2/15}

\bibitem[{{Tacconi} {et~al.}(2020){Tacconi}, {Genzel}, \&
  {Sternberg}}]{Tacconi2020}
{Tacconi}, L.~J., {Genzel}, R., \& {Sternberg}, A. 2020, \araa, 58, 157,
  \dodoi{10.1146/annurev-astro-082812-141034}

\bibitem[{{Tacconi} {et~al.}(2006){Tacconi}, {Neri}, {Chapman}, {Genzel},
  {Smail}, {Ivison}, {Bertoldi}, {Blain}, {Cox}, {Greve}, \&
  {Omont}}]{Tacconi2006}
{Tacconi}, L.~J., {Neri}, R., {Chapman}, S.~C., {et~al.} 2006, \apj, 640, 228,
  \dodoi{10.1086/499933}

\bibitem[{{Tacconi} {et~al.}(2008){Tacconi}, {Genzel}, {Smail}, {Neri},
  {Chapman}, {Ivison}, {Blain}, {Cox}, {Omont}, {Bertoldi}, {Greve},
  {F{\"o}rster Schreiber}, {Genel}, {Lutz}, {Swinbank}, {Shapley}, {Erb},
  {Cimatti}, {Daddi}, \& {Baker}}]{Tacconi2008}
{Tacconi}, L.~J., {Genzel}, R., {Smail}, I., {et~al.} 2008, \apj, 680, 246,
  \dodoi{10.1086/587168}

\bibitem[{{Tacconi} {et~al.}(2013){Tacconi}, {Neri}, {Genzel}, {Combes},
  {Bolatto}, {Cooper}, {Wuyts}, {Bournaud}, {Burkert}, {Comerford}, {Cox},
  {Davis}, {F{\"o}rster Schreiber}, {Garc{\'{\i}}a-Burillo}, {Gracia-Carpio},
  {Lutz}, {Naab}, {Newman}, {Omont}, {Saintonge}, {Shapiro Griffin}, {Shapley},
  {Sternberg}, \& {Weiner}}]{Tacconi2013}
{Tacconi}, L.~J., {Neri}, R., {Genzel}, R., {et~al.} 2013, \apj, 768, 74,
  \dodoi{10.1088/0004-637X/768/1/74}

\bibitem[{{Tacconi} {et~al.}(2018){Tacconi}, {Genzel}, {Saintonge}, {Combes},
  {Garc{\'\i}a-Burillo}, {Neri}, {Bolatto}, {Contini}, {F{\"o}rster Schreiber},
  {Lilly}, {Lutz}, {Wuyts}, {Accurso}, {Boissier}, {Boone}, {Bouch{\'e}},
  {Bournaud}, {Burkert}, {Carollo}, {Cooper}, {Cox}, {Feruglio}, {Freundlich},
  {Herrera-Camus}, {Juneau}, {Lippa}, {Naab}, {Renzini}, {Salome}, {Sternberg},
  {Tadaki}, {{\"U}bler}, {Walter}, {Weiner}, \& {Weiss}}]{Tacconi2018}
{Tacconi}, L.~J., {Genzel}, R., {Saintonge}, A., {et~al.} 2018, \apj, 853, 179,
  \dodoi{10.3847/1538-4357/aaa4b4}

\bibitem[{{Tagore}(2014)}]{Tagore2014}
{Tagore}, A.~S. 2014, PhD thesis, Rutgers The State University of New Jersey -
  New Brunswick

\bibitem[{{THE CASA TEAM} {et~al.}(2022){THE CASA TEAM}, {Bean}, {Bhatnagar},
  {Castro}, {Donovan Meyer}, {Emonts}, {Garcia}, {Garwood}, {Golap}, {Gonzalez
  Villalba}, {Harris}, {Hayashi}, {Hoskins}, {Hsieh}, {Jagannathan},
  {Kawasaki}, {Keimpema}, {Kettenis}, {Lopez}, {Marvil}, {Masters},
  {McNichols}, {Mehringer}, {Miel}, {Moellenbrock}, {Montesino}, {Nakazato},
  {Ott}, {Petry}, {Pokorny}, {Raba}, {Rau}, {Schiebel}, {Schweighart},
  {Sekhar}, {Shimada}, {Small}, {Steeb}, {Sugimoto}, {Suoranta}, {Tsutsumi},
  {van Bemmel}, {Verkouter}, {Wells}, {Xiong}, {Szomoru}, {Griffith},
  {Glendenning}, \& {Kern}}]{CASA}
{THE CASA TEAM}, {Bean}, B., {Bhatnagar}, S., {et~al.} 2022, arXiv e-prints,
  arXiv:2210.02276.
\newblock \doarXiv{2210.02276}

\bibitem[{{Tokuoka} {et~al.}(2022){Tokuoka}, {Inoue}, {Hashimoto}, {Ellis},
  {Laporte}, {Sugahara}, {Matsuo}, {Tamura}, {Fudamoto}, {Moriwaki},
  {Roberts-Borsani}, {Shimizu}, {Yamanaka}, {Yoshida}, {Zackrisson}, \&
  {Zheng}}]{Tokuoka2022}
{Tokuoka}, T., {Inoue}, A.~K., {Hashimoto}, T., {et~al.} 2022, \apjl, 933, L19,
  \dodoi{10.3847/2041-8213/ac7447}

\bibitem[{{Toomre}(1964)}]{Toomre1964}
{Toomre}, A. 1964, \apj, 139, 1217, \dodoi{10.1086/147861}

\bibitem[{{Turner} {et~al.}(2017){Turner}, {Cirasuolo}, {Harrison}, {McLure},
  {Dunlop}, {Swinbank}, {Johnson}, {Sobral}, {Matthee}, \&
  {Sharples}}]{Turner2017}
{Turner}, O.~J., {Cirasuolo}, M., {Harrison}, C.~M., {et~al.} 2017, \mnras,
  471, 1280, \dodoi{10.1093/mnras/stx1366}

\bibitem[{{{\"U}bler} {et~al.}(2017){{\"U}bler}, {F{\"o}rster Schreiber},
  {Genzel}, {Wisnioski}, {Wuyts}, {Lang}, {Naab}, {Burkert}, {van Dokkum},
  {Tacconi}, {Wilman}, {Fossati}, {Mendel}, {Beifiori}, {Belli}, {Bender},
  {Brammer}, {Chan}, {Davies}, {Fabricius}, {Galametz}, {Lutz}, {Momcheva},
  {Nelson}, {Saglia}, {Seitz}, \& {Tadaki}}]{Ubler2017}
{{\"U}bler}, H., {F{\"o}rster Schreiber}, N.~M., {Genzel}, R., {et~al.} 2017,
  \apj, 842, 121, \dodoi{10.3847/1538-4357/aa7558}

\bibitem[{{{\"U}bler} {et~al.}(2018){{\"U}bler}, {Genzel}, {Tacconi},
  {F{\"o}rster Schreiber}, {Neri}, {Contursi}, {Belli}, {Nelson}, {Lang},
  {Shimizu}, {Davies}, {Herrera-Camus}, {Lutz}, {Plewa}, {Price}, {Schuster},
  {Sternberg}, {Tadaki}, {Wisnioski}, \& {Wuyts}}]{Ubler2018}
{{\"U}bler}, H., {Genzel}, R., {Tacconi}, L.~J., {et~al.} 2018, \apjl, 854,
  L24, \dodoi{10.3847/2041-8213/aaacfa}

\bibitem[{{{\"U}bler} {et~al.}(2019){{\"U}bler}, {Genzel}, {Wisnioski},
  {F{\"o}rster Schreiber}, {Shimizu}, {Price}, {Tacconi}, {Belli}, {Wilman},
  {Fossati}, {Mendel}, {Davies}, {Beifiori}, {Bender}, {Brammer}, {Burkert},
  {Chan}, {Davies}, {Fabricius}, {Galametz}, {Herrera-Camus}, {Lang}, {Lutz},
  {Momcheva}, {Naab}, {Nelson}, {Saglia}, {Tadaki}, {van Dokkum}, \&
  {Wuyts}}]{Ubler2019}
{{\"U}bler}, H., {Genzel}, R., {Wisnioski}, E., {et~al.} 2019, \apj, 880, 48,
  \dodoi{10.3847/1538-4357/ab27cc}

\bibitem[{{{\"U}bler} {et~al.}(2021){{\"U}bler}, {Genel}, {Sternberg},
  {Genzel}, {Price}, {F{\"o}rster Schreiber}, {Shimizu}, {Pillepich}, {Nelson},
  {Burkert}, {Davies}, {Hernquist}, {Lang}, {Lutz}, {Pakmor}, \&
  {Tacconi}}]{Ubler2021}
{{\"U}bler}, H., {Genel}, S., {Sternberg}, A., {et~al.} 2021, \mnras, 500,
  4597, \dodoi{10.1093/mnras/staa3464}

\bibitem[{Virtanen {et~al.}(2020)Virtanen, Gommers, Oliphant, Haberland, Reddy,
  Cournapeau, Burovski, Peterson, Weckesser, Bright, {van der Walt}, Brett,
  Wilson, Millman, Mayorov, Nelson, Jones, Kern, Larson, Carey, Polat, Feng,
  Moore, {VanderPlas}, Laxalde, Perktold, Cimrman, Henriksen, Quintero, Harris,
  Archibald, Ribeiro, Pedregosa, {van Mulbregt}, \& {SciPy 1.0
  Contributors}}]{scipy}
Virtanen, P., Gommers, R., Oliphant, T.~E., {et~al.} 2020, Nature Methods, 17,
  261, \dodoi{10.1038/s41592-019-0686-2}

\bibitem[{{Ward} {et~al.}(2003){Ward}, {Zmuidzinas}, {Harris}, \&
  {Isaak}}]{Ward2003}
{Ward}, J.~S., {Zmuidzinas}, J., {Harris}, A.~I., \& {Isaak}, K.~G. 2003, \apj,
  587, 171, \dodoi{10.1086/368175}

\bibitem[{{Wisnioski} {et~al.}(2015){Wisnioski}, {F{\"o}rster Schreiber},
  {Wuyts}, {Wuyts}, {Bandara}, {Wilman}, {Genzel}, {Bender}, {Davies},
  {Fossati}, {Lang}, {Mendel}, {Beifiori}, {Brammer}, {Chan}, {Fabricius},
  {Fudamoto}, {Kulkarni}, {Kurk}, {Lutz}, {Nelson}, {Momcheva}, {Rosario},
  {Saglia}, {Seitz}, {Tacconi}, \& {van Dokkum}}]{Wisnioski2015}
{Wisnioski}, E., {F{\"o}rster Schreiber}, N.~M., {Wuyts}, S., {et~al.} 2015,
  \apj, 799, 209, \dodoi{10.1088/0004-637X/799/2/209}

\bibitem[{{Wuyts} {et~al.}(2011){Wuyts}, {F{\"o}rster Schreiber}, {Lutz},
  {Nordon}, {Berta}, {Altieri}, {Andreani}, {Aussel}, {Bongiovanni}, {Cepa},
  {Cimatti}, {Daddi}, {Elbaz}, {Genzel}, {Koekemoer}, {Magnelli}, {Maiolino},
  {McGrath}, {P{\'e}rez Garc{\'\i}a}, {Poglitsch}, {Popesso}, {Pozzi},
  {Sanchez-Portal}, {Sturm}, {Tacconi}, \& {Valtchanov}}]{Wuyts2011a}
{Wuyts}, S., {F{\"o}rster Schreiber}, N.~M., {Lutz}, D., {et~al.} 2011, \apj,
  738, 106, \dodoi{10.1088/0004-637X/738/1/106}

\bibitem[{{Wuyts} {et~al.}(2012){Wuyts}, {F{\"o}rster Schreiber}, {Genzel},
  {Guo}, {Barro}, {Bell}, {Dekel}, {Faber}, {Ferguson}, {Giavalisco}, {Grogin},
  {Hathi}, {Huang}, {Kocevski}, {Koekemoer}, {Koo}, {Lotz}, {Lutz}, {McGrath},
  {Newman}, {Rosario}, {Saintonge}, {Tacconi}, {Weiner}, \& {van der
  Wel}}]{Wuyts2012}
{Wuyts}, S., {F{\"o}rster Schreiber}, N.~M., {Genzel}, R., {et~al.} 2012, \apj,
  753, 114, \dodoi{10.1088/0004-637X/753/2/114}

\bibitem[{{Wuyts} {et~al.}(2016){Wuyts}, {F{\"o}rster Schreiber}, {Wisnioski},
  {Genzel}, {Burkert}, {Bandara}, {Beifiori}, {Belli}, {Bender}, {Brammer},
  {Chan}, {Davies}, {Fossati}, {Galametz}, {Kulkarni}, {Lang}, {Lutz},
  {Mendel}, {Momcheva}, {Naab}, {Nelson}, {Saglia}, {Seitz}, {Tacconi},
  {Tadaki}, {{\"U}bler}, {van Dokkum}, {Wilman}, \& {Wuyts}}]{Wuyts2016}
{Wuyts}, S., {F{\"o}rster Schreiber}, N.~M., {Wisnioski}, E., {et~al.} 2016,
  \apj, 831, 149, \dodoi{10.3847/0004-637X/831/2/149}

\bibitem[{{Yuan} {et~al.}(2020){Yuan}, {Elagali}, {Labb{\'e}}, {Kacprzak},
  {Lagos}, {Alcorn}, {Cohn}, {Tran}, {Glazebrook}, {Groves}, {Freeman},
  {Spitler}, {Straatman}, {Fisher}, \& {Sweet}}]{Yuan2020}
{Yuan}, T., {Elagali}, A., {Labb{\'e}}, I., {et~al.} 2020, Nature Astronomy, 4,
  957, \dodoi{10.1038/s41550-020-1102-7}

\bibitem[{{Zhang} {et~al.}(2018){Zhang}, {Romano}, {Ivison}, {Papadopoulos}, \&
  {Matteucci}}]{Zhang2018}
{Zhang}, Z.-Y., {Romano}, D., {Ivison}, R.~J., {Papadopoulos}, P.~P., \&
  {Matteucci}, F. 2018, \nat, 558, 260, \dodoi{10.1038/s41586-018-0196-x}

\end{thebibliography}
